\documentclass[a4paper, traditabstract]{aa}   
\usepackage{graphicx}
\usepackage{comment}
\usepackage{txfonts}
\usepackage{placeins}
\usepackage{float}
\usepackage{multirow}
\usepackage[switch]{lineno}
\usepackage{natbib,twoopt}
\usepackage[breaklinks=true]{hyperref} 
\hypersetup{
  colorlinks   = true, %Colours links instead of ugly boxes
  urlcolor     = red, %Colour for external hyperlinks
  linkcolor    = blue, %Colour of internal links
  citecolor    = blue, %Colour of citations
  breaklinks   = true % to avoid \citeads line fills
}
\usepackage{bm}
\bibpunct{(}{)}{;}{a}{}{,}       
\usepackage[dvipsnames]{xcolor}

\newcommand{\eqreff}{Eq.~\eqref}
\newcommand{\secreff}{Sect.~\ref}

\newcommand{\figreff}{Figure~\ref}
\newcommand{\tabreff}{Table~\ref}

\begin{document} 
    \titlerunning{Weak lensing mass-richness relation of redMaPPer clusters in DC2}
    \authorrunning{Constantin Payerne, Zhuowen Zhang, et al. (LSST DESC)}
   \title{Weak lensing mass-richness relation of redMaPPer clusters\\ in the LSST DESC DC2 simulations}
\author{
Constantin Payerne,\inst{1, 2} \fnmsep\thanks{Corresponding author: \email{constantin.payerne@gmail.com}}
Zhuowen Zhang,\inst{3}
Michel Aguena,\inst{4, 5}
Céline Combet,\inst{2}
Thibault Guillemin,\inst{6}
Marina Ricci,\inst{4}
Nathan Amouroux,\inst{6}
Camille Avestruz,\inst{7, 8}
Eduardo J. Barroso,\inst{6}
Arya Farahi,\inst{9, 10}
Eve Kovacs,\inst{11}
Calum Murray,\inst{4, 12}
Markus M. Rau,\inst{13, 14}
Eli S. Rykoff,\inst{15, 16}
Samuel J. Schmidt,\inst{17}
and the LSST Dark Energy Science Collaboration}

\institute{
Université Paris-Saclay, CEA, IRFU, 91191, Gif-sur-Yvette, France
\and
Université Grenoble Alpes, CNRS, IN2P3, LPSC, 38000 Grenoble, France
\and
Kavli Institute for Cosmological Physics, University of Chicago, Chicago, IL 60637, USA
\and
Université Paris Cité, CNRS, IN2P3, APC, 75013 Paris, France
\and
Italian National Institute of AstroPhysics, Osservatorio Astronomico di Trieste, Italy
\and
Université de Savoie, CNRS, IN2P3, LAPP, Annecy-le-Vieux, France
\and
Department of Physics, University of Michigan, Ann Arbor, MI 48109, USA
\and
Leinweber Center of Theoretical Physics, University of Michigan, Ann Arbor, MI 48109, USA
\and
Department of Statistics and Data Sciences, The University of Texas at Austin, TX 78712, USA
\and
The NSF-Simons AI Institute for Cosmic Origins, University of Texas at Austin, Austin, TX 78712, USA
\and
HEP Division, Argonne National Laboratory, 9700 S. Cass Ave., Lemont, IL 60439, USA
\and
Université Paris-Saclay, Université Paris Cité, CEA, CNRS, AIM, 91191, Gif-sur-Yvette, France
\and
School of Mathematics, Statistics and Physics, Newcastle University, Newcastle upon Tyne, NE17RU, United Kingdom
\and
High Energy Physics Division, Argonne National Laboratory, Lemont, IL 60439, USA
\and
Kavli Institute for Particle Astrophysics \& Cosmology, P. O. Box 2450, Stanford University, Stanford, CA 94305, USA
\and
SLAC National Accelerator Laboratory, Menlo Park, CA 94025, USA
\and
Department of Physics and Astronomy, University of California, One Shields Avenue, Davis, CA 95616, USA
}      
   \date{Received 12 February 2025/ Accepted 08 June 2025}
  \abstract
   {Cluster scaling relations are key ingredients in cluster abundance-based cosmological studies.
In optical cluster cosmology, where clusters are detected through their richness, cluster-weak gravitational lensing has proven to be a powerful tool to constrain the cluster mass-richness relation. This work is conducted as part of the Dark Energy Science Collaboration (DESC), which aims to analyze the Legacy Survey of Space and Time (LSST) of the Vera C. Rubin Observatory, starting in 2026.}
{Weak lensing-inferred cluster properties, such as mass, suffer from several sources of bias. In this paper, we aim to test the impact of modeling choices and observational systematics in cluster lensing on the inference of the mass-richness relation.}
   {We constrain the mass-richness relation of 3,600 clusters detected by the redMaPPer algorithm in the cosmoDC2 extra-galactic mock catalog of the LSST DESC DC2 simulation, covering $440$~deg$^2$, using number count measurements and either stacked weak lensing profiles or mean cluster masses in several intervals of richness ($20 \leq \lambda \leq 200$) and redshift ($0.2 \leq z \leq 1$).}
   { We provide the first constraints of the redMaPPer cluster mass-richness relation detected in cosmoDC2. We find that, for an LSST-like source galaxy density, our constraints are robust to changes in the concentration–mass relation and dark matter density profile modeling choices, when source redshifts and shapes are perfectly known. We find that photometric redshift uncertainties can introduce bias at the $1\sigma$ level, which can be mitigated by an overall correction factor, fitted jointly with the scaling parameters. We find that including positive shear–richness covariance in the fit shifts the results by up to $0.5\sigma$. Our constraints also compare fairly to a fiducial mass-richness relation, obtained from matching cosmoDC2 halo masses to redMaPPer-detected cluster richnesses.}
   {}
   \keywords{Galaxies: clusters: general - Gravitational lensing: weak –methods: statistical}
\maketitle
 \section{Introduction}
\label{sec:introduction}
Galaxy clusters have been essential in the construction of the standard model of cosmology, providing some of the first evidence of dark matter \citep{Zwicky1937Coma} through the motions of galaxies within clusters. Their spatial distribution has also provided evidence for the primordial origin of density fluctuations \citep{kaiser1984spatial}.
They originate from the gravitational collapse of large matter overdensities that decouple from the cosmic expansion and form galaxy clusters. They represent the most massive structures in the Universe held together by gravity. Their formation history, as well as their spatial and mass distributions, are strongly dependent on the nature of gravity, the growth rate of large-scale structures, and the Universe's expansion history \citep{Bartlett1997clusterabundance,Allen2011review,Kravtsov2012review}.

Over the years, cluster abundance has proven to provide competitive constraints\footnote{especially on the total amount of matter in the Universe $\Omega_m$ and the amplitude of matter fluctuations $\sigma_8$.}, complementary to other large-scale structure and geometrical probes, as well as with the analysis of the Cosmic Microwave Background, e.g., using X-ray clusters detected by the ROSAT All-Sky Survey \citep{Mantz2014WTGCL}, XMM-Newton \citep{Pacaud2018XXLCL}, or eROSITA \citep{Ghirardini2024eROSITACL}; clusters detected through their Sunyaev–Zeldovich effect by the \textit{Planck} satellite \citep{Ade2014PlanckCL,Ade2016PlanckCL}, the South Pole Telescope \citep[SPT,][]{Bocquet2024SPTCL,Vogt2024SPTCL}, or the Atacama Cosmology Telescope  \citep[ACT,][]{Hasselfield2013ACTCL}; and also using clusters detected through their galaxy member populations, such as by the Dark Energy Survey \citep{Abbott2020DESCL,DEScollabY3CL}, the Kilo-Degree Survey \citep[KiDS,][]{Lesci2022KIDSCL}, the Sloan Digital Sky Survey \citep[SDSS,][]{Fumagalli2023SDSS}, or with shear-selected clusters from HSC \citep{Hiu2024HSCCL} (see Table 2 in \citet{Payerne2024unbinnedSSC} that recaps the cluster abundance-based cosmological analyses before June 2024).

The constraining power of cluster number counts is currently limited by our understanding of the cluster scaling relations \citep[e.g.,][]{Pratt2019massrichness,Costanzi2019SDSSCL,Abbott2020DESCL}, which denote the statistical relationship between cluster observables and their total masses. At optical wavelengths, clusters are often detected through the color and brightness of their member galaxies \citep{Rykoff2014redmapper}. Still, clusters of galaxies can also be identified as high signal-to-noise ratio peaks on weak-lensing aperture mass maps \citep{Hetterscheidt2005shearselectedclusters,Chen2024shearselectedclusters}.

Weak gravitational lensing has become a robust tool for constraining cluster masses \citep[e.g.,][]{McClintock2019masscalibration,Umetsu2020clusterlensing, Murray2022lensingmasses,Mistele2024lensing,Grandis2024lensing}, through the coherent distortion of the shapes of background galaxies, caused by the bending of the light path due to the cluster's gravitational field. 
Cluster abundance and cluster weak lensing information (using either masses or profiles directly) are usually combined, as they display different degeneracies on scaling relation parameters and cosmology, which allows for tighter constraints on both cosmological parameters and cluster scaling relations \citep{Mantz2014WTGCL,Murata2019HSCrichnessmassrelation,Mulroy2019locuss,Abbott2020DESCL,Lesci2022KIDSCL,Sunayama2023HSClensing}. However, the mapping between the measurements of weak gravitational lensing and cluster masses is not well understood, as it is affected by several sources of statistical noise and systematic uncertainties \citep[e.g.,][]{Becker2011modeling,Kohlinger2015lensingbias,Grandis2021lensingbias}.

First, weak lensing cluster mass reconstruction is statistically limited by the finite galaxy number density. Ongoing surveys such as DES, KiDS, HSC, and the \textit{Euclid} mission \citep{Euclid2011whitepaper}, as well as future wide optical surveys like the Legacy Survey of Space and Time\footnote{\url{http://www.lsst.org}} \citep[LSST,][]{LSST2009whitepaper} and the Nancy Grace Roman Space Telescope \citep{Spergel2015roman}, provide or will provide a large influx of data to reduce statistical uncertainties in lensing measurements to an unprecedented level. Other sources of statistical noise affect the measurement of the cluster lensing signal, including intrinsic variability of cluster morphology, line-of-sight correlated structures, and galaxy shape intrinsic variations.

Second, several systematic biases may affect cluster mass inference. Some are related to "theoretical uncertainties", such as unknowns in the modeling of the dark matter density in the cluster field \citep{Becker2011modeling,Lee2018wlmass}, or the modeling of the presence of baryons \citep{Cromer2022lensingmassbaryons}. Other biases are more directly related to the observations themselves and may arise from shear calibration \citep{Hernandez2020shape}, the calibration of the photometric redshift distribution of the background galaxy sample \citep{Wright2020photoz}, contamination of the source galaxy sample by foreground/cluster member galaxies \citep{Varga2019contamination}, miscentering — possible offset between the cluster's detected center and the center of its gravitational potential \citep{Becker2011modeling,Lee2018wlmass,Zhang2019redmappermiscentering,Sommer2022miscentering} — or selection/projection effects, when galaxies physically unassociated with clusters are counted as members by cluster finders \citep{Myles2021projection_redmapper,Lee2024OpticalGC,Myles2025projectioneffects}.
All of these sources of scatter and bias must be carefully controlled, given the unprecedented volume and quality of lensing data provided by the upcoming large-footprint lensing surveys.

The Vera Rubin Observatory in Chile will conduct the Legacy Survey of Space and Time (LSST), a 10-year wide-field imaging survey that will cover 18,000 deg$^2$ of the southern sky, starting at the end of 2025 \citep{LSST2009whitepaper}. The LSST aims to measure the shapes and redshifts of a few billion galaxies up to $z \sim 3$, and will enable the measurement of number counts and weak gravitational lensing for $\sim 100,000$ galaxy clusters up to $z \sim 1$ \citep{LSST2009whitepaper}. In this context, the \citet{DESC2012wpaper}\footnote{\url{http://lsstdesc.org}} -- DESC -- is preparing the cosmological analysis of the upcoming LSST data.

In this work, we make use of the LSST DESC Data Challenge 2 \citep[DC2,][]{Korytov2019cosmoDC2,Abolfathi2021DC2} simulated dataset. We aim to infer the mass-richness relation of redMaPPer-detected galaxy clusters in DC2, using a combination of cluster weak lensing and abundance. 

This paper is organized as follows. \secreff{sec:formalism} introduces the theoretical background of this work, describing how cluster number counts, stacked lensing profile, and mean mass of a cluster ensemble are intrinsically linked to the cluster mass-richness relation that we aim at constraining. The DESC cosmoDC2 simulated dataset used in this work is presented in \secreff{sec:DC2dataset}, that describes in detail the galaxy catalog (used to estimate the weak lensing signal) and the DC2 redMaPPer cluster catalog; the fiducial mass-richness relation to which the results will be compared to is also derived in that section. \secreff{sec:method_scaling_relation} presents the methodology to build the data vectors (counts, stacked lensing profiles, or mean lensing masses) and the likelihoods that combine them with the models of \secreff{sec:formalism} to infer the cluster mass-richness relation. In \secreff{sec:results}, we present the results of the inference procedure and discuss the stability of the derived redMaPPer-detected cluster mass-richness relation in cosmoDC2 when considering a variety of systematic effects. Finally, we summarize our findings and conclude in \secreff{sec:conclusion}.
\section{Theoretical background}
\label{sec:formalism}
Since we aim to constrain the mass-richness relation of redMaPPer-selected clusters in the LSST DESC DC2 simulations, we first present the general theoretical framework for modeling cluster number counts (\secreff{sec:cluster_count}) and cluster weak gravitational lensing (\secreff{sec:cluster_lensing}). We then introduce the parametrization of the mass-richness relation that we consider in this work (see \secreff{sec:scaling_relation_formalism}).
\subsection{Cluster number count}
\label{sec:cluster_count}

For clusters detected by their observed richness, the redshift-richness cluster number density is given by
\begin{equation}
    \frac{d^2N(\lambda, z)}{d\Omega dzd\lambda}=  \int_{m_{\rm min}}^{m_{\rm max}}dm\ \frac{dn(m, z)}{dm}\frac{d^2V(z)}{dzd\Omega} \Phi(\lambda, m, z) P(\lambda|m,z),
    \label{eq:number_density}
\end{equation}
where $\Omega$ is the survey sky area\footnote{Here, we assume the survey depth is uniform across the sky.}, $P(\lambda|m,z)$ corresponds to the mass-richness relation (see hereafter), $dn(m,z)/dm$ is the halo mass function (which predicts the comoving number density of dark matter halos per mass interval), $d^2V(z)/dzd\Omega$ is the differential comoving volume, and $\Phi(\lambda, m, z)$ is the cluster selection function. The differential comoving volume is given by
\begin{equation}
    \frac{d^2V(z)}{dzd\Omega} = d_H \frac{D_C^2(z)}{H(z)/H_0},
\end{equation}
where $H(z)$ is the Hubble parameter at redshift $z$ (with the Hubble constant given by $H_0 = H(z=0)$), $d_H = c/H_0$ is the Hubble distance, and $D_C$ is the radial comoving distance \citep{Hogg1999distances}. The masses $m_{\rm min}$ and $m_{\rm max}$ correspond to the minimum and maximum mass accessible in the survey by the cluster finder, which are generally set by the survey strategy.

The selection function $\Phi$ denotes our ability to detect clusters. It is associated with the performance of the cluster finder algorithm and the survey strategy, and its calibration plays a crucial role in cluster-based cosmological analyses. The selection function of a cluster finder can be assessed by (i) using end-to-end simulated data, geometrically matching the observed cluster catalog to the underlying dark matter halo population \citep{Euclis2019clusterselection,Lesci2022KIDSCL,Bulbul2024selection}, (ii) injecting mock clusters into the real dataset \citep{Rykoff2014redmapper,Rykoff2016selection,Planck2016selection}, or (iii) cross-validating the cluster finder’s performance by comparing datasets at different wavelengths \citep{Sadibekova2014redmapperX,Saro2015massrichness}. We follow the prescription presented in \citet{Aguena2018completenesspurity}, where the selection function is given by
\begin{equation}
    \Phi(\lambda, m, z) = \frac{c(m,z)}{p(\lambda, z)}.
    \label{eq:selection_function}
\end{equation}
First, the completeness $c(m,z)$ gives the fraction of true underlying dark matter halos detected by the cluster finder algorithm. If the completeness is less than 1, then a portion of the underlying dark matter halo population is systematically missing from our dataset \citep{Mantz2019selection}. Second, the purity $p(\lambda, z)$ denotes the fraction of "spurious" detections in the cluster catalog (false positives or mis-identified structures along the line-of-sight), where $\lambda$ and $z$ are the cluster richness and redshift\footnote{Purity is generally defined as $p(\lambda, z_{\rm obs})$ instead of $p(\lambda, z)$, where $z_{\rm obs}$ is the cluster finder-inferred redshift. For simplicity, we consider here that observed and true redshifts (i.e., the redshift of the underlying dark matter halo) are equal.}.

With all these elements, the predicted cluster number count $N_{ij}$ in the $i$-th redshift and $j$-th richness bin is finally given by
\begin{equation}
    N_{ij} = \Omega \int_{z_{i}}^{z_{i+1}} dz \int_{\lambda_{j}}^{\lambda_{j+1}} d\lambda \frac{d^2N(\lambda, z)}{d\Omega dzd\lambda}.
    \label{eq:N_count}
\end{equation}

\subsection{Cluster weak gravitational lensing}
\label{sec:cluster_lensing}
The observed ellipticity $\epsilon^{\rm obs}$ of a lensed source galaxy is related to its intrinsic (unlensed) shape $\epsilon^{\rm int}$, as given by \citep{Schneider1992wl}
\begin{equation}
    \epsilon^{\rm obs} =
        \frac{\epsilon^{\rm int} + g}{1 + g^*\epsilon^{\rm int}},
        \label{eq:e_obs_e_int}
\end{equation}
where $g = \gamma/(1 - \kappa)$ is the reduced shear, with $\gamma$ being the shear and $\kappa$ the convergence. In the weak lensing regime, we assume that $\kappa \ll 1$, so the first-order Taylor expansion in $\gamma$ of the observed ellipticity gives $\epsilon^{\rm obs} \approx \gamma + \epsilon^{\rm int}$. Taking the average of this expression and assuming that $\langle \epsilon^{\rm int} \rangle = 0$ in the absence of large-scale galaxy intrinsic alignments, we find that the cluster's local shear can be estimated statistically through $\langle\epsilon^{\rm obs}\rangle \approx \gamma$. In practice, we decompose galaxy shapes into tangential and cross components, defined respectively by
\begin{align}
    \epsilon_+ + i\epsilon_\times = - \epsilon^{\rm obs} \exp\{-i2\phi\},
    \label{eq:epsilon_+x}
\end{align}
where $\epsilon^{\rm obs}$ is the observed galaxy's ellipticity and $\phi$ is the polar angle of the source galaxy relative to the center of the galaxy cluster\footnote{The cluster center is typically identified from observational counterparts, such as the position of the brightest central galaxy. However, there may be an offset -- called miscentering radius -- between the identified cluster center and the center of its gravitational potential.}. For any mass distribution, we obtain that $\langle \epsilon_+ \rangle_{\mathcal{C}} = \gamma_+$, where $\gamma_+$ is the tangential shear, and $\langle \epsilon_\times \rangle_{\mathcal{C}} = \gamma_\times = 0$ where $\langle \cdot \rangle_{\mathcal{C}}$ denotes the average along a closed (circular) loop \citep{Bernstein2009multipole,Umetsu2020clusterlensing}.

The excess surface density $\Delta\Sigma(R)$ is commonly used as the cluster weak-lensing shear estimator \citep{Mandelbaum2005HSM,Murata2019HSCrichnessmassrelation,McClintock2019masscalibration}. At a given projected radius $R$ from the cluster center, the excess surface density is related to the local tangential shear $\gamma_+(R, z_l, z_s)$ -- where $z_l$ is the redshift of the cluster (denoted as "lens") and $z_s$ is the redshift of the source galaxy -- by
\begin{equation}
    \Delta\Sigma(R) = \Sigma_{{\rm crit}}(z_s, z_l)\gamma_+(R, z_l, z_s),
\end{equation}
where the critical surface mass density $\Sigma_{\rm crit}(z_s, z_l)$ is a geometrical factor given by
\begin{equation}
    \Sigma_{{\rm crit}}(z_s, z_l) = \frac{c^2}{4 \pi G} \frac{D_A(z_s)}{D_A(z_l) D_A(z_s, z_l)},
    \label{eq:critical_surface_dens}
\end{equation}
where $D_A(z_l)$, $D_A(z_s)$, and $D_A(z_{l},z_{s})$ are respectively the physical angular diameter distances to the lens, to the source, and between the lens and the source \citep{Hogg1999distances}.

To model the stacked lensing signal, $\Delta\Sigma(R)$ can be related to the surface mass density of the cluster $\Sigma(R)$. For a perfectly centered cluster, it is given by
\begin{equation}
\Sigma(R) = \int_{-\infty}^{+\infty} dy\rho \left(\sqrt{y^2 + R^2}\right),
\label{eq:sigma}
\end{equation}
where $\rho$ is the three-dimensional dark matter density. Then, $\Delta\Sigma$ is given by
\begin{equation}
    \Delta\Sigma(R) = \Sigma(<R) - \Sigma(R),
    \label{eq:dsigma_diff}
\end{equation}
where
\begin{align}
    \Sigma(<R) &= \frac{2}{R^2}\int_0^R dR'R'\Sigma(R').
\end{align}
The total matter content around clusters is parametrized by the three-dimensional density $\rho(r)$ (and by extension $\Sigma(R)$ through \eqreff{eq:sigma}). 

The matter density around clusters has two contributions. The first one is the 1-halo term, denoting the intrinsic cluster matter density. We generally assume that the 1-halo term is well described by the Navarro-Frank-White \citep[NFW, ][]{Navarro1997nfw} profile\footnote{The NFW profile, as well as other profiles used in the literature, are fitting functions calibrated on N-body simulations.}, set by the spherical overdensity mass
\begin{equation}
    M_{\rm 200c} = \frac{4\pi r_{\rm 200c}^3}{3} 200 \rho_{\rm c}(z),
    \label{eq:M200c}
\end{equation}
where $r_{200c}$ is the radius of the sphere that contains an average matter density 200 times higher than the critical density $\rho_{\rm c}(z)$. Such profiles also depend on a concentration parameter $c_{\rm 200c}$, which quantifies the level of concentration of mass in the innermost regions of the cluster (we discuss in detail in \secreff{sec:impact_cM_relation} the different concentration-mass relations used in the literature, and their impact on the inference of the cluster mass). The 1-halo term dominates the lensing signal for $R < 3.5$ Mpc, while the 2-halo term -- associated with the contribution to the matter density field from surrounding, correlated halos -- becomes increasingly more important at larger scales \citep{Oguri2011lensing}.

The collapsed dark matter halos are not expected to be spherical, due to the non-spherical initial density peaks from which they form, and also due to their complex individual accretion history in the cosmic web \citep{Sheth2001hmf}. The underlying dark matter halos have been shown to have complex triaxial structures in simulations \citep{Jing2002dmhalostriaxial,Schneider2012triaxial,Despali2014triaxialsim}, as well as from the observation of their member galaxies \citep{BBinggeli1982triaxialclustergalaxies} or their weak lensing shear field \citep{Oguri2003triaxial,Oguri2010ellipticity}. In this analysis, we will consider stacks of individual cluster lensing profiles. Not only does stacking profiles enhance the signal-to-noise ratio, which is particularly beneficial for low-mass clusters (where the gravitational lensing signal is weaker), but the stacking method also averages out the intrinsic triaxiality of individual halos and substructures \citep{Corless2009lensingstack}, effectively recovering a spherical profile.

As for the cluster count, accounting for the selection function of the cluster finder is essential to predict the average cluster lensing profile for an ensemble of clusters. 
In this work, we neglect the contributions from "spurious"\footnote{e.g., low-mass galaxy halos, misidentified structures along the line-of-sight.} redMaPPer detections -- encoded in the purity -- to the measured cluster lensing signal, as we consider it to be negligible compared to the cluster lensing induced by properly identified halos -- encoded in the completeness. However, we recall that it is a strong assumption, since any redMaPPer-identified structure would inevitably induce positive lensing signal; however, this has not been explored extensively in the literature, and this is not the purpose of this work, so we keep using the aforementioned assumption\footnote{In the latter, we measure the cluster lensing profile for $\lambda > 20$, such as the redMaPPer purity is $\sim 99\%$.}.  
Then, the predicted stacked excess surface density profile is given by
\begin{align}
\begin{split}
    \Delta\Sigma_{ij}(R) = \frac{1}{N_{ij}}\int_{z_{i}}^{z_{i+1}} dz \int_{\lambda_{j}}^{\lambda_{j+1}} d\lambda \int_{m_{\rm min}}^{+\infty}dm\times\\ \frac{d^2N(m, z)}{dzdm} c(m,z) P(\lambda|m,z)\Delta\Sigma(R|m,z),
    \end{split}
    \label{eq:DS_stack_th}
\end{align}
where we use the completeness $c(m,z)$, i.e., the fraction of true detected halos relative to the underlying halo population.

Alternatively, instead of using the stacked profile directly, one can use the mean mass of the clusters within the $ij$-th redshift-richness bin — derived from the corresponding excess surface density profile — to investigate the cluster scaling relation (we will provide more details in \secreff{sec:method_scaling_relation}). For this purpose, using the same arguments as above, the mean mass can be modeled by
\begin{align}
\begin{split}
    M_{ij} = \frac{1}{N_{ij}}\int_{z_{i}}^{z_{i+1}} dz \int_{\lambda_{j}}^{\lambda_{j+1}} d\lambda \int_{m_{\rm min}}^{+\infty}dm\times\\ \frac{d^2N(m, z)}{dzdm} c(m,z) P(\lambda|m,z)m.
    \end{split}
    \label{eq:M_lensing}
\end{align}
In this analysis, either equation~(\ref{eq:DS_stack_th}) or (\ref{eq:M_lensing}) will be used to model the cluster lensing information, in combination with cluster counts, to constrain the mass-richness relation (see the next section).
\subsection{Cluster mass-richness relation}
\label{sec:scaling_relation_formalism}
In this paper, we aim to infer the mass-richness relation of galaxy clusters detected by the redMaPPer \citep{Rykoff2014redmapper} cluster finder in the LSST DESC DC2 simulations using a combination of cluster lensing and cluster counts. The observed cluster richness $\lambda$ traces the number of member galaxies inside a galaxy cluster. Due to (i) the complex cluster formation history and baryon physics, and (ii) observational effects in richness measurement \citep[see e.g.][]{Costanzi2018proj}, the observed cluster richness $\lambda$ is a statistical variable for a halo with a fixed mass $m$. In this paper, we consider the log-normal scaling relation $P(\lambda|m,z)$ \citep[see e.g.][]{Mantz2008cluster,Evrard2014massobservable,Saro2015massrichness,Farahi2018cov,Murata2019HSCrichnessmassrelation,Anbajagane2020stellat} given by
\begin{equation}
    P(\ln\lambda|m,z) = \frac{1}{\sqrt{2\pi}\sigma_{\ln\lambda|m,z}}\exp\left\{-\frac{[\ln\lambda - \langle \ln \lambda|m, z\rangle]^2}{2\sigma_{\ln\lambda|m,z}^2}\right\},
    \label{eq:p_lambda_m}
\end{equation}
where
\begin{equation}
    \langle \ln \lambda|m, z\rangle = \ln\lambda_0 + \mu_z\ln\left(\frac{1 + z}{1 + z_0}\right) + \mu_m\log_{10}\left(\frac{m}{m_0}\right),
\label{eq:richness_mass}
\end{equation}
and
\begin{equation}
    \sigma_{\ln \lambda|m, z} = \sigma_{\ln\lambda_0} + \sigma_z\ln\left(\frac{1 + z}{1 + z_0}\right) + \sigma_m\log_{10}\left(\frac{m}{m_0}\right).
    \label{eq:sigma_forward_modelling}
\end{equation}
In the above equation, $\sigma_{\ln \lambda|m, z}$ is the total error, i.e., including contributions from the richness measurement errors and the intrinsic sources of scatter \citep[see e.g.][]{Murata2019HSCrichnessmassrelation}. It is possible to split the two contributions by adding a Poisson variance term \citep{Zhang2023triax} in \eqreff{eq:sigma_forward_modelling} (see Appendix \ref{app:fiducial}). Let us note that the forward modeling $P(\lambda|m,z)$ is different from the backward formalism $P(m|\lambda,z)$, which has been used in several other works \citep[see e.g.][]{Baxter2016massrichness,Melchior2017logslope,Simet2017SDSSmassrichness,Jimeno2018massrichness,McClintock2019masscalibration}.

From the above, the cluster scaling relation has 6 free parameters: $\ln\lambda_0$, $\mu_z$, and $\mu_m$ denote respectively the offset richness, the redshift dependence, and the mass dependence of the mean cluster scaling relation, where the offset variance of the relation is given by $\sigma_{\ln\lambda_0}$, and $\sigma_z$ and $\sigma_m$ denote the redshift and mass dependencies of the variance.

\section{Datasets}
\label{sec:DC2dataset}
\begin{figure*}[t]
\centering\includegraphics[width=1\textwidth]{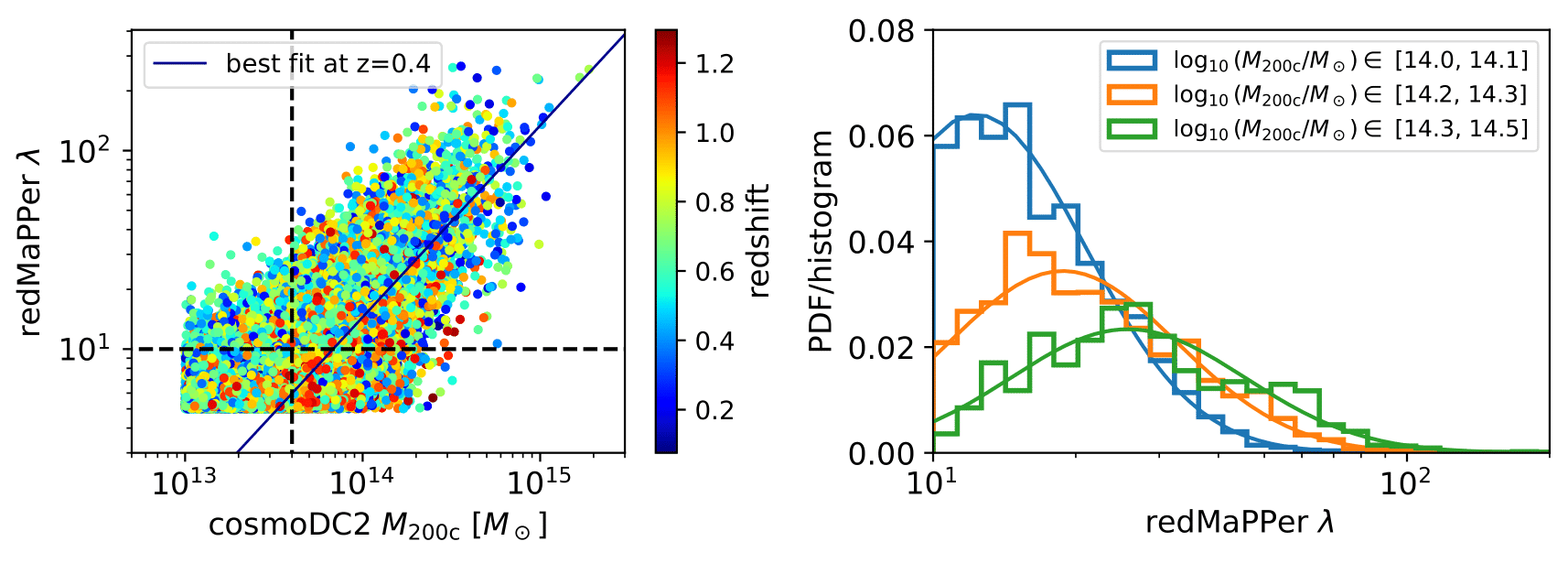}
    \caption{Left: cosmoDC2 halo $M_{\rm 200c}$ masses (from the simulation) versus the redMaPPer cluster richnesses. The points are colored with the redMaPPer cluster redshift. The full line is the best-fitted mean richness-mass relation in \eqreff{eq:richness_mass} at $z=0.4$. The dashed lines represent the low mass and low richness cut used on the cosmoDC2-redMaPPer matched catalog for the fit of the fiducial scaling relation. Right: Histogram of richnesses in bins of mass, and best-fit probability distributions $P(\ln\lambda|M,z)$ from \eqreff{eq:p_lambda_m} (full lines) for the different mass bins.}
    \label{fig:fiducial_relation}
\end{figure*}

In this section, we detail the datasets used in this paper. We first present the DESC Data Challenge 2 (\secreff{sec:DC2}), from which the cosmoDC2 galaxy catalog and associated the photometric redshift catalogs (\secreff{sec:cosmoDC2_sources}), and the catalog of clusters detected by redMaPPer \citep{Rykoff2014redmapper} in cosmoDC2 (\secreff{sec:redmapper_cluster_catalog}), are derived. Matching these clusters to dark matter halos in cosmoDC2 allows us to estimate a fiducial mass-richness relation (that will serve as reference for the remaining of the paper) and the selection function of the cosmoDC2 redMaPPer clusters, as detailed in \secreff{sec:matched_catalog}.

\subsection{The simulated catalogs of the DESC Data Challenge 2 (DC2)}
\label{sec:DC2}
DC2 is a vast simulated astronomical dataset covering $\Omega_{\rm DC2} = 440$ square degrees, designed to help develop and test the pipeline and analysis tools of the DESC for interpreting LSST data \citep[see full details in ][]{Abolfathi2021DC2}.

The cosmoDC2 galaxy catalog detailed in the next \secreff{sec:cosmoDC2_sources} is built upon the {\sc OuterRim} N-body (gravity-only) simulation \citep{Heitmann2019Outerim}. Each dark matter halo has been identified by a friend-of-friend (FoF) halo finder and assigned both a mass $M_{\rm FoF}$ (the sum of the individual dark matter particles associated through the FoF algorithm) and a spherical overdensity mass $M_{\rm 200c}$ (obtained by fitting an NFW profile to the distribution of dark matter particles). Galaxies within halos were generated using the Galacticus semi-analytic model of galaxy formation \citep{Benson2012Galacticus}, and were placed into halos using {\sc GalSampler} \citep{Hearin2020GalSampler}. The resulting galaxy catalog includes properties such as stellar mass, morphology, spectral energy distributions, broadband filter magnitudes, and host halo information. Weak lensing shears and convergences at each galaxy position were computed using a ray-tracing algorithm applied to past light-cone particles from the simulation. We note that the performance of the ray-tracing procedure is degraded in the innermost regions of cluster fields (typically at $R \leq 1$ Mpc); this resolution effect was observed in \citet{Korytov2019cosmoDC2} and discussed in \citet{Kovacs2022cosmoDC2} for galaxy-galaxy and cluster-galaxy lensing. We describe in more detail in \secreff{sec:data_vector_ds_profiles} how this issue is handled in our data analysis pipeline.

\subsection{The cosmoDC2 extra-galactic catalog and photometric redshift add-on catalogs}
\label{sec:cosmoDC2_sources}

The cosmoDC2 extragalactic catalog \citep{Korytov2019cosmoDC2,Kovacs2022cosmoDC2} contains $\sim 2.26$ billion galaxies, providing an inventory of $\sim 550$ properties per galaxy, including "true" galaxy attributes (e.g., true magnitudes in the six LSST bands, true redshift, true shapes, etc.) as well as ray-tracing quantities (shear and convergence). The catalog reaches a magnitude depth of 28 in the $r$-band and extends out to redshift $z \sim 3$. In this sense, it represents an idealized LSST dataset, as it includes only galaxies and does not account for observational contaminants such as dust extinction, stars, or instrumental systematics.

This work also uses two cosmoDC2 photometric redshift add-on catalogs, produced by DESC using two representative, well-established approaches in the literature: FlexZBoost \citep{IzBicki2017Flexzboost}\footnote{\url{https://github.com/rizbicki/FlexCoDE}} and BPZ \citep[Bayesian Photometric Redshifts,][]{Benitez2011BPZ}\footnote{\url{https://github.com/LSSTDESC/rail_bpz}}. FlexZBoost is an empirical, machine-learning-based technique that has been shown to yield highly accurate conditional photometric redshift density estimates in prior studies \citep{Schmidt2020pzlsst}. It uses spatially overlapping data from LSST and accurate spectroscopic redshifts to estimate the conditional redshift distribution $p(z|m)$, where $m$ corresponds to LSST photometric magnitudes. BPZ is a template-fitting method that formulates a likelihood of a galaxy’s observed colors based on a set of redshifted Spectral Energy Distribution (SED) models. This approach incorporates informative priors, such as the expected redshift distribution and galaxy type.

The DC2 simulation suite has two key limitations that affect the accuracy of photometric redshift estimates. First, the simulation does not model spectroscopic surveys or their galaxy selection functions. In cosmoDC2, FlexZBoost was trained on a complete subsample of galaxies down to $i<25$. While this depth is representative of LSST, it extends beyond the range where real spectroscopic surveys are complete. As a result, the photometric redshift estimates obtained with FlexZBoost should be considered optimistic in terms of accuracy. Second, the color–redshift space of the simulated galaxies is not continuous, but discrete, due to the mock photometry construction process \citep{Korytov2019cosmoDC2}\footnote{Specifically, a very low-resolution internal dust model was used, which introduced artificial "breaks" in some galaxy spectra, especially at high redshift.}. While template-fitting methods are generally less sensitive to incomplete calibration data, they depend more heavily on accurate SED models\footnote{Template-fitting techniques are typically less affected by incomplete calibration data, but are more agnostic than empirical techniques, relying more on accurate SED modeling rather than the specificity of the calibration dataset.}. This artificial discreteness in the data leads to poor fits for template-based methods, which inherently assume that galaxy spectral energy distributions evolve smoothly with redshift\footnote{This is not a fundamental limitation of template-fitting methods, but stems from the strong prior assumptions imposed by empirical techniques when generalizing $p(z|m)$ from the calibration set to the LSST data.}. Therefore, the accuracy of BPZ photometric redshifts in DC2 should be regarded as pessimistic.
\begin{table}[t]
\begin{center}
\caption{Fiducial values of the parameters of the redMaPPer cluster mass-richness relation.}
\label{tab:best_fits_scaling_relation_matching}
\begin{tabular}{ c|c|c} 
  Parameters & Priors & $p\pm\Delta p$\\
\hline
\hline
 $\ln\lambda_0$ & $\mathcal{U}(0, +\infty)$ & 3.35 $\pm$ 0.01\\ 
  $\mu_z$ & $\mathcal{U}(-2, 2)$& 0.06 $\pm$ 0.08\\ 
   $\mu_m$ & $\mathcal{U}(0, +\infty)$ & 2.23 $\pm$ 0.05\\ 
 $\sigma_{\ln\lambda_0}$ & $\mathcal{U}(0, +\infty)$&0.56 $\pm$ 0.01\\ 
  $\sigma_z$ & $\mathcal{U}(-2,2)$ & -0.05 $\pm$ 0.04 \\
   $\sigma_m$ & $\mathcal{U}(-2,2)$& 0.10 $\pm$ 0.02 
\end{tabular}
\end{center}
\tablefoot{Second column: Prior we use in this analysis, in addition to $\sigma_{\ln \lambda|m, z} > 0$ (the dispersion is set to be positive). Third column: Posterior means for the six parameters of the scaling relation, by using individual dark matter halo masses and redMaPPer richnesses. }
\end{table}

We expect realistic photometric redshift estimates to lie somewhere between the optimistic FlexZBoost and the pessimistic BPZ results from the DC2 runs, although they are likely to be closer to FlexZBoost, given that the pessimism in BPZ primarily arises from non-physical artifacts in the simulation.

\subsection{The redMaPPer cluster catalog}
\label{sec:redmapper_cluster_catalog}
As mentioned in the introduction, galaxy clusters can be identified through their member galaxies using galaxy catalogs. The redMaPPer cluster finder \citep{Rykoff2014redmapper} identifies clusters via the presence of red-sequence galaxies, leveraging the multiple optical bands available in the survey. redMaPPer has already been widely used on SDSS \citep{Abdullah2020SDSSCL} and DES \citep{Abbott2020DESCL}, and has been run on cosmoDC2 using the six LSST-like bands $u$, $g$, $r$, $i$, $z$, and $y$ (see Ricci et al., in prep. for details). We note that since redMaPPer was applied directly to the cosmoDC2 galaxy catalog\footnote{In addition to redMaPPer, other cluster detection methods have been applied to the cosmoDC2 data, including WaZP \citep{Aguena2021WAZP}, AMICO \citep{Bellagamba2017AMICO}, and YOLO-CL \citep{Grishin2023yolocl}. The corresponding catalogs are currently available within DESC or are expected to be released soon.}, the resulting cluster catalog is free from atmospheric and instrumental systematics. The redMaPPer catalog provides galaxy cluster positions, redshifts, and richness values \citep[computed as the sum of membership probabilities of galaxies around the cluster, see][]{Rykoff2014redmapper}, along with per-cluster membership galaxy catalogs, for 880,000 clusters within the ranges $5 < \lambda < 270$ and $0.1 < z < 1.15$.

\subsection{The matched cosmoDC2 halo - redMaPPer cluster catalog}
\label{sec:matched_catalog}

To infer the 'true' cluster scaling relation, we associate the redMaPPer-detected clusters with halos from the cosmoDC2 dark matter halo catalog using the DESC software ClEvaR\footnote{\url{https://github.com/LSSTDESC/ClEvaR}}. The association is based on membership matching \citep[see e.g.][]{Farahi2016matchingdispersion}, where we use the member galaxies\footnote{Specifically, we use the redMaPPer member galaxies for each cluster (i.e., the output of redMaPPer), and the cosmoDC2 galaxies flagged as “members” for each cosmoDC2 dark matter halo in the simulation.} of the detected redMaPPer clusters and those of the cosmoDC2 halos, respectively. Each halo is associated with all detected clusters with which it shares galaxies, and vice versa.

The one-to-one catalog is then obtained by selecting the associated system (redMaPPer cluster–cosmoDC2 halo pair) with the highest "membership fraction" (i.e., the overlap between the member galaxies of the cosmoDC2 halo and the redMaPPer cluster) and retaining only reciprocal matches (i.e., two-way matches).

This matching procedure was performed using a cut of $M_{\rm FoF} > 10^{13}M_\odot$ for the dark matter halo catalog and $\lambda > 5$ for the redMaPPer cluster catalog. We verified that applying a cut of $M_{\rm 200c} > 10^{13}M_\odot$ instead yields identical catalogs. The resulting redMaPPer–cosmoDC2 halo masses and richnesses are shown in \figreff{fig:fiducial_relation} (left panel). 

Then, each redMaPPer cluster with richness $\lambda_k$ is assigned a "true" spherical overdensity mass $M_k = M_{\rm{200c}, k}$. The matched redMaPPer cluster–cosmoDC2 halo catalog objects in the mass–richness plane are shown in \figreff{fig:fiducial_relation} (left panel), color-coded according to their redshifts.

\subsubsection{The fiducial mass-richness relation}
\label{sec:fiducial_relation}
From the cosmoDC2–redMaPPer matched catalog, we can infer a "fiducial" scaling relation. We consider the fiducial set to be the set of clusters with $\lambda > 10$ (horizontal dashed line in \figreff{fig:fiducial_relation}, left panel), as well as $M > 4 \times 10^{13} M_\odot$ (vertical dashed line) and with $0.2 < z < 1$. The fiducial likelihood is given by
\begin{equation}
    \mathcal{L}_{\rm fid} = \prod_{k=1}^{N_{\rm tot}}P_{\lambda_k > 10}(\ln\lambda_k| M_k, z_k)\;,
    \label{eq:fiducial_likelihood}
\end{equation}
where $P(\ln\lambda_k | M_k, z_k)$ is the mass-richness relation given in \eqreff{eq:p_lambda_m} for the $k$-th cluster-halo match (we consider the pivot redshift $z_0 = 0.5$ and pivot mass $m_0 = 10^{14.3} M_\odot$), and the subscript denotes truncated Gaussians with $\lambda_k > \lambda_{\rm min} = 10$. We use the \texttt{emcee} package \citep{ForemanMackey2013emcee}, with flat priors inspired by \citet{Murata2019HSCrichnessmassrelation}, as listed in \tabreff{tab:best_fits_scaling_relation_matching} (second column). The best-fitted fiducial relation is shown in the third column of  \tabreff{tab:best_fits_scaling_relation_matching}. These values can serve as a reference against which we will compare the results obtained using cluster weak lensing and abundance.

The low mass/richness cuts applied to the matched catalog are deliberately conservative. These cuts are chosen to ensure that the completeness of the redMaPPer cluster catalog (see \secreff{sec:selection_function}) does not influence the inference of the fiducial cluster scaling relation. In Appendix \ref{app:fiducial}, we examine the impact of alternative cuts, motivated by the modeling of the mass-richness relation used in this work. While this topic warrants further investigation, for this study, we adopt the fiducial values listed in \tabreff{tab:best_fits_scaling_relation_matching}.
\begin{figure*}[t]
    \centering
    \includegraphics[width=1\textwidth]{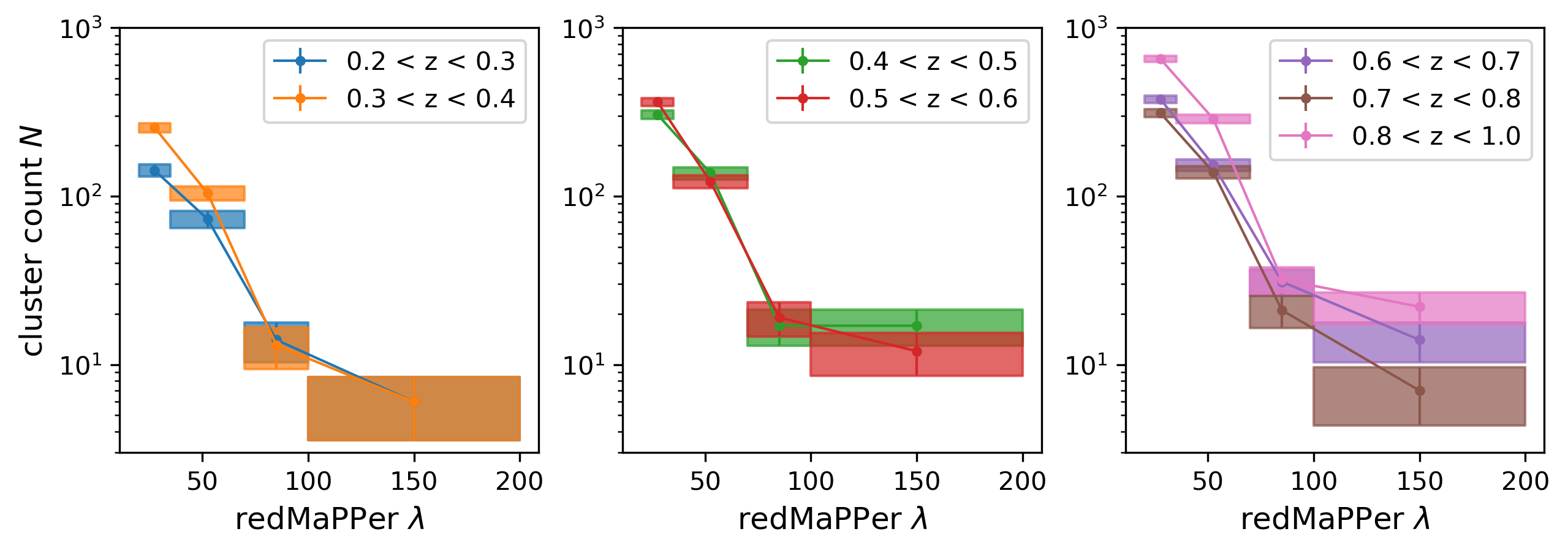}
    \caption{Measured count of redMaPPer cluster as a function of richness, for different richness bins. For each richness-redshift bin, the width of the shaded area represent the width of the richness bin, and the height correspond to the Poisson noise $\sqrt{N}$. The steeper slope of cluster counts at intermediate richness (between the second and third richness bin) for all redshift bins is primarily due to the choice of richness binning, which is not perfectly log-spaced. This steeper inner slope would be mitigated if a log-spaced richness binning were used.}
    \label{fig:cluster_count}
\end{figure*}

We show, over-plotted in black in the left panel of \figreff{fig:fiducial_relation}, the mean scaling relation for the best-fit parameters at $z = 0.4$. The right panel of \figreff{fig:fiducial_relation} presents the normalized histograms of redMaPPer richness in three different mass bins (within the redshift range $z \in [0.2, 1]$). Over-plotted on these histograms is the best-fit probability density function, $P(\lambda | m_{\rm center}, z_{\rm center})$, where $z_{\rm center}$ and $m_{\rm center}$ represent the center values of each redshift-mass bin. We observe that the predicted distribution matches the observed one quite well.

\subsubsection{The redMaPPer selection function in DC2}
\label{sec:selection_function}
To properly model cluster number counts, we must account for the selection function $\Phi$ of the redMaPPer algorithm \citep[see e.g.][]{Euclis2019clusterselection,Lesci2022KIDSCL}.
Defined in \eqreff{eq:selection_function}, the selection function reflects the fact that the cluster-finding algorithm may miss a fraction of true galaxy clusters (completeness) as well as detect "false" clusters that are not associated with underlying collapsed dark matter structures (purity).

From the matched catalog, we can disentangle the effects of purity and completeness within narrow mass-redshift-richness bins. The measured completeness $\widehat{c}_{ij}$ in the mass and redshift bins $[m_i, m_{i+1}]$ and $[z_j, z_{j+1}]$, and the measured purity $\widehat{p}{kl}$ in the richness and observed redshift bins $[\lambda_k, \lambda_{k+1}]$ and $[z_{\mathrm{obs}, l}, z_{\mathrm{obs}, l+1}]$, are given by 
\begin{equation}
    \widehat{c}_{ij} = \frac{N_{\mathrm{match}, ij}^{\rm halos}}{N_{\mathrm{tot}, ij}^{\rm halos}}\hspace{0.5cm}\mathrm{and}\hspace{0.5cm}\widehat{p}_{kl} = \frac{N_{\mathrm{match}, kl}^{\rm halos}}{N_{\mathrm{obs}, kl}^{\rm clusters}},
\end{equation}
where $N_{\mathrm{tot}, ij}^{\rm halos}$ is the total number of halos in the $ij$ mass-redshift bin, and $N_{\mathrm{obs}, kl}^{\rm clusters}$ is the number of redMaPPer clusters in the $kl$ richness-redshift bin. The quantity $N_{\mathrm{match}, ij}^{\rm halos}$ (respectively $N_{\mathrm{match}, kl}^{\rm halos}$) is the number of halos that have been matched to redMaPPer clusters in the $ij$ mass-redshift bin (respectively in the $kl$ richness-redshift bin). We model the observed completeness $\widehat{c}_{ij}$ and purity $\widehat{p}_{kl}$, allowing them to be included in the count/lensing formalism as detailed in \secreff{sec:formalism}.
The performance of the DC2 redMaPPer cluster catalog is analyzed in detail in Ricci et al. (in prep.). We note that the redMaPPer cluster catalog is complete at the $80\%$ level for $M_{\rm 200c} > 10^{14} M_\odot$, with a small redshift dependence, and is pure at the $> 90\%$ level for $\lambda > 12$.

\section{Methodology to constrain the mass-richness relation}
\label{sec:method_scaling_relation}
As shown in \secreff{sec:formalism}, the modeling of cluster number counts and/or cluster lensing (with stacked excess surface density profiles or mean cluster masses) in bins of redshift and richness depends on the mass-richness relation. To constrain the parameters of the latter using this observables requires building the corresponding data vectors (\secreff{sec:data_vectors}) and defining the likelihood functions that link the data vectors, the covariances, and the models for each of these observables, as described in (\secreff{sec:likelihoods}). These likelihoods may then be used individually or in combination to constrain the parameters of the mass-richness relation. To implement the methodology described in this section, we rely on a variety of software tools that are briefly listed in \secreff{sec:softwares}.
\subsection{Data vectors}
\label{sec:data_vectors}

We now turn to the construction of the data vectors that will be used for the inference of the redMaPPer cluster mass–richness relation. For the stacking strategy we employ, we have considered the redshift bin edges $z_i = {0.2, 0.3, 0.4, 0.5, 0.6, 0.7, 0.8, 1}$ and the richness bin edges $\lambda_i = {20, 35, 70, 100, 200}$. We use $\lambda > 20$ and $z > 0.2$, inspired by other redMaPPer-based cluster analyses \citep[see e.g.][]{McClintock2019masscalibration,Abbott2020DESCL}. The $\lambda > 20$ cut ensures a high-purity cluster sample \citep{Costanzi2019SDSSCL}, close to 100$\%$ pure in our case, and restricting to $z > 0.2$ mimics the conservative cut used in DES cluster-based analyses \citep{Abbott2020DESCL,DEScollabY3CL}. This cut was implemented to prevent the degradation of redMaPPer performance at low redshifts\footnote{For $z<0.2$ galaxies, the distinctive Balmer Break at 4000 \r{A} falls in the $u$-band.}, where the red-sequence galaxy population becomes harder to isolate due to the lack of $u$-band data in DES (which used $g$, $r$, $i$, and $z$ bands). Let us note that robust detection of low-redshift clusters below $z = 0.2$ will be feasible with LSST, since $u$-band imaging will be available.

\begin{figure*}
    \centering
    \includegraphics[width=1\textwidth]{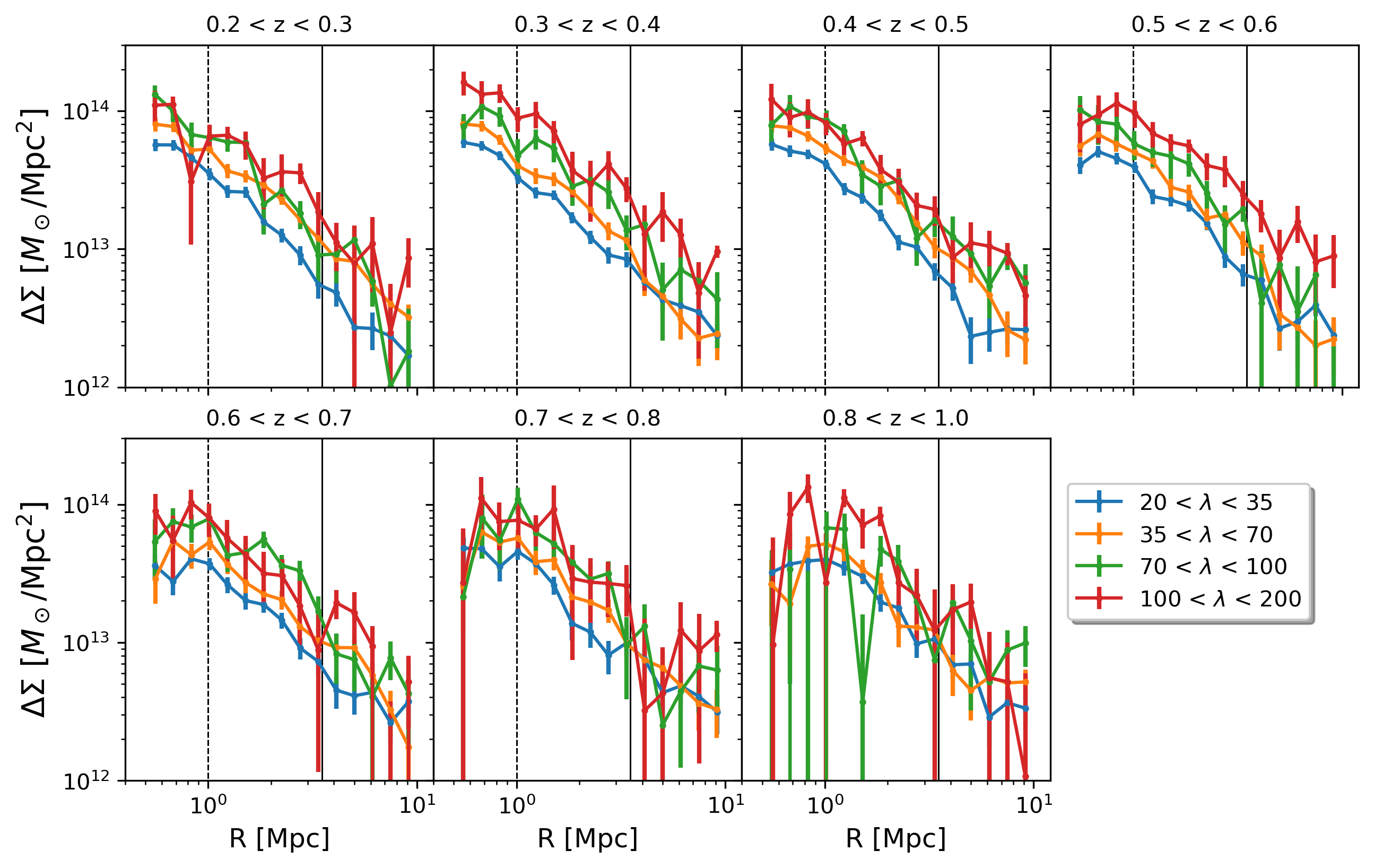}
\caption{Stacked excess surface density profile as a function of the distance to the cluster center, for different richness bins (colors) and different redshift bins (from top left to bottom right). The error bars are the diagonal elements of the bootstrap covariance matrices. The black dashed vertical line (resp. black filled vertical line) represents the $R > 1$ Mpc cut (resp. $R < 3.5$ Mpc cut). }
    \label{fig:excess_surface_density_profiles}
\end{figure*}

\subsubsection{Cluster number counts}
We show in \figreff{fig:cluster_count} the measured count of redMaPPer clusters as a function of richness for different redshift bins. As expected, there are fewer clusters at higher richness than at lower richness.

\subsubsection{Stacked excess surface density profiles}
\label{sec:data_vector_ds_profiles}

Considering an ensemble of $N_l$ clusters (or "lenses"), where each cluster has $N_{ls}$ background source galaxies in the radial bin $[R - \Delta R / 2, R + \Delta R / 2]$, the maximum likelihood estimator of the stacked $\Delta\Sigma(R)$ profile is given by \citep{Shirasaki2018DeltaSigmaEstimator, Sheldon2004DeltaSigmaestimator}
\begin{equation}
     \widehat{\Delta\Sigma}_+(R) = \frac{1}{\sum\limits_{l= 1}^{N_{l}}\sum\limits_{s= 1}^{N_{ls}}w_{ls}}
     \sum\limits_{l= 1}^{N_{l}}\sum\limits_{s= 1}^{N_{ls}}w_{ls}\widehat{\Sigma_{{\rm crit}}}(z_s, z_l)\epsilon_+^{l,s},
\label{eq:deltasigma_stack}
\end{equation}
where the sum runs over all lens-source pairs, and only for sources located within the physical projected radius interval $[R - \Delta R / 2, R + \Delta R / 2]$ from the lens $l$. Here, $\epsilon_+^{l,s}$ is the tangential ellipticity of the source galaxy $s$ relative to the position of lens $l$, as given by \eqreff{eq:epsilon_+x}. The quantity $\widehat{\Sigma_{\rm crit}}(z_s, z_l)$ is the effective critical surface mass density of the lens-source system, averaged over the photometric redshift probability density function $p(z_s)$ of the galaxy with index $s$, such that
\begin{equation}
    \widehat{\Sigma_{\rm crit}}(z_s, z_l)^{-1} = \int_{z_l}^{+\infty} d z_s\ p(z_s)\ \Sigma_{\rm crit}(z_s,z_l)^{-1}.
    \label{eq:sigma_crit_photoz}
\end{equation}
The weights $w_{ls}$ maximize the signal-to-noise ratio for this estimator \citep{Sheldon2004DeltaSigmaestimator} and can be written as the product $w_{ls} = w_{ls}^{\rm geo} w_{ls}^{\rm shape}$, such that
\begin{align}
    \label{eq:weights}
    w_{ls}^{\rm geo} &= \langle\widehat{\Sigma_{{\rm crit}}}(z_s, z_l)\rangle^{-2} ,\\
    w_{ls}^{\rm shape} &= \frac{1}{\sigma^2_{\rm rms}(\epsilon^+_s) + \sigma^2_{\rm meas}(\epsilon^+_s)}.
\end{align}
The quantity $\sigma_{\rm rms}(\epsilon^+_s) = \sigma_{\rm SN}$ is the intrinsic galaxy shape noise of the ellipticity's tangential component, arising from the fact that unlensed galaxies have an intrinsic distribution of ellipticities and position angles \citep[see e.g.][]{Pranjal2023wlshapenoise,Chang2013pdfgalaxies}. In cosmoDC2, the intrinsic shape noise for each galaxy's ellipticity component is $\sigma_{\rm SN} = 0.25$\footnote{To infer the intrinsic cosmoDC2 galaxy shape noise, we compute $\sigma_{\rm SN} = [\text{stddev}(\epsilon_1^{\rm int}) + \text{stddev}(\epsilon_2^{\rm int})]/2$, where $\epsilon_{\{1,2\}}^{\rm int}$ are the two components of the intrinsic galaxy ellipticities, identified as \texttt{ellipticity$\_\{$1,2$\}\_$true} in the cosmoDC2 catalog.}, which is comparable to what has been found in the literature \citep{Chang2013pdfgalaxies}. In contrast, $\sigma_{\rm meas}(\epsilon^+_s)$ represents the uncertainty originating from galaxy ellipticity estimation on images \citep[see e.g.][]{HIRATA2003HSM,Mandelbaum2005HSM,Sheldon2017metacalibration}. In the ideal case of cosmoDC2, where galaxy redshifts and shapes are perfectly measured, these weights reduce to $w_{ls}^{\rm geo} = \Sigma^{-2}_{{\rm crit}}(z_s, z_l)$ and $w_{ls}^{\rm shape} = \sigma^{-2}_{\rm rms}(\epsilon^+_s)$, where only the intrinsic variation of galaxy shapes remains.

The evaluation of the $\widehat{\Delta\Sigma}_+(R)$ data vector relies on a selection of background sources. Source selection is initially based on an $r$-band magnitude cut of $r < 28$ and an $i$-band magnitude cut of $i < 24.25$, aiming to reach a number density of galaxies $n_{\rm gal} \approx 25$ arcmin$^{-2}$, comparable to the number density expected in the context of LSST after 10 years of data \citep{Chang2013pdfgalaxies}. For each cluster in the redMaPPer catalog, we then extract the galaxy source catalog in a circular aperture of $R = 10$ Mpc, applying the source selection $z_s > z_l + 0.2$\footnote{We adopt a redshift separation of 
$\Delta z=0.2$ to mimic the common mitigation techniques of projection effects due to e.g., contamination from foregrounds or cluster member galaxies. Additionally, since the lensing strength increases with the separation between the lens and the source, this threshold ensures that selected sources are more strongly lensed, thereby enhancing the signal-to-noise ratio of the measurement. We do not explore less conservative cuts (e.g., $\Delta z = 0.1$ as in \citealp{McClintock2019masscalibration}) in this work, as we expect them to have minimal impact on the results aside from increasing the per-cluster source galaxy number density, thereby enhancing the precision of the lensing measurement.}, where $z_s$ is the true cosmoDC2 galaxy redshift and $z_l$ is the redMaPPer cluster redshift\footnote{We will present in \secreff{sec:results} other methodologies we consider for source selection when using photometric redshifts.}. 

We construct each source galaxy lensed ellipticity $\epsilon_{\rm obs}$ from \eqreff{eq:e_obs_e_int} using (i) its intrinsic shape $\epsilon_{\rm int}$ and (ii) the shear and convergence values at the galaxy's location that are provided in the cosmoDC2 catalog. 
For each stack of $N_l$ clusters, we consider 10 log-spaced radial bins from 0.5 Mpc to 10 Mpc and estimate the stacked lensing profile using \eqreff{eq:deltasigma_stack}.

\begin{figure*}
    \centering
    \includegraphics[width=0.95\textwidth]{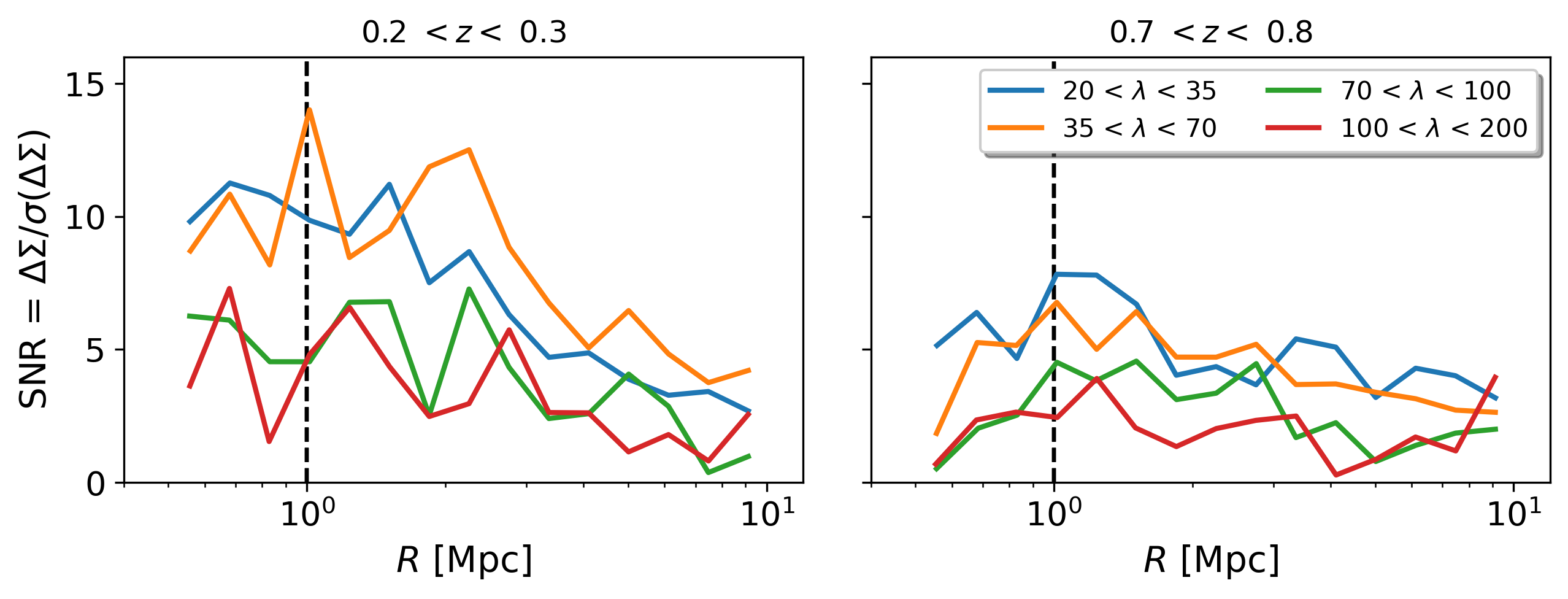}
    \caption{Signal-to-noise ratio of the stacked excess surface density profiles, for two different richness bins (left and right panel) and different richness bins (colors). }
\label{fig:excess_surface_density_profiles_snr}
\end{figure*}

The corresponding excess surface density profiles are displayed in \figreff{fig:excess_surface_density_profiles} as a function of the distance to the cluster center, for different richness bins and different redshift bins. At fixed richness, we do not observe a particular trend with redshift at scales larger than 1 Mpc, as denoted by the vertical dashed lines. 

However, for $R < 1$ Mpc, we observe that the stacked lensing profiles are progressively attenuated in the innermost regions with redshift. This attenuation has already been observed in DC2 galaxy-galaxy lensing \citep{Korytov2019cosmoDC2} and cluster-galaxy lensing \citep{Kovacs2022cosmoDC2} and is associated with the limited resolution of ray tracing used to compute the lensing shear and convergence at each galaxy position in the simulation. As a consequence, we chose to use only the $R > 1$ Mpc region for each stack in the analysis. This choice is valid across the full richness and redshift range. However, it also means that we cannot use the innermost regions with the largest signal-to-noise ratio (SNR) values. This intrinsic limitation will constrain the cluster-related forecasts of what could be achieved with the LSST data. We show in \figreff{fig:excess_surface_density_profiles_snr} the signal-to-noise ratio of some excess surface density profiles for a low (left panel) and a high (right panel) redshift bin. In the considered redshift range, the SNR oscillates between 2 and 13 for the low redshift bin, and between 0.5 and 8 for the high redshift bin. The profiles in the last richness bin have the lowest SNR, due to the low statistics in these bins (with $\sim 10$ times fewer clusters than in the first richness bin). Let us note that the stacking strategy allows us to significantly increase the SNR, as the SNR of individual cluster excess surface density profiles (not shown here) ranges from 0.1 to 3\footnote{To go from stacked profile SNR to individual profile SNR, a good approximation is to rescale them by $1/\sqrt{N_l}$.}.

\subsubsection{Mean cluster masses}
\label{sec:stacked_mass}

As mentioned earlier, instead of directly using the stacked profiles for weak-lensing information to constrain the mass-richness relation, one can use the corresponding mean cluster lensing masses as an alternative data vector. To do so, for each $ij$-th redshift-richness bin, we first need to infer the mean cluster mass $\log_{10}\widehat{M}_{ij}$ and its corresponding error $\sigma(\log_{10}\widehat{M}_{ij})$ from the corresponding stacked lensing profile $\widehat{\Delta\Sigma}_{ij}$. 

This is done by fitting, for each bin, a dark matter density profile\footnote{Many parameterizations exist in the literature, such as the Navarro-Frank-White \citep{Navarro1997nfw}, as well as those by \citet{Einasto1965haloprofile} and \citet{Hernquist1990haloprofile}.} to the corresponding stacked lensing profile obtained in \secreff{sec:data_vector_ds_profiles} using the \texttt{Minuit}\footnote{We use \texttt{minuit.migrad()} and \texttt{minuit.minos()} sequentially to (i) determine the best-fit mean cluster mass $\log_{10}M$ and (ii) compute the asymmetric error bars $\log_{10}M^{+\sigma_+}_{-\sigma_-}$. In \eqreff{eq:likelihood_cluster_mass}, we adopt a Gaussian likelihood with symmetric errors, for which we use the mean error $(\sigma_+ + \sigma_-)/2$. However, asymmetric errors can be more accurately modeled using, for example, a split normal distribution.} minimizer \citep{James1975minuit}, which is much faster than an MCMC. We ensured that the recovered mass and error from \texttt{Minuit} coincide at the $<0.1\sigma$ level with the mean mass and errors inferred from an MCMC procedure.

\begin{figure*}
    \centering
    \includegraphics[width=1\textwidth]{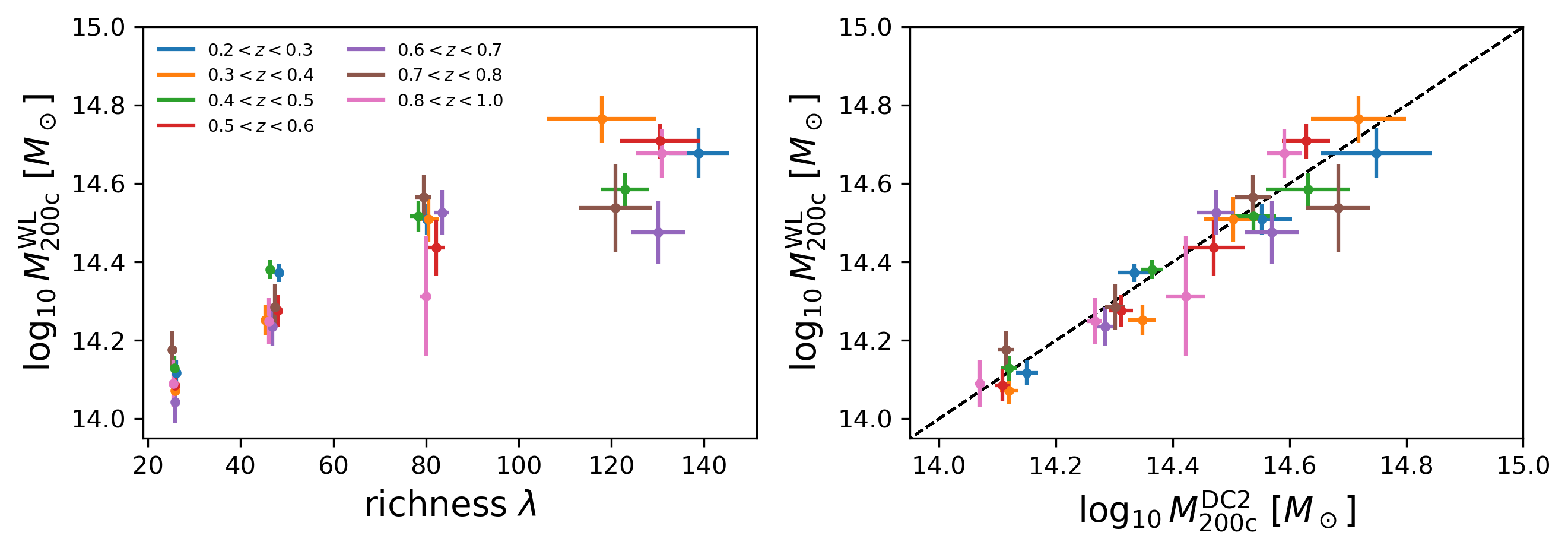}
    \caption{Left: Mean cluster lensing mass as a function of mean richness within the stack, the different colors correspond to the different redshift bins. Right: Mean cluster lensing masses as a function of the DC2 matched masses from the cosmoDC2 dark matter halo catalog. The different colors correspond to the different redshift bins. The dashed line corresponds to $y=x$.}
    \label{fig:stacked_masses}
\end{figure*}

We use a Navarro-Frank-White profile \citep[NFW,][]{Navarro1997nfw} with the concentration-mass relation of \citet{Duffy2008cM}\footnote{These modeling choices are referred to as the "baseline" analysis in the latter \ref{sec:baseline_analysis}.}. After fitting the stacked profiles within the radial range\footnote{We discuss the upper radial cut $R=3.5$ Mpc in \secreff{sec:stacked_lensing_profile_likelihood}, which is set to use only the 1-halo term lensing signal.} $R \in [1, 3.5]\ $Mpc, we show in the left panel of \figreff{fig:stacked_masses} the inferred mean cluster masses as a function of richness for the different redshift bins. The error bars on the $y$-axis correspond to the statistical error obtained from our fitting procedure of the cluster mass, while the $x$-axis error represents the dispersion of richness values in each bin.

For validation, we have compared the recovered mean lensing mass to the true mean mass in each redshift-richness bin in the right panel of \figreff{fig:stacked_masses}, where the true masses are obtained by (i) matching the redMaPPer cluster catalog to dark matter halos (ii) computing the corresponding cosmoDC2 mass in each redshift-richness bin. We see a clear correlation between weak lensing mean masses and cosmoDC2 masses (we added the one-to-one line $x = y$ to allow for a visual comparison).

\subsection{Likelihoods and priors}
\label{sec:likelihoods}
After building the data vectors, we may now turn our attention to inferring the parameters of the mass-richness relation defined in \secreff{sec:scaling_relation_formalism}, namely $\theta = (\ln\lambda_0, \mu_z, \mu_m, \sigma_{\ln \lambda_0}, \sigma_z, \sigma_m)$. We adopt the pivot redshift $z_0 = 0.5$ and pivot mass $m_0 = 10^{14.3} M_\odot$ in the mass-richness relation in \eqreff{eq:richness_mass}.

\subsubsection{Cluster count likelihood}
We consider the cluster count likelihood \citep{Hu2003SampleVariance} given by 
\begin{equation}
    \mathcal{L}_{\rm N} \propto |\Sigma_{\rm N}|^{-1/2}\exp  -\frac{1}{2}\sum_{ijkl}(N-\widehat{N})_{ij}[\Sigma_{\rm N}]^{-1}[ij,kl](N-\widehat{N})_{kl} .
    \label{eq:likelihood_cluster_count}
\end{equation}
In the above equation, the index $ij$ (resp. $kl$) refers to the $i$-th redshift bin and the $j$-th richness bin (respectively the $k$-th redshift bin and the $l$-th richness bin). $\widehat{N}_{ij}$ (resp. $N_{ij}$) denotes the measured (resp. predicted) cluster count in the $ij$-th redshift–richness bin. For the count prediction, we adopt $m_{\rm min} = 10^{12}  M_\odot$ and $m_{\rm max} = 10^{15.5}  M_\odot$ as the minimum and maximum halo masses for the integration of the halo mass function of \citet{Despali2015hmf} in \eqreff{eq:number_density}. The best-fit selection function $\Phi$, as described in \secreff{sec:redmapper_cluster_catalog}, is used in \eqreff{eq:number_density}. The quantity $\Sigma_{\rm N}$ is the cluster count covariance matrix, which accounts for two contributions: the Poisson shot noise and the Super-Sample Covariance (SSC). First, the Poisson noise affects the counting of clusters in uncorrelated bins \citep{Poisson1837}. Second, the SSC denotes the contribution from the intrinsic fluctuations of the matter overdensity field, within and beyond the survey volume \citep{Hu2003SampleVariance}. As a result, SSC introduces a covariance between the counts in different bins and increases their variance. SSC is particularly important in cluster abundance-based likelihoods, as it impacts the precision of the recovered cosmological parameters by approximately $20\%$ for surveys such as the Vera C. Rubin Observatory’s LSST or the \textit{Euclid} mission \citep{Fumagalli2021pinochhio,Payerne2023testinglikelihoodaccuracy,Payerne2024unbinnedSSC}, which are each expected to detect hundreds of thousands of clusters. The shot-noise and SSC contributions to the total covariance can, in principle, be estimated through resampling techniques directly on the data, such as jackknife resampling \citep{Escoffier2016jackknife}. However, due to the relatively small number of clusters detected in the DC2 galaxy catalog, jackknife estimates of the cluster count covariance matrix suffer from high noise. We therefore opt to use a theoretical prediction for the covariance instead, given by 
\begin{equation}
     \Sigma_{\mathrm{N}}[ij,kl] = N_{ij}\delta^K_{ik}\delta^K_{jl}+ N_{ij}N_{kl}\langle b\rangle_{ij}\langle b\rangle_{kl} S_{ik}. 
     \label{eq:covariance}
\end{equation}
The first term in \eqreff{eq:covariance} represents the Poisson shot noise, with $\delta^K_{ik}$ denoting the Kronecker delta function. The second term corresponds to SSC, where $S_{ik} = \langle \delta_{m,i} \delta_{m,k} \rangle$ is the covariance of the smoothed matter overdensities $\delta_{m}$ in the $i$-th and $k$-th redshift bins, respectively \citep{Lacasa2018SSCpartialsky}. The coupling between the matter overdensity field and the halo number density field is encoded by the halo bias $b(m,z)$ \citep{Tinker10halobias}, which depends on mass and redshift and is averaged over the $ij$-th (resp. $kl$-th) redshift–richness bin to yield $\langle b \rangle_{ij}$ (resp. $\langle b \rangle_{kl}$). Finally, $N_{ij}$ is the model prediction for the number of clusters in the $ij$-th redshift-richness bin, as defined in \eqreff{eq:N_count}. This likelihood model, which incorporates both Poisson noise and SSC, has been adopted in previous cluster analyses such as \citet{Abbott2020DESCL,Costanzi2019SDSSCL,Sunayama2023HSClensing}.

\subsubsection{Stacked lensing profile likelihood}
\label{sec:stacked_lensing_profile_likelihood}
We can constrain the mass–richness relation by using the stacked excess surface density profiles directly \citep{Murata2019HSCrichnessmassrelation,Park2023lensingabundance,Sunayama2023HSClensing}. The likelihood for stacked excess surface density profiles, denoted $\mathcal{L}_{\rm WLp}$, is given by
    \begin{equation}
    \mathcal{L}_{\Delta\Sigma}\propto \exp  -\frac{1}{2}\sum_{ij}(\Delta\Sigma-\widehat{\Delta\Sigma})_{ij}\Sigma^{-1}_{\Delta\Sigma}[ij](\Delta\Sigma-\widehat{\Delta\Sigma})_{ij}.
    \label{eq:likelihood_cluster_profile}
\end{equation}
Here, $\widehat{\Delta\Sigma}_{ij}$ corresponds to the stacked excess surface density profile estimated in Eq.~(\ref{eq:deltasigma_stack}) for the $ij$-th redshift–richness bin, $\Delta\Sigma_{ij}$ is the corresponding theoretical prediction from Eq.~(\ref{eq:DS_stack_th}), and $\Sigma_{\Delta\Sigma}[ij]$ denotes the associated covariance matrix. This covariance originates from a variety of phenomena, as discussed in \citet{Hoekstra2003uncorLSS,Wu2019covarianceDeltaSigma,Gruen2015covDeltaSigma,McClintock2019masscalibration}. A primary contribution is the intrinsic scatter in the shapes of background galaxies, combined with the limited number of clusters and source galaxies within each bin. This term, often referred to as shape noise, dominates the small-scale regime but can be reduced by increasing the sample size. Another significant contribution comes from uncorrelated large-scale structures along the line of sight, which introduce stochastic fluctuations in the lensing signal unrelated to the lensing cluster. In addition, correlated structures around the cluster—such as nearby halos and filaments—affect the lensing profile through variations in the two-halo term, thereby contributing an additive component to the covariance. At smaller scales, intrinsic variations in the properties of halos within each stack, including the scatter in concentration at fixed mass, halo ellipticity, and orientation, also contribute to the overall uncertainty. Finally, the spread in true halo masses among richness-selected clusters within a given bin introduces further variance in the measured lensing profiles.

As for cluster counts, the covariance of the cluster lensing profiles can be predicted analytically \citep[see e.g.][]{Wu2019covarianceDeltaSigma}\footnote{\citet{Wu2019covarianceDeltaSigma} developed the \texttt{cluster-lensing-cov} package \footnote{\url{https://github.com/hywu/cluster-lensing-cov}}, which provides analytical predictions for the excess surface density profile covariance.} or using semi-analytical approaches \citep[see e.g.][]{McClintock2019masscalibration}. These methods require modeling a wide range of data properties, including the redshift distributions of the source and lens samples, the dispersion of intrinsic galaxy shapes, the dark matter halo density profile, and the level of scatter in concentration, miscentering, and morphology of clusters in the redMaPPer catalog. In our case, we choose a simpler, data-driven approach that estimates all these contributions simultaneously without relying on model assumptions, while also keeping computational costs relatively low. Specifically, we use bootstrap resampling \citep[see e.g.][]{Simet2017SDSSmassrichness,Parroni2017CLlensingcov} with $N_{\rm boot} = 400$ resampled cluster ensembles. We retain only the diagonal terms of the covariance matrices, as the off-diagonal elements are too noisy due to the limited cluster count statistics in each richness–redshift bin within the DC2 footprint. \citet{Wu2019covarianceDeltaSigma} pointed out that neglecting off-diagonal terms at large scales ($R \gtrsim 5$–6 Mpc) can lead to an underestimation of parameter uncertainties inferred from weak-lensing profiles -- such as cluster mass and concentration -- especially when shape noise (i.e., uncorrelated sources of noise) is subdominant. However, since our analysis focuses exclusively on the $R < 3.5$ Mpc regime, this impact is expected to be minimal. Therefore, we assume diagonal covariance matrices throughout, as the off-diagonal terms remain extremely noisy \citep{Phriksee2019WLcov}.

\subsubsection{Mean cluster mass likelihood}
The cluster mass-richness relation can also be inferred using the information from mean mass estimates \citep{McClintock2019masscalibration,Abbott2020DESCL,Lesci2022KIDSCL}. Then, the mean mass likelihood is given by
    \begin{equation}
    \mathcal{L}_{M_{\rm WL}}\propto \exp  -\frac{1}{2}\sum_{ij}\left(\frac{\log_{10}M_{ij}-\log_{10}\widehat{M}_{ij}}{\sigma(\log_{10}\widehat{M}_{ij})}\right)^2,
    \label{eq:likelihood_cluster_mass}
\end{equation}
where $\log_{10}M_{ij}$ is the mass prediction that is given in \eqreff{eq:M_lensing} in the $ij$-th redshift-richness bin, and $\log_{10}\widehat{M}_{ij} \pm \sigma(\log_{10}\widehat{M}_{ij})$ is the measured mean mass. 

It is important to highlight the differences between the one-step approach (described in \secreff{sec:stacked_lensing_profile_likelihood}) and the two-step approach (discussed in this section). The one-step approach offers greater flexibility. Specifically, it allows for more comprehensive modeling of various systematic effects that influence the lensing signal, including cluster miscentering, projection effects from both correlated and uncorrelated large-scale structures, contamination by cluster member galaxies, and deviations from spherical symmetry in the cluster mass distribution compared to the two-step approach. However, the one-step approach presents a significant computational challenge, as it necessitates marginalizing over numerous theoretical and observational effects \citep{Aguena2023DESpipeline}, compared to the two-step approach. 

\subsubsection{Total likelihood and priors}
To infer the parameters of the mass-richness relation,  the total likelihood $\mathcal{L}_{\rm tot}$ is given by considering either counts, cluster lensing profiles, or cluster lensing masses, or the combination between counts and cluster lensing, i.e.
\begin{equation}
    \mathcal{L}_{\rm tot} = 
\begin{cases}
  \mathcal{L}_{\rm N} & \text{if: Abundance alone, }  \\
  \mathcal{L}_{\Delta\Sigma} & \text{if: Lensing profiles alone, }  \\
  \mathcal{L}_{M_{\rm WL}} & \text{if: Lensing masses alone, }\\
  \mathcal{L}_{\Delta\Sigma}\times\mathcal{L}_{\rm N} & \text{if: Lensing profiles + Abundance,}  \\
  \mathcal{L}_{M_{\rm WL}}\times\mathcal{L}_{\rm N} & \text{if: Lensing masses + Abundance}.
\end{cases}
\label{eq:total_likelihood}
\end{equation}
It is important to note that the way the lensing and count likelihoods are combined assumes no covariance between the two. As shown by \citet{Costanzi2019SDSSCL}, the cross-correlation between the abundance and weak lensing inferred quantities is consistent with zero.

We draw samples from the parameter posterior distribution given by
\begin{equation}
    \mathcal{P}(\theta|\mathrm{data}) = \frac{\mathcal{L}_{\rm tot}(\mathrm{data}|\theta)\times\pi(\theta)}{\mathcal{L}_{\rm tot}(\mathrm{data})},
    \label{eq:posterior}
\end{equation}
where $\theta = (\ln\lambda_0, \mu_z, \mu_m, \sigma_{\ln \lambda_0}, \sigma_z, \sigma_m)$, $\mathcal{L}_{\rm tot}(\mathrm{data}|\theta)$ is the total likelihood in \eqreff{eq:total_likelihood} ($\mathcal{L}_{\rm tot}(\mathrm{data})$ is the likelihood evidence) where “data” refers to either the stacked lensing profiles, the inferred mean lensing masses, the cluster counts, or combinations of abundance and lensing information. 

To compute the prior $\pi(\theta)$, we consider the choices made for the fiducial mass-richness relation and summarized in the second column of \tabreff{tab:best_fits_scaling_relation_matching}. All fits are performed at a fixed cosmology, chosen to be the fiducial cosmology of the DC2 simulation, which is close to the seven-year \textit{Wilkinson Microwave Anisotropy Probe} best-fit $\Lambda$CDM values \citep[WMAP,][]{Komatsu2011wmap} given by $h = H_0/100 = 0.71$ (the reduced Hubble constant $H_0$), $\Omega_{\rm cdm}h^2 = 0.1109$ ($\Omega_{\rm cdm}$ is the current fractional energy density associated with cold dark matter), $\Omega_{\rm b}h^2 = 0.02258$ ($\Omega_{\rm b}$ is the current fractional energy density associated with baryons), $n_s = 0.963$ (spectral index of the primordial power spectrum), $\sigma_8 = 0.8$ (dispersion of matter density fluctuations smoothed over $8\ h^{-1}$ Mpc), and $w = -1$ (dark energy equation of state). 

\subsection{Softwares}
\label{sec:softwares}
This work allows us to demonstrate the use of some of the software tools developed in DESC (in combination with others) for concrete cluster analysis.
For this work, we use several DESC software tools that are currently in development as part of the DESC pipeline for analyzing the upcoming LSST data. To extract the cluster, halo, and background galaxy catalogs from the DC2 dataset, we use the DESC package \texttt{GCRCatalogs}\footnote{\url{https://github.com/LSSTDESC/gcr-catalogs}}, along with \texttt{Qserv}\footnote{\url{https://qserv.lsst.io/index.html}}, an open-source SQL database system (Massively Parallel Processing) originally designed to host LSST data. For the estimation of the stacked lensing profiles, we use the DESC Cluster Lensing Mass Modeling \citep[CLMM,][]{Aguena2021CLMM} package\footnote{\url{https://github.com/LSSTDESC/CLMM}}, which provides various tools for estimating cluster lensing profiles as well as for halo modeling. The Core Cosmology Library\footnote{\url{https://github.com/LSSTDESC/CCL}} \citep[CCL,][]{Chisari2019CCL} is used to predict the halo mass function, halo bias, and concentration-mass relations. The combination of these codes is carried out within the {\tt LSSTDESC/CLCosmo$\_$Sim} repository\footnote{\url{https://github.com/LSSTDESC/CLCosmo_Sim}}. To compute the binned $S_{ij}$ terms for the cluster count covariance in \eqreff{eq:covariance}, we use the \texttt{PySSC}\footnote{\url{ https://github.com/fabienlacasa/PySSC}} package \citep{Lacasa2018SSCpartialsky,Gouyou2022SSC}. Cluster scaling relation parameters were inferred using the Markov Chain Monte Carlo (MCMC) implementation in the {\tt emcee} package \citep{ForemanMackey2013emcee}, with visualization performed via {\tt getdist} \citep{Lewis2019getdist}. Finally, we also use the \texttt{Minuit}\footnote{\url{https://scikit-hep.org/iminuit/about.html}} minimizer \citep{James1975minuit} to fit the mean cluster mass per redshift-richness bin in the two-step procedure.

\section{Results}
\label{sec:results}

All the ingredients are now in place to perform the analysis of the scaling relation of redMaPPer clusters. In \secreff{sec:baseline_analysis}, we begin by presenting the setup that serves as our baseline analysis. Subsequent sections present variations of this baseline, changing one ingredient at a time.\secreff{sec:impact_modeling} explores how the modeling choices for the halo density and mass-concentration relation impact the results, while \secreff{subsec:results_obs} focuses on observational systematic effects, specifically the impact of source photometric redshifts and the shear-richness covariance. \tabreff{tab:summary_table} summarizes the tables and figures associated with the different study cases.
\begin{table*}
\begin{center}
\caption{List of the different study cases explored in \secreff{sec:results}.}
\label{tab:summary_table}
\begin{tabular}{ c|c|c|c} 
  Analysis &Section& Numerical values & Posterior distribution\\
\hline
\hline
 Fiducial relation &\secreff{sec:fiducial_relation} &\tabreff{tab:best_fits_scaling_relation_matching}&-\\ 
 Baseline &\secreff{sec:baseline_analysis}& \tabreff{tab:params_WLN_mass_richness} (first block)& \figreff{fig:baseline_posterior_scaling_relation}\\
  Modeling - $c(M)$ relation &\secreff{sec:impact_cM_relation}& \tabreff{tab:params_WLN_mass_richness} (second block)& \figreff{fig:impact_cM_posterior_scaling_relation} (left panel) \\ 
Modeling - dark matter density profile &\secreff{sec:impact_halo_model}& \tabreff{tab:params_WLN_mass_richness} (third block)& \figreff{fig:impact_cM_posterior_scaling_relation} (right panel)\\ 
Observational - photometric redshifts & \secreff{sec:impact_photoz}& \tabreff{tab:params_WLN_mass_richness} (fourth block)& \figreff{fig:impact_photoz_posterior_scaling_relation} \\ 
Observational - shear-richness covariance &\secreff{sec:impact_shear_richness}& \tabreff{tab:params_WLN_mass_richness} (fifth block)& \figreff{fig:impact_shear_richness_cov_posterior_scaling_relation} 
\end{tabular}
\end{center}
\end{table*}

\subsection{Defining the count/weak lensing baseline analyses: $N+M_{\rm WL}$ versus $N+\Delta\Sigma$}
\label{sec:baseline_analysis}
We model the per-cluster excess surface density, $\Delta\Sigma(R|m,z)$, using a three-dimensional dark matter density profile given by
\begin{equation}
    \rho(r) = \rho_s\left(\frac{r}{r_s}\right)^{-1}\left( 1  + \frac{r}{r_s}\right)^{-n}.
    \label{eq:nfw_density}
\end{equation}
For $n=2$, the \eqreff{eq:nfw_density} corresponds to the Navarro-Frank-White profile \citep{Navarro1997nfw}. Then, $r_s = r_{\rm 200c}/c_{\rm 200c}$ is the scale radius, and $\rho_s$ is the scale density\footnote{The scale density is given by 
\begin{equation}
    \rho_s = \frac{200}{3}\frac{c_{\rm 200c}^3}{f(c_{\rm 200c})} \rho_{c}(z)\
\hspace{0.3cm}\mathrm{where}\hspace{0.3cm} f(x) = \ln(1 + x) - x/(1 + x).
\end{equation}
.}. We use the concentration mass-relation of \citet{Duffy2008cM} (see \ref{sec:impact_cM_relation} hereafter). 

We restrict the fitting to $1 < R < 3.5$ Mpc. The typical value of $R_{\rm max} = 3.5$ Mpc is inspired by \citet{Lee2018wlmass},  \citet{Giocoli2021stackedlensingAMICO},  \citet{Cromer2022lensingmassbaryons},  \citet{Murray2022lensingmasses},  \citet{Bocquet2024SPTCL} and is typically the distance beyond which the two-halo term increasingly dominates the cluster lensing signal and cannot be neglected (this choice is discussed further in Appendix \ref{app:impact_rmax} and later in the text).

\begin{figure*}
    \centering
    \includegraphics[width=.49\textwidth]{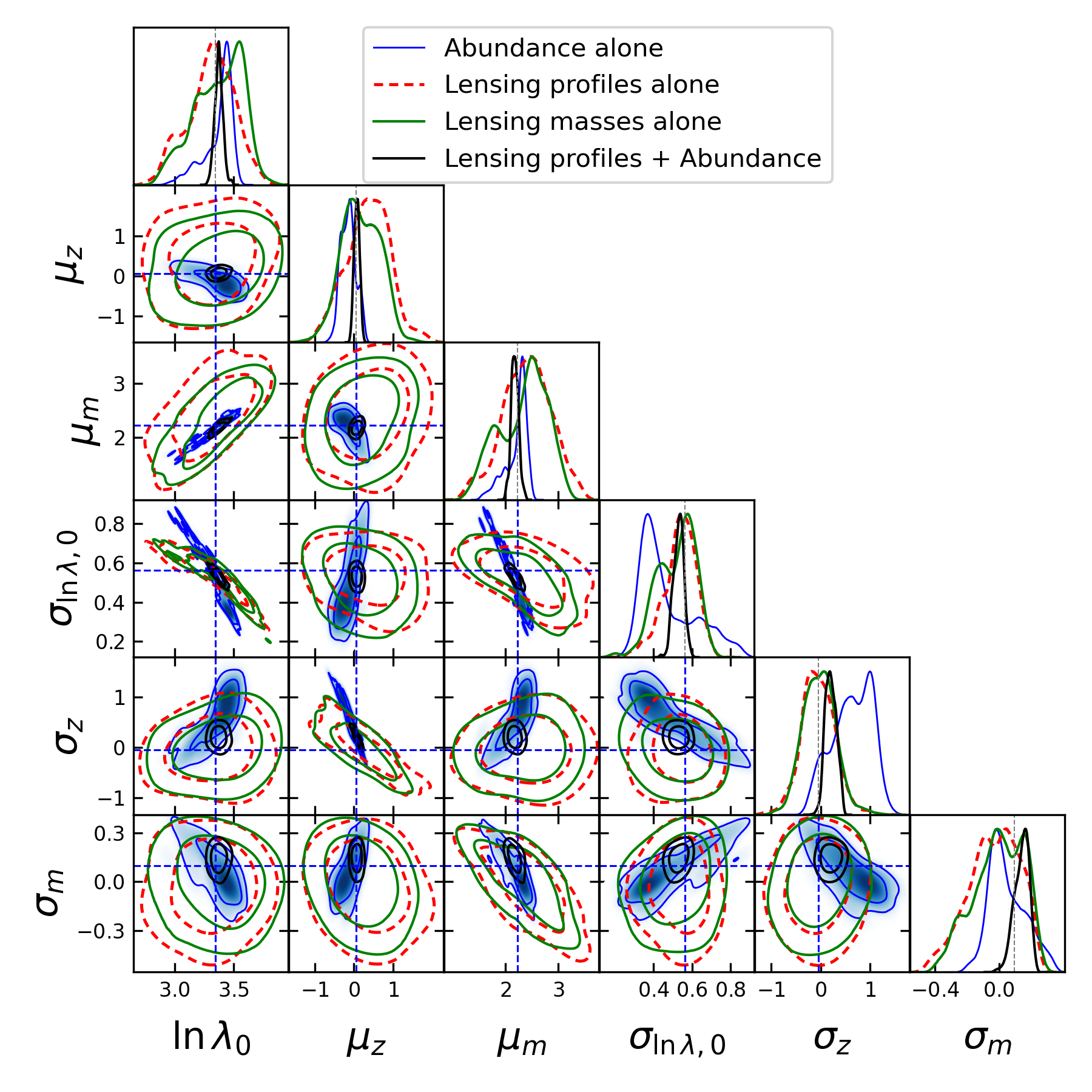}
    \includegraphics[width=.49\textwidth]{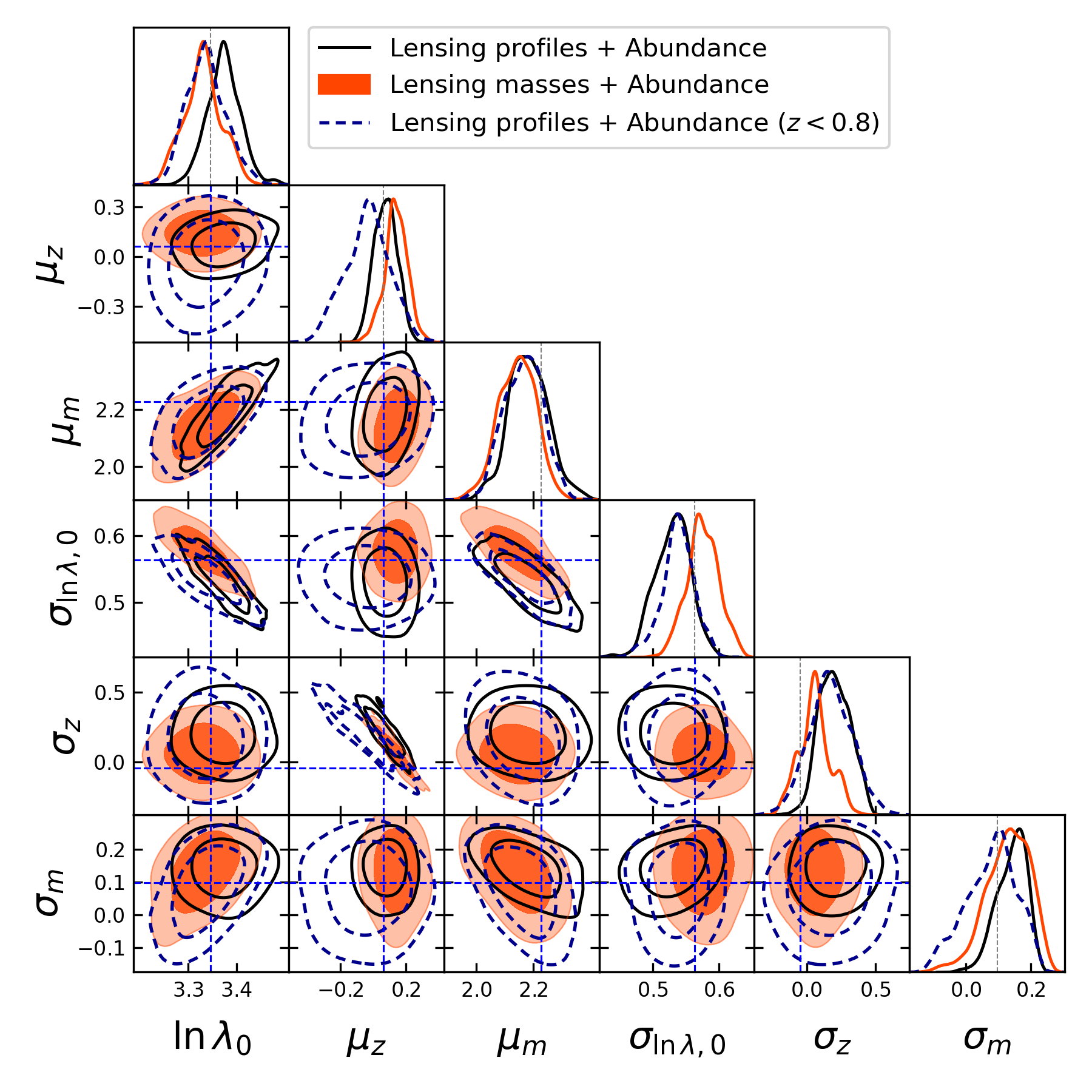}
    \caption{Left: Posterior distribution of the scaling relation parameters using abundance alone (blue), lensing profiles alone (dashed red), or masses alone (green, full lines). The joint abundance and lensing profiles posterior is displayed in black. Right: Posterior distribution of scaling relation parameters when combining abundance and lensing profiles (blue, full lines) and combining abundance and cluster masses (orange-filled contours). The dashed contours are obtained by removing the last redshift bin for both abundance and lensing. }
\label{fig:baseline_posterior_scaling_relation}
\end{figure*}

We show in the left panel of \figreff{fig:baseline_posterior_scaling_relation} the 68$\%$ and 95$\%$ credible intervals on the cluster scaling relation when using, separately, the cluster abundance likelihood, stacked lensing profiles, or mean cluster masses. The vertical/horizontal lines represent the fiducial values of the redMaPPer cluster scaling relation inferred in \secreff{sec:fiducial_relation}. We observe that the abundance provides tighter constraints on the mean parameters, but has different degeneracies compared to lensing (either stacked profiles or mean masses). The contours exhibit characteristic degeneracies and significant overlap, which suggests they can be combined. Additionally, we note that the posteriors are fairly compatible with the fiducial constraints at the 1-2$\sigma$ level.

The different degeneracy directions between the abundance and lensing contours provide tight final constraints on the scaling parameters when combined\footnote{Again, we recall that the cosmological parameters are fixed to the DC2 values in the MCMC.}. By combining abundance with lensing, we obtain the posteriors shown in the right panel of  \figreff{fig:baseline_posterior_scaling_relation}, either with stacked profiles or mean masses\footnote{To compare joint constraints with abundance-alone or lensing-alone contours, we show in black in the left panel of \figreff{fig:baseline_posterior_scaling_relation} the joint abundance and lensing profiles posterior.}. We observe that using either mean masses or stacked profiles provides roughly the same constraints with comparable precision. The errors on the mean scaling parameters ($\ln\lambda_0$, $\mu_z$, and $\mu_m$) are reduced by a factor of $\sim 7$, while the errors on the scatter terms ($\sigma_{\ln\lambda_0}$, $\sigma_z$, and $\sigma_m$) are reduced by a factor of $\sim 3-4$. The one-dimensional posteriors obtained from our joint lensing and abundance likelihood fairly recover the fiducial values at the $<2\sigma$ level; however, we observe a $>2\sigma$ tension with the projected posterior on $\mu_z-\sigma_z$.

We check whether the stacked lensing profiles in the last redshift bin are still impacted by the ray-tracing shear resolution beyond $R=1$ Mpc, despite our conservative radial cut, since the attenuation increases with redshift and is therefore maximal in the last bin. To assess this, we remove the last redshift bin from the estimation of the posterior (from both abundance and lensing profiles). This result is shown in dashed lines in \figreff{fig:baseline_posterior_scaling_relation} (right panel), and yields results compatible with our baseline analysis, albeit with wider error bars (the last redshift bin contains 30$\%$ of the full cluster catalog). This provides a sanity check for our lensing and count pipeline, which compares reasonably well to the 'fiducial' constraints of the redMaPPer mass–richness relation. Residual systematics inherent to our analysis -- such as the calibration of the halo mass function and uncertainties in the selection function -- may still induce some bias in the recovered parameters.

We also examine in Appendix \ref{app:impact_rmax} the impact of alternative $R_{\rm max}$ cuts on the scaling relation parameters. We find that using the full available radial range (up to $R_{\rm max} = 10$ Mpc) without modeling the two-halo term yields results comparable to those obtained with the baseline $R_{\rm max} = 3.5$ Mpc cut, showing only small biases ($<1\sigma$) and similar error bars. This is attributed to the noisy lensing measurements at larger scales, which diminish the impact of the two-halo term and do not provide additional constraining power. We adopt the $R_{\rm max} = 3.5$ Mpc cut throughout this work. However, the two-halo term cannot be neglected for larger datasets, which will provide a larger signal-to-noise ratio in the two-halo regime. 

In the following, we consistently use the combination of lensing (either mean masses or stacked profiles) and cluster counts to probe the underlying cluster scaling relation, leveraging their complementarity in parameter space.

\subsection{Impact of modeling choices (in the $N+M_{\rm WL}$ framework)}
\label{sec:impact_modeling}

The inference of the cluster scaling relation is impacted by several systematic effects, stemming from the many degrees of freedom in the modeling of cluster observables. The choice of a particular mass distribution model can significantly affect the inferred cluster lensing mass and, consequently, have a non-negligible impact on the mass–richness relation. We restrict our analysis to the one-halo regime, i.e., at scales below $R = 3.5$ Mpc, which is more sensitive to the internal halo properties (such as cluster mass and concentration) than the two-halo regime, which is primarily influenced by cosmological quantities through the power spectrum and halo bias \citep{Oguri2011lensing}.

\subsubsection{Concentration-mass relation}
\label{sec:impact_cM_relation}
As mentioned in \secreff{sec:formalism}, the cluster density profile depends on the concentration parameter $c_{\rm 200c}$.
Cosmological simulations indicate that cluster concentration and mass are statistically correlated. This correlation is typically quantified by a fitting formula known as the concentration-mass relation, labeled $c(M)$. The concentration $c$ of an individual cluster with mass $M$ is then scattered around $c(M)$, with a dispersion denoted $\sigma_{c|M}$. Several $c(M)$ relations exist in the literature \citep[see, e.g.,][]{Diemer2014haloprofile,Duffy2008cM,Prada2012cM,Bhattacharya2013cM,Diemer2019concentration}. For example, our baseline choice in \secreff{sec:baseline_analysis} adopts the $c(M)$ relation of \citet{Duffy2008cM}, calibrated using N-body simulations based on the 5-year WMAP cosmology \citep{Komatsu2009wmap5}, which differs by at most 5$\%$ from the 7-year WMAP cosmology \citep{Komatsu2011wmap} used in cosmoDC2. The alternative relations examined in this section \citep{Prada2012cM,Bhattacharya2013cM,Diemer2014haloprofile} have been shown to accurately describe the relationship between halo mass and concentration across a variety of $\Lambda$CDM cosmologies. These relations, despite some intrinsic scatter, capture the general trend observed in simulations. Employing such a relation can reduce the number of free parameters in cluster-based analyses and thus improves constraints on other relevant parameters.

To test the impact of the $c(M)$ relation on the scaling relation parameters, we (i) fit the mean cluster masses using various $c(M)$ relations from the literature (where $M$ denotes the mean mass, considered as a free parameter), (ii) use these mean mass estimates in \eqreff{eq:likelihood_cluster_mass}, and (iii) combine them with the cluster count likelihood from \eqreff{eq:likelihood_cluster_count} to infer the scaling relation parameters. Alternatively, the concentration can be treated as a free parameter in each redshift–richness bin. In this case, we jointly fit the concentrations and mean masses in each bin and use the resulting mass estimates in the mass likelihood.

\begin{figure*}
    \centering
\includegraphics[width=.48\textwidth]{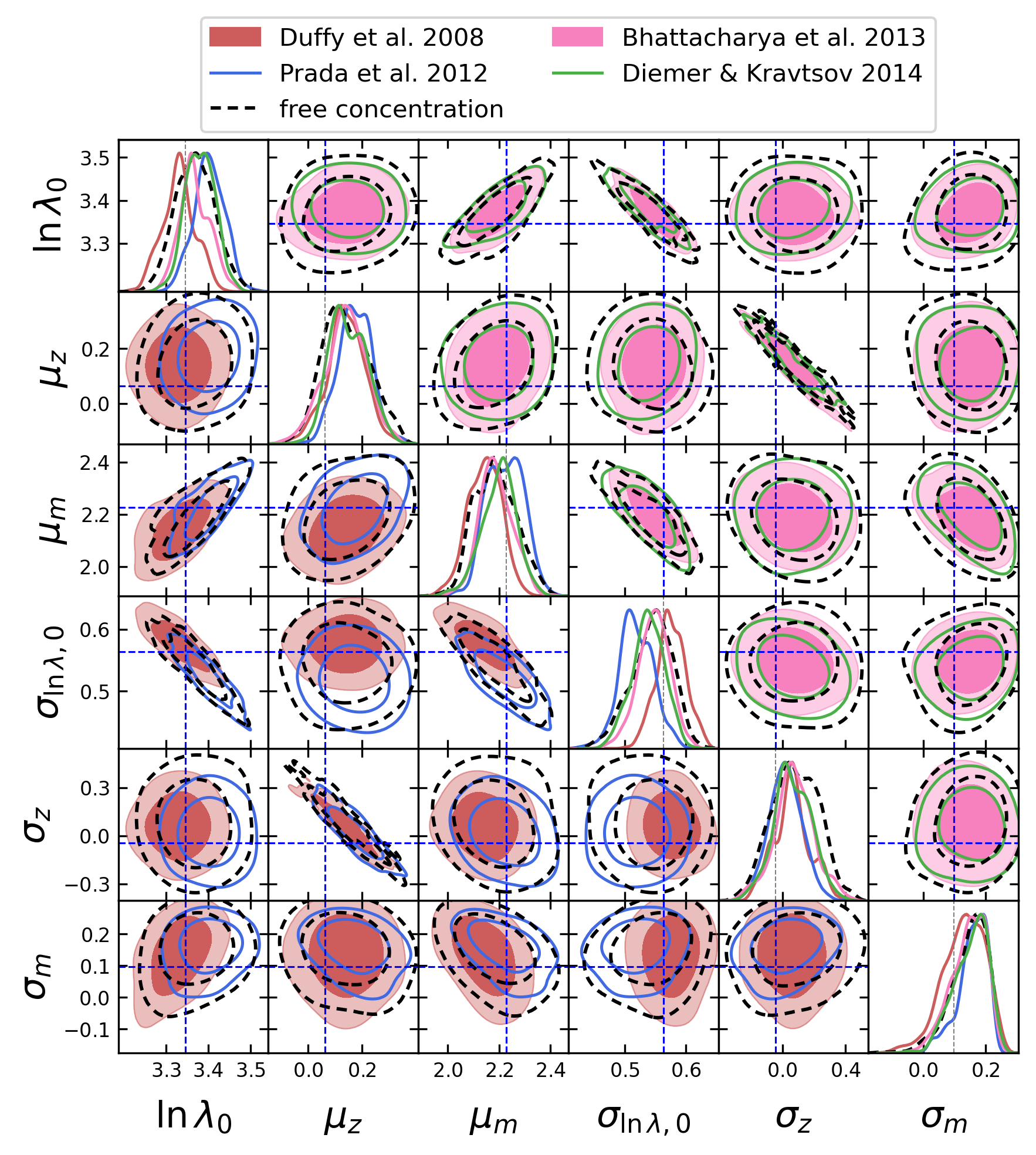}
\includegraphics[width=.48\textwidth]{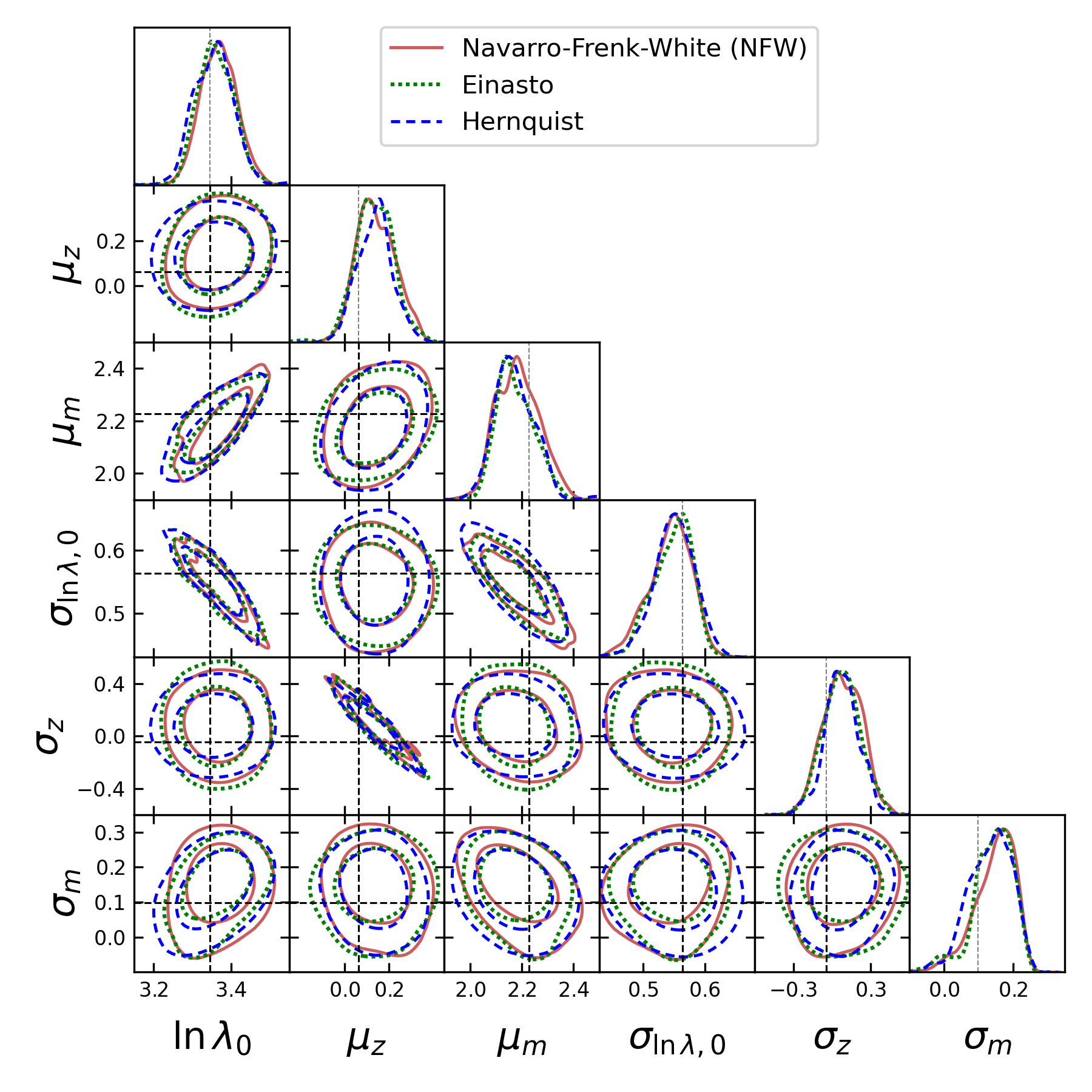}
    \caption{Left: Posterior distribution of the scaling relation parameters using count/mass likelihood. The concentration-free case is shown with the dashed lines, the other contours are obtained using various concentration-mass relations from the literature. Right: Posterior distribution of the scaling relation parameters using count/mass likelihood, without using a concentration-mass relation, and using different modeling of the cluster density profile, namely NFW (red filled contours), Einasto (full lines), Hernquist (dashed lines). }
\label{fig:impact_cM_posterior_scaling_relation}
\end{figure*}

In the left panel of \figreff{fig:impact_cM_posterior_scaling_relation} the concentration-free case shows reasonable agreement with the fiducial constraints. The posteriors obtained using different concentration-mass relations are fully consistent with the free-concentration case. When adopting a fixed $c(M)$ relation, the inferred posteriors exhibit the same level of parameter correlation, with slightly reduced error bars due to the smaller number of free parameters in the fitting procedure. We note that the choice of concentration model is expected to have a larger impact in the inner regions ($R < 1$ Mpc), which are not probed here due to the limitations imposed by ray-tracing resolution.

\subsubsection{Modeling of the matter density profile} 
\label{sec:impact_halo_model}
Besides the NFW parameterization, \citet{Einasto1965haloprofile} suggested an empirical profile that has been shown to provide a slightly better fit to N-body simulations of galaxy clusters compared to the NFW profile—particularly at small scales and over a broad range of halo masses and redshifts \citep{Klypin2016dmprofile,Sereno2016dmprofiles,Wang2020profile}. The corresponding three-dimensional density profile is given by
\begin{equation}
    \rho^{\rm ein}(r) = \rho_{-2}\exp\left(-\frac{2}{\alpha}\left[\left(\frac{r}{r_{-2}}\right)^\alpha - 1\right]\right),
\end{equation}
where $r_{-2} = r_{\rm 200c}/c_{\rm 200c}$ and the $\rho_{-2}$ is given in the footnote\footnote{
The density $\rho_{-2}$ is given by
\begin{equation}
    \rho_{-2} = 200\rho_{c}(z)\frac{2}{3}\gamma^{-1}\left[\frac{3}{\alpha}, \frac{2}{\alpha}c_{\rm 200c}^\alpha \right] \frac{c_{\rm 200c}^3}{e^{2/\alpha}} \left(\frac{2}{\alpha}\right)^{\frac{3-\alpha}{\alpha}},
\end{equation}
where $\gamma$ is the lower incomplete gamma function.}. In the above equation, $\alpha$ is the shape parameter. We adopt a typical constant value of $\alpha = 0.25$, though it is worth noting that more sophisticated models incorporate mass, redshift, and cosmology dependence in $\alpha$, as observed in simulations \citep{Gao2008einasto}. In such models, $\alpha$ typically ranges from 0.15 to 0.3 as a function of mass.

Less commonly, the Hernquist density model (corresponding to $n = 1$ in \eqreff{eq:nfw_density}) has been used, for example, in \citet{Buote2004Hernquist,Sanderson2009HennquidtX} to model the mass density of X-ray-detected galaxy clusters, though it has not been extensively used in the lensing literature.

Concentration–mass relations for the Einasto and Hernquist models have not been extensively studied in the literature. Therefore, we consider the fit without assuming a specific $c(M)$ relation, prioritizing the count/mass cluster likelihood instead. In the right panel of \figreff{fig:impact_cM_posterior_scaling_relation}, we show the posterior distribution of the scaling relation parameters when considering the count/mass likelihood, using the NFW, Einasto, and Hernquist profiles. The results are in good agreement with each other, likely due to the radial range used in the fit, where the profiles are similar and only start to diverge at smaller scales.

By fitting the mean cluster mass from 1 Mpc to 3.5 Mpc, where the one-halo regime is dominant, and using either the NFW, Einasto, or Hernquist profile, we found that the inferred scaling relation remains fairly stable.

\subsection{Observational systematic effects (in $N+\Delta\Sigma$ framework)}
\label{subsec:results_obs}
We addressed several potential sources of bias arising from uncertainties in the modeling of the cluster lensing observable. Another major source of bias in cluster lensing analyses comes from the data itself, including the calibration of galaxy shapes \citep{Hernandez2020shape}, the redshift distribution of the background galaxy sample \citep{Wright2020photoz}, contamination of the source galaxy sample by foreground galaxies \citep{Varga2019contamination}, and miscentering in the cluster catalog \citep{Sommer2022miscentering}.

Using the cosmoDC2 catalog, we rely on the true shapes of galaxies (i.e., accounting only for the intrinsic shape and local shear), and thus cannot investigate the effects of shape measurement errors with this dataset. 
In the following, we focus instead on the impact of the calibration of the background galaxy photometric redshift distribution on the cluster scaling relation (\secreff{sec:impact_photoz}), as well as on the impact of shear–richness covariance (\secreff{sec:impact_shear_richness}). With this simulated dataset and the analysis choices we made, the effects of miscentering and contamination of the source sample by cluster members, which generally need to be accounted for, are not significant; they are discussed in Appendix \ref{app:miscentering} and Appendix \ref{app:contamination}, respectively.

\subsubsection{Source galaxy photometric redshifts} 
\label{sec:impact_photoz}
In the previous \secreff{sec:baseline_analysis} and \secreff{sec:impact_modeling}, we used the true galaxy redshifts to compute the geometrical lensing weights $w^{\rm geo}_{ls}$. In contrast, with imaging data, each source galaxy behind clusters is assigned a photometric redshift distribution, derived from galaxy magnitudes measured in several optical bands. The uncertainty in the measured redshift can be incorporated into the lens-source geometrical weights through \eqreff{eq:sigma_crit_photoz}.

\begin{figure*}
   \centering
\includegraphics[width=0.5\textwidth]{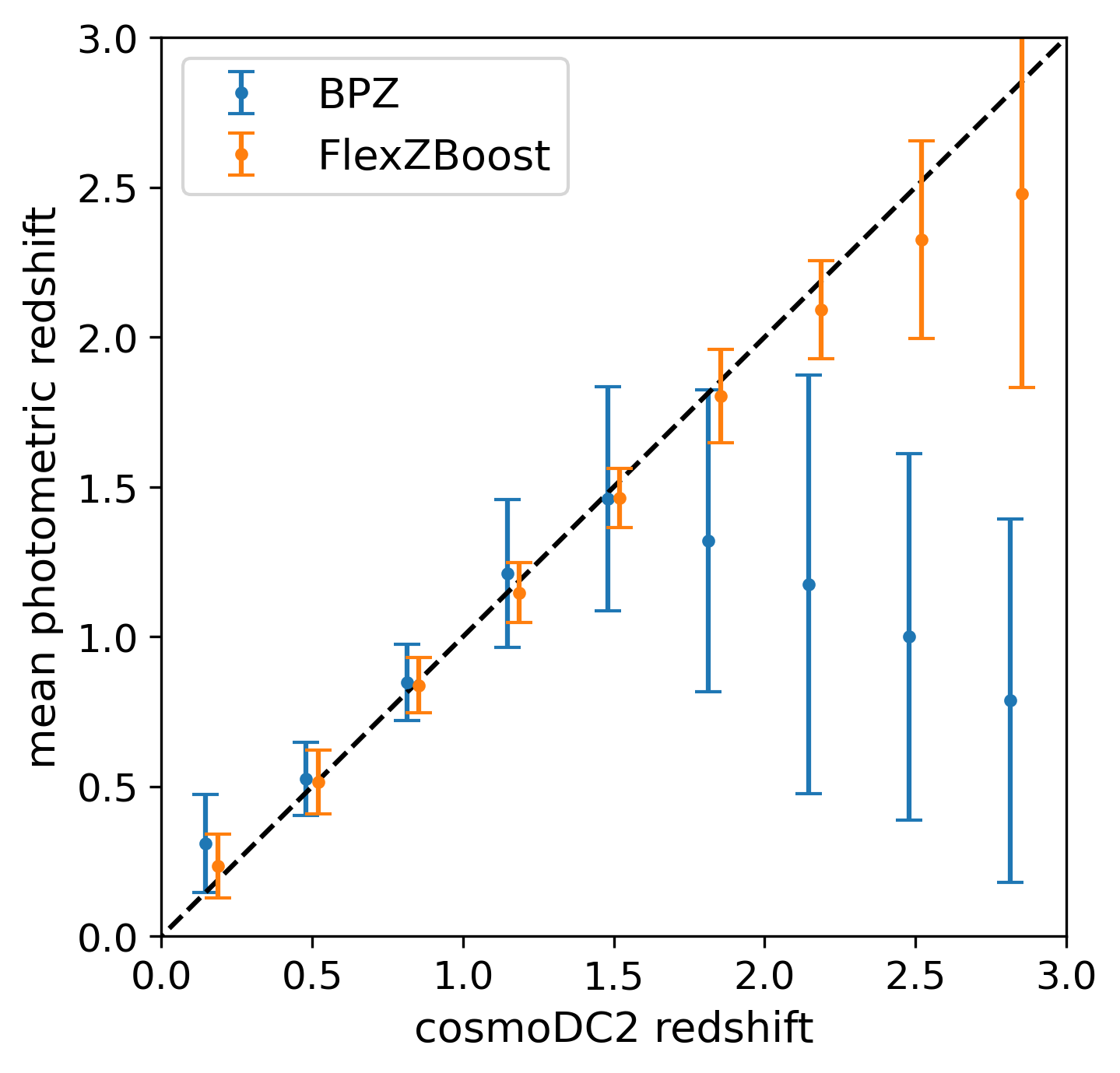}
\includegraphics[width=0.47\textwidth]{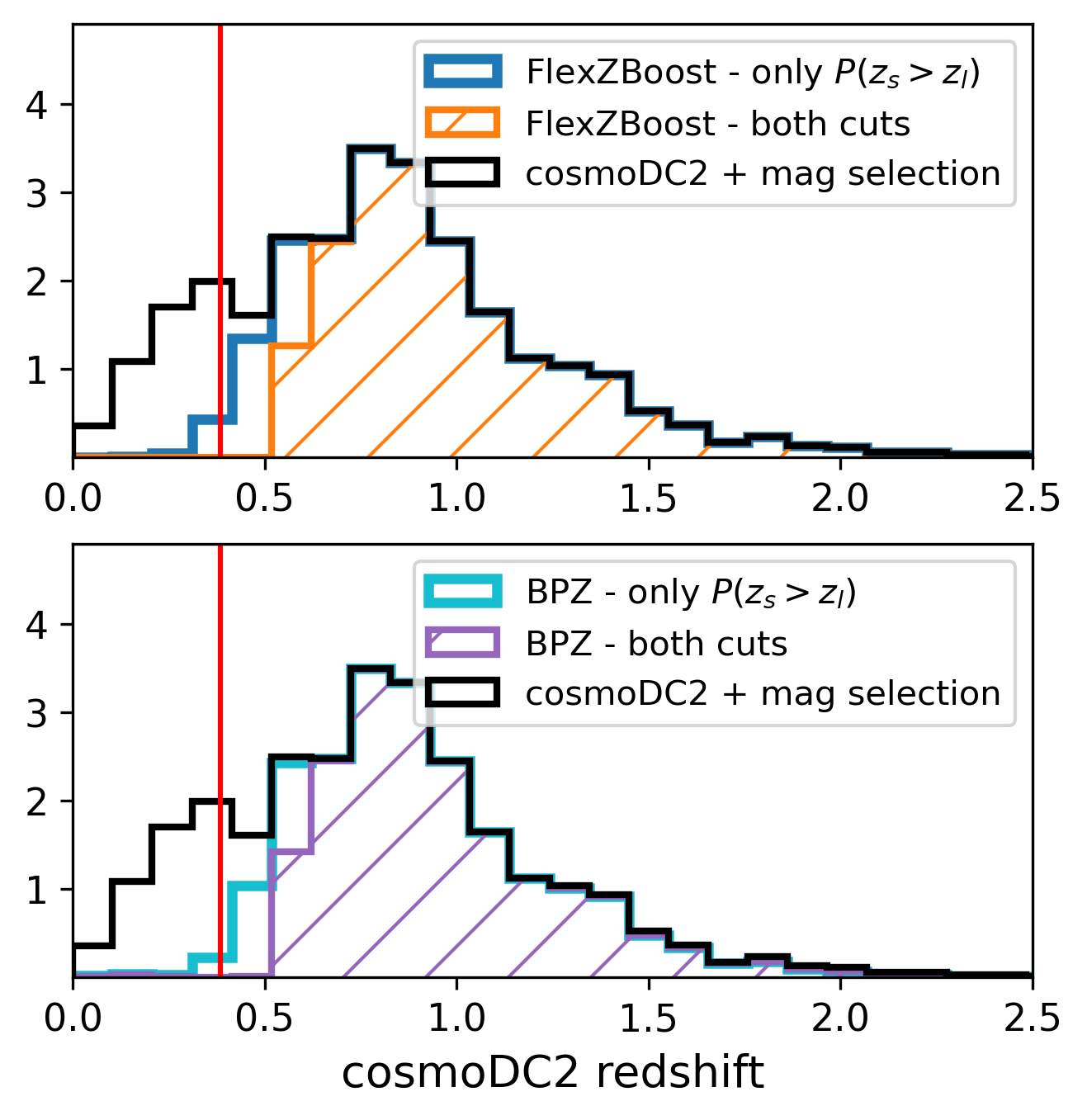}
    \caption{Left: Mean photometric redshift against true cosmoDC2 redshift for BPZ and FlexZBoost methods. Right: distribution of true cosmoDC2 redshifts after our magnitude cuts (black). Upper panel: The blue distribution represents the same distribution after a cut on the probability $P(z_s > z_l)$ using FlexZBoost photometric redshift, for a cluster at $z = 0.38$. The orange distribution is obtained after combining the $P(z_s > z_l)$ cut and the mean redshift cut. Lower panel: same as upper panel but for BPZ redshifts. On both panels, the vertical red line indicates the position of the cluster redshift.}
\label{fig:photoz_diagnostic_zz}
\end{figure*}

In this section, we test the impact of using the BPZ \citep{Benitez2011BPZ} and FlexZBoost \citep{IzBicki2017Flexzboost} output photometric redshifts on the inference of the cluster scaling relation. To do this, we modify the source selection, which is now based on photometric-related quantities. \figreff{fig:photoz_diagnostic_zz} shows the mean photometric redshift and its dispersion as a function of the true cosmoDC2 redshift (no quality cut is applied to the photometric-derived quantities). As FlexZBoost provides robust estimates up to $z = 3$, BPZ yields more dispersed and biased results, making the use of BPZ redshifts without quality cuts problematic above $z \sim 2$, and potentially biasing the constraint of the mass-richness relation. The increased dispersion and bias of BPZ estimates at higher redshifts is not unexpected, as the model photometry in the simulation included a non-physical dust feature in some SEDs at high redshift that was not present in the SED templates used to compute theoretical fluxes by BPZ, leading to a bias and increased dispersion.

As discussed in \secreff{sec:DC2dataset}, none of these estimations are realistic due to (i) the deep selection of the training sample up to $i < 25$ for FlexZBoost, which makes the results optimistic, and (ii) the 'discreteness' in the color-redshift space of the modeled galaxies \citep{Korytov2019cosmoDC2}, which negatively impacts the performance of BPZ. Updated redshifts are currently being produced by DESC. These updated redshifts were not available at the time of this work, so we proceeded with the first version of the photometric redshift catalogs. The actual LSST data will likely fall between the pessimistic BPZ and optimistic FlexZBoost runs we use. By considering both approaches, we can bracket the impact of photometric redshifts on the results.

For the source sample selection, we apply two cuts simultaneously: (i) based on the mean photometric redshift, i.e. $\langle z_s \rangle > z_l + 0.2$, and (ii) based on the probability density function, with $P(z_s > z_l) > 0.8$, where
\begin{align}
    \ P(z_s > z_l) = \int_{z_l}^{\rm +\infty} dz_{s}\ p(z_{s})
\end{align}
is the probability for a galaxy to be located behind the cluster. In the above equation, $p(z_s)$ is the galaxy photometric redshift distribution (obtained either from BPZ or FlexZBoost). Moreover, we do not apply any quality cut on the photometric redshifts, allowing us to test the robustness of the cluster scaling relation with respect to photometric redshifts in the 'worst' case scenario—i.e., with no mitigation of the impact of outliers on the final results. We show in \figreff{fig:photoz_diagnostic_zz} (right, upper panel) the distribution of true galaxy redshifts in cosmoDC2 after our magnitude cuts. We also show the background source sample after applying the $P(z_s > z_l)$ cut alone and after combining the $P(z_s > z_l)$ and mean redshift cuts on FlexZBoost redshifts, for a redMaPPer cluster at redshift $z = 0.38$. We see that our photometric source selection efficiently removes foreground and member galaxies. The same distributions are shown in \figreff{fig:photoz_diagnostic_zz} (right, lower panel) for BPZ redshifts.

\begin{figure*}
    \centering
\includegraphics[width=.47\textwidth]{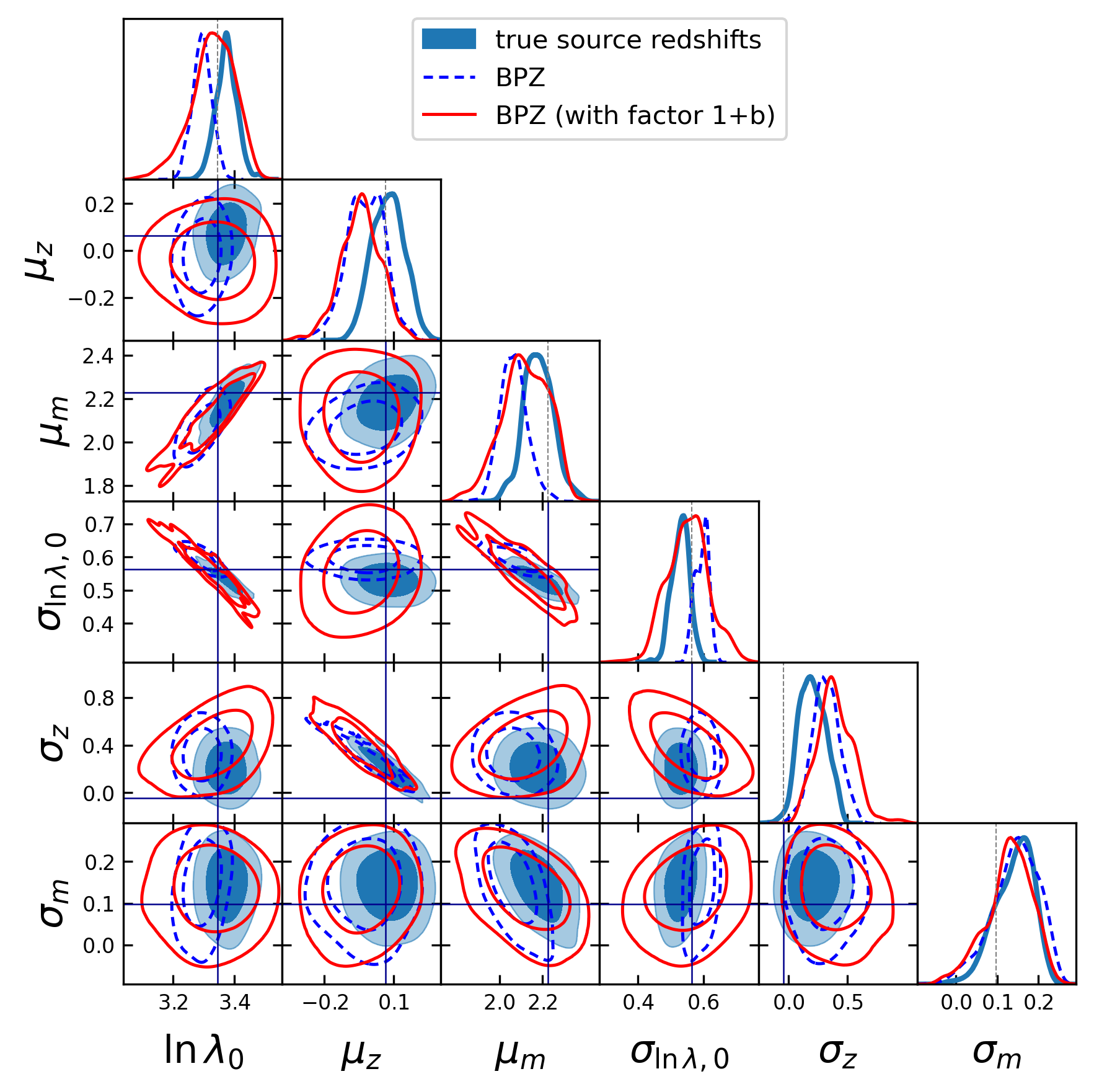}
\includegraphics[width=.47\textwidth]{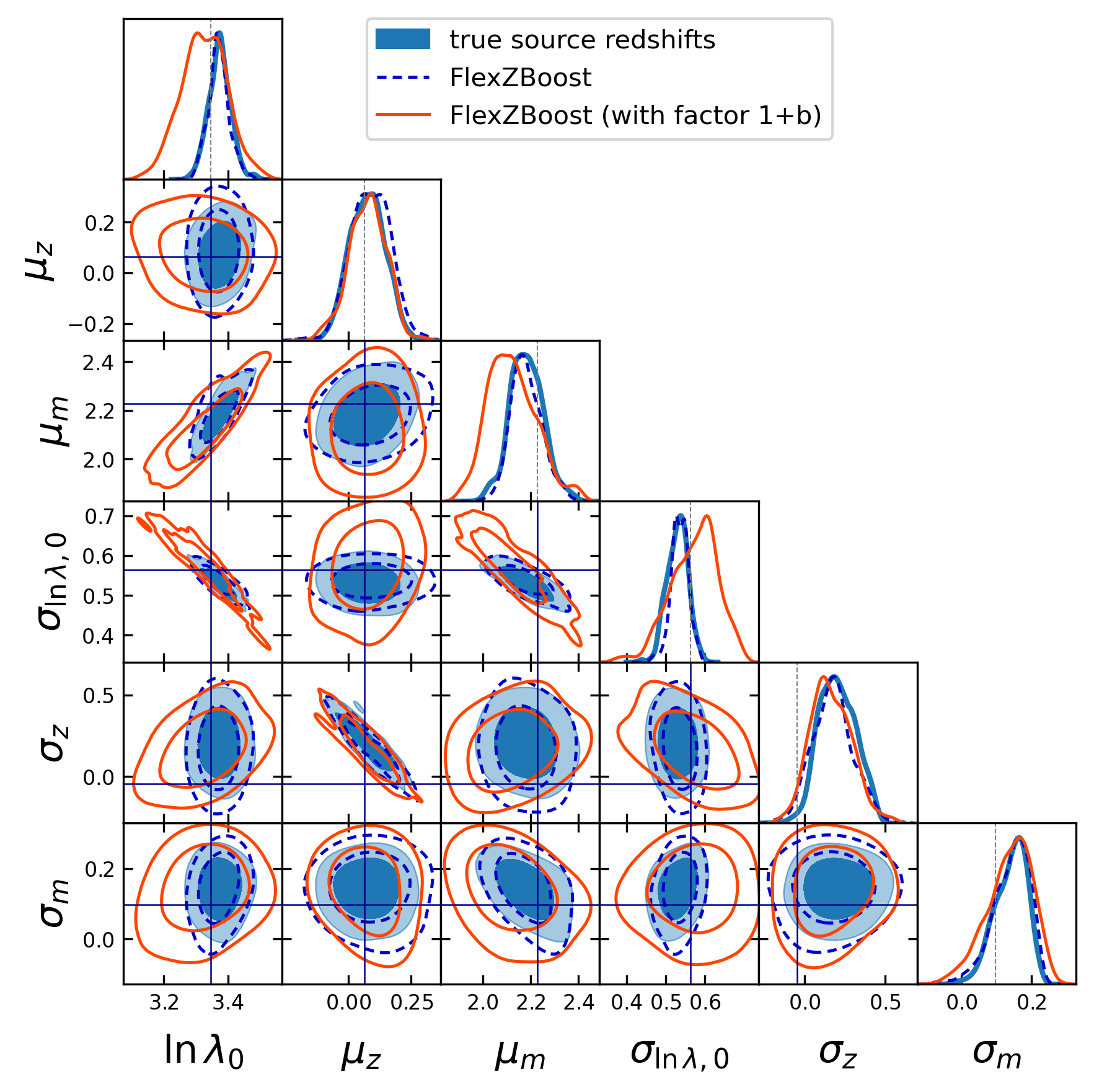}
    \caption{ Impact of photometric redshifts of source galaxies on scaling relation parameter estimation. Left: The filled contours show the constraints when using true redshifts of source galaxies. Dashed contours are obtained when considering the BPZ photometric redshifts. Empty contours are obtained using BPZ photometric redshifts, after marginalizing over the $(1+b)$ correcting factor, considered as a free parameter, used to correct from photometric redshift calibration issues. Right: same as left panel, but for FlexZBoost. }
\label{fig:impact_photoz_posterior_scaling_relation}
\end{figure*}

We show in \figreff{fig:impact_photoz_posterior_scaling_relation} the constraints on the scaling relation when considering photometric redshifts from either BPZ (left panel) or FlexZBoost (right panel). Using BPZ introduces a small negative bias of $\sim 1\sigma$ in the normalization of the scaling relation, as well as in the mass dependence, compared to the true redshift case.

We incorporate a simple model to account for a possible bias in photometric redshift calibration at the level of the stacked lensing profile model, as proposed by \citet{Simet2017SDSSmassrichness}. We correct the modeling of the stacked excess surface density profile in \eqreff{eq:DS_stack_th} in each redshift-richness bin by a unique multiplicative factor $(1+b)$, common to all redshift-richness bins. This factor $(1+b)$ is fitted jointly with the scaling relation parameters, using a flat prior of $[-0.5, 0.5]$ for $b$. The results are shown in \figreff{fig:impact_photoz_posterior_scaling_relation}. For BPZ, we see that adding a free parameter to the fit increases the size of the posteriors as expected and helps decrease the tension between the photometric redshift case and the true redshift case. We recover a negative bias $b = -0.02 \pm 0.03$, which is compatible with 0 at the $1\sigma$ level. For FlexZBoost (right panel), we first observe that without correction, the constraints are fully compatible with the true redshift case. Adding $(1+b)$, we recover $b = 0.02 \pm 0.03$, and although the error bars increase, the posteriors do not shift significantly.

\subsubsection{Impact of shear-richness covariance}
\label{sec:impact_shear_richness}

In this section, we discuss a systematic bias arising from selection biases in cluster finder algorithms. For optically selected clusters, one relatively unexplored category of cluster systematics is the covariance between the two cluster observables, first, the shear estimated from background galaxies and second the richness, as an outcome of cluster finders. This covariance was left unconstrained in previous cluster cosmology analyses \citep{McClintock2019masscalibration,Abbott2020DESCL}. This property covariance can induce additive biases that cannot be mitigated with increased sample size and reduced shape noise \citep{Nord2008selection,Evrard2014massobservable,Farahi2018cov,Wu2019covarianceDeltaSigma}. To achieve percent-level mass calibration of clusters, this new category of systematics must be accurately and precisely quantified \citep{Rozo2014scalingrel,Zhou2024selectionbias}. In this work, we follow the prescription of \citet{Zhang2023cov} in modeling and quantifying the systematic uncertainty induced by the property covariance of optically selected clusters on the cluster scaling relation. 

We make a note of the difference between intrinsic and extrinsic covariance. In \citet{Zhang2023cov} the authors measured the intrinsic covariance of cluster weak lensing observables by encircling galaxies within a three-dimensional physical radius around the halo to measure the 'true' richness and second, measured the integrated dark matter density around the halo. This intrinsic contribution to the covariance originates from the galaxy assembly bias, the degree of which can be quantified by secondary halo properties (e.g. concentration, mass accretion rate, kinetic-to-potential energy ratio) that relate to the formation history of the cluster-sized halo. On the other side, the extrinsic covariance arises from systematic biases introduced by the cluster-finding algorithm, such as projection effects \citep[see e.g.][]{Myles2021projection_redmapper,Myles2025projectioneffects}. In this work, we measure the total covariance (intrinsic + extrinsic) using the redMaPPer cluster finder to determine the observed richness and the galaxy shear as a proxy for the surface density. This realistic mock catalog introduces observational systematics such as photometric redshift uncertainty \citep{Graham2017photoz} and shape noise \citep{Wu2019covarianceDeltaSigma} for the galaxy shear signal and projection and percolation effects \citep{Costanzi2018proj}, triaxiality bias \citep{Zhang2023triax} and a set of other observational and modeling systematics \citep{McClintock2019masscalibration} related to a realistic cluster finder as redMaPPer. 

Our goal is to model and correct for the systematic uncertainty in the estimated stacked lensing profiles in \eqreff{eq:deltasigma_stack}. From \citet{Evrard2014massobservable,Aihara2018HSC,Wu2019covarianceDeltaSigma, Zhang2023cov}, the corrected stacked excess surface density profile is given by
\begin{align} 
    \Delta\Sigma_{ij}^{\rm corr} &= \Delta\Sigma_{ij} + \ln(10)\frac{[\beta_1]_{ij}}{\mu_m} \times \langle\mathrm{Cov}(\Delta\Sigma, \ln\lambda|m,z)\rangle_{ij}, \label{eq:1st_order_correction} 
\end{align}
where $\Delta\Sigma_{ij}$ is given in \eqreff{eq:DS_stack_th}, and $[\beta_1]_{ij}$ is the average logarithmic slope of the \citet{Despali2015hmf} halo mass function in the $ij$-th redshift-richness bin. For the latter, we compute $[\beta_1]{ij}$ once before running MCMC, using the fiducial cluster scaling relation, and we consider the denominator $\mu_m$ to be free. $\langle \mathrm{Cov}(\gamma, \ln\lambda | m, z) \rangle_{ij}$ is the averaged shear-richness covariance (see calculation details in Appendix \ref{app:selection_bias} and also in \citet{Zhou2024selectionbias}), associated with potential selection bias in the cluster finder algorithms. The factor $\ln(10)$ is used to match the definition in \citet{Evrard2014massobservable}. In the latter, we will present the estimation of $\langle \mathrm{Cov}(\Delta\Sigma, \ln\lambda | m, z) \rangle_{ij}$, to be used in the estimation of the cluster scaling relation parameters.

While \citet{Zhang2023cov} observed a radially dependent covariance at small scales $R < R_{200c}$ due to cluster formation physics, because of the attenuation of the cluster lensing signal at small scales due to the ray tracing resolution, we will rather focus on the $R > 1$ Mpc scales.
From mock-cluster studies analyzing selection effects such as \citet{Sunayama2020_projection,Wu2022selectionbiases,Zhou2024selectionbias}, the total bias (intrinsic+extrinsic) is close to null at small scales and increases at larger scales for $R > R_{200c}$. For this reason, the covariance after applying the scale cut due to lensing attenuation may be sufficient for describing to impact of the total covariance.

We apply a Kernel Local Linear Regression (KLLR) \citep{Farahi2022kllr} to find the best fit local mean with mass as the independent variable and $\Delta\Sigma(R)$ and $\ln\lambda$ as the dependent variables. Specifically, the residuals of $\Delta\Sigma$ and $\ln\lambda$ are taken around their local mean quantities $\langle \Delta\Sigma \mid m, z, R \rangle$ and $\langle \ln\lambda \mid m, z \rangle$ such as
\begin{align}
{\rm res}_{\Delta\Sigma}(m, z, R) &= \Delta\Sigma - \langle \Delta\Sigma \mid m, z, R \rangle,  \\
{\rm res}_{\ln\lambda}(m, z) &= \ln\lambda- \langle \ln\lambda \mid m, z \rangle,
\end{align}
and the covariance is measured around the local residuals, i.e.
\begin{equation}
\langle\mathrm{Cov}(\Delta\Sigma, \ln\lambda|m,z)\rangle_{ij} = {\rm Cov}({\rm res}_{\Delta\Sigma}(m, z, R), {\rm res}_{\ln\lambda}(m, z)).
\label{eq:cov_kllr}
\end{equation}

It is important to note that as we remove the residual dependence of the mean mass on the covariance, the size of the covariance itself can still nonetheless be a function of $(m,z)$, i.e. the scatter and correlation are modeled as $\sigma_{\ln\lambda|M,z}(M,z)$ and $\sigma_{\Delta\Sigma|m,z}(m,z)$ and $r_{\ln\lambda-\Delta\Sigma|m,z}(m,z)$. For this reason, we test the covariance on small enough $(m,z)$ bins that the scatter can be considered homoskedastic, or in other words, approximated as a constant. 
In our benchmark tests, we choose to bin the covariance either in mass or redshift but not by both to retain enough statistical constraint, especially for high mass, low redshift bins with a low cluster count. 

\begin{figure*}
    \centering
    \includegraphics[width=1\textwidth]{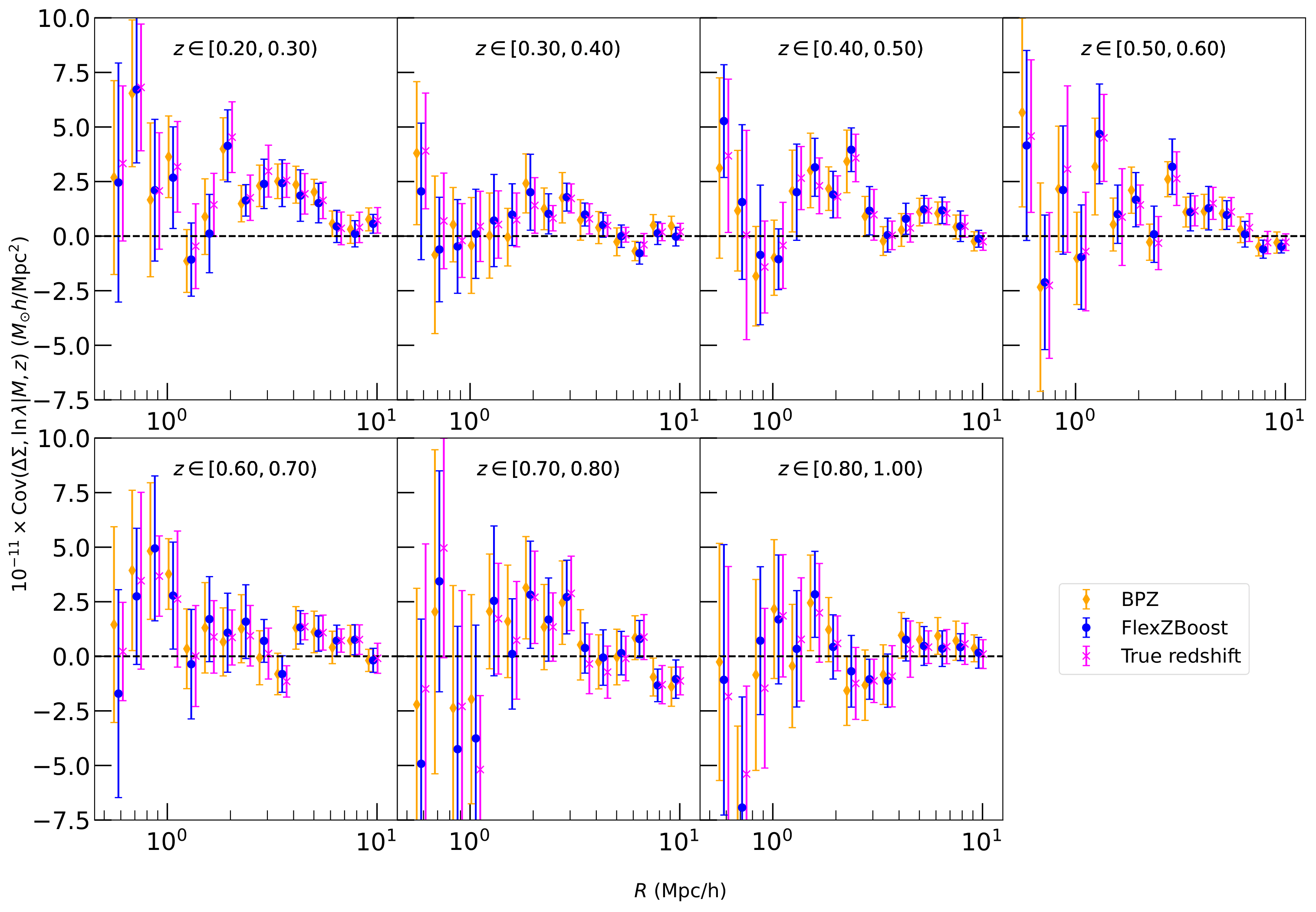}
    \caption{Binned excess surface density-richness covariance in \eqreff{eq:cov_kllr}, as a function of the radius from the cluster center. We stack all clusters between $\lambda \in [20,70)$ and subdivide by different redshift bins from the top left panel to the bottom right panel. The true redshift case is displayed in pink, the BPZ and FlexZBoost cases are represented in orange and blue, respectively. }
    \label{fig:covariance_redshift_binned}
\end{figure*}

\begin{figure*}
    \centering
\includegraphics[width=.47\textwidth]{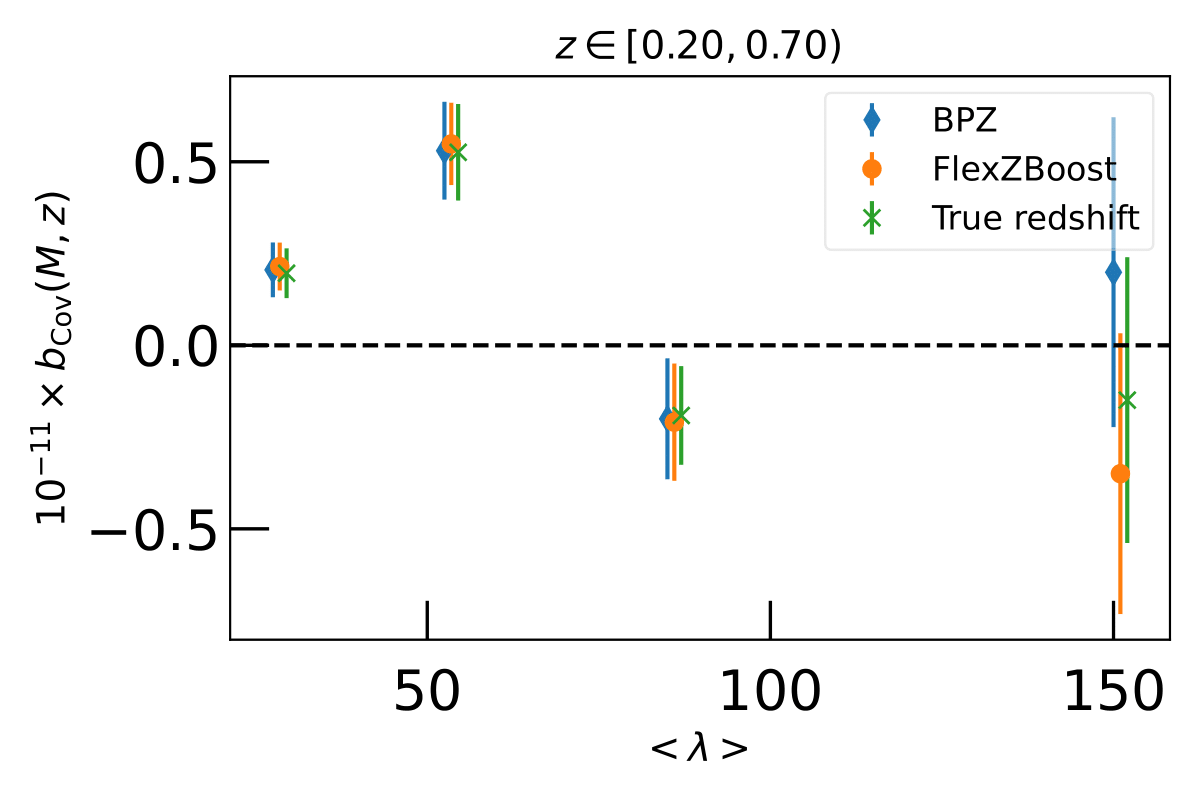}
\includegraphics[width=.47\textwidth]{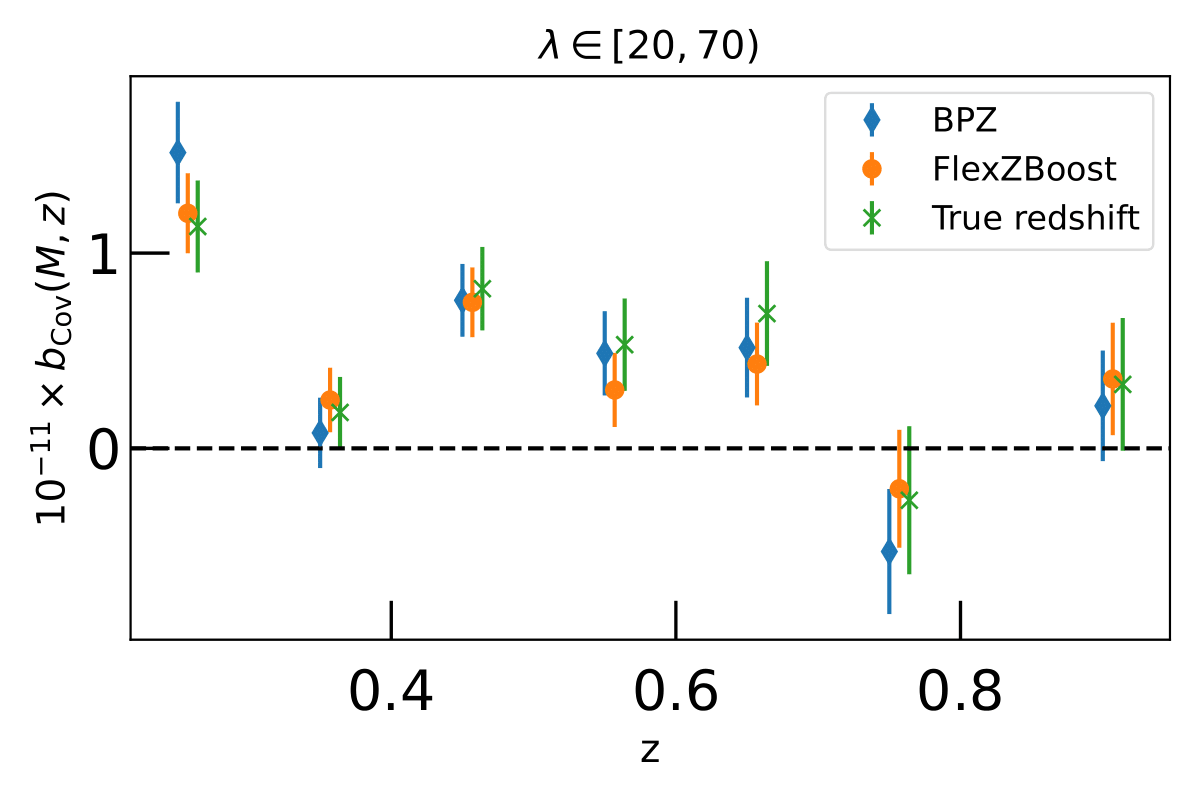}
    \caption{The covariance in \eqreff{eq:cov_kllr} modeled as a constant bias term $b_{\rm Cov}(M,z)$ across radius when binned by richness (left panel) or redshift (right panel). The positive covariance at large scales at lower redshift and richness bins is consistent with expectations from projection effects.}
\label{fig:cov_mass_redshift_trend}
\end{figure*}

We plot the covariance merged by mass and binned in redshift in \figreff{fig:covariance_redshift_binned}. Visually, we observe a slight positive covariance in most bins. A similar trend (not shown) is found in a benchmark test for covariances binned by richness and merged over $z \in [0.2, 0.7]$ (it can be shown that the covariances in $z \in [0.7, 1.0]$ are negligible).
To explicitly show the mass and redshift dependence, we model the covariance as a constant bias across radius, which we denote as $b_{\rm Cov}(M,z)$. The covariance trends with respect to richness (left panel) and redshift (right panel) are plotted in \figreff{fig:cov_mass_redshift_trend}. From the figure, we see that the amplitude of the effect is smaller by a factor of 10 compared to the amplitude of the stacked lensing profile (i.e., $10^{12}$ versus $10^{13}$).

\begin{figure*}
    \centering
\includegraphics[width=.47\textwidth]{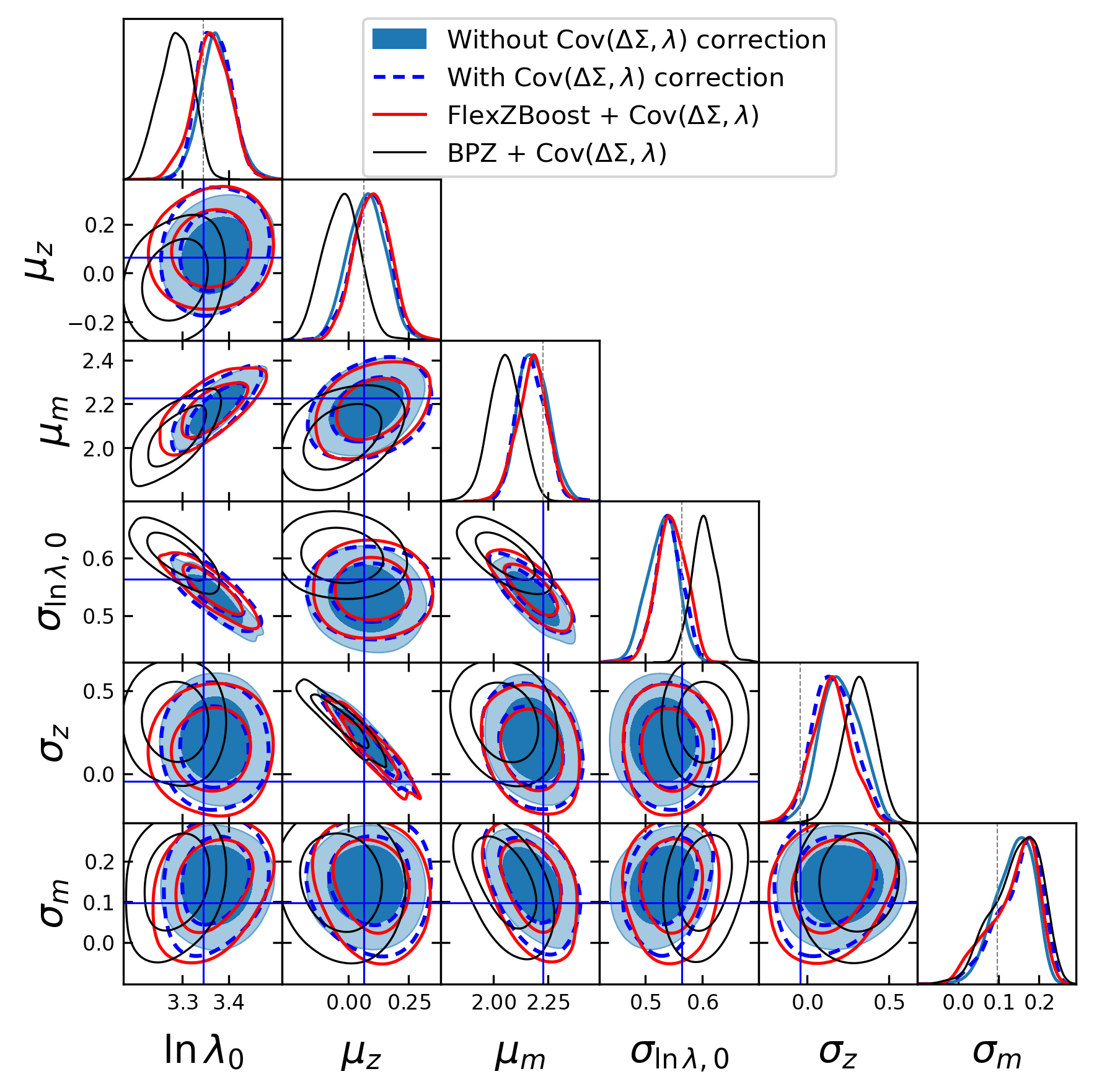}
\includegraphics[width=.47\textwidth]{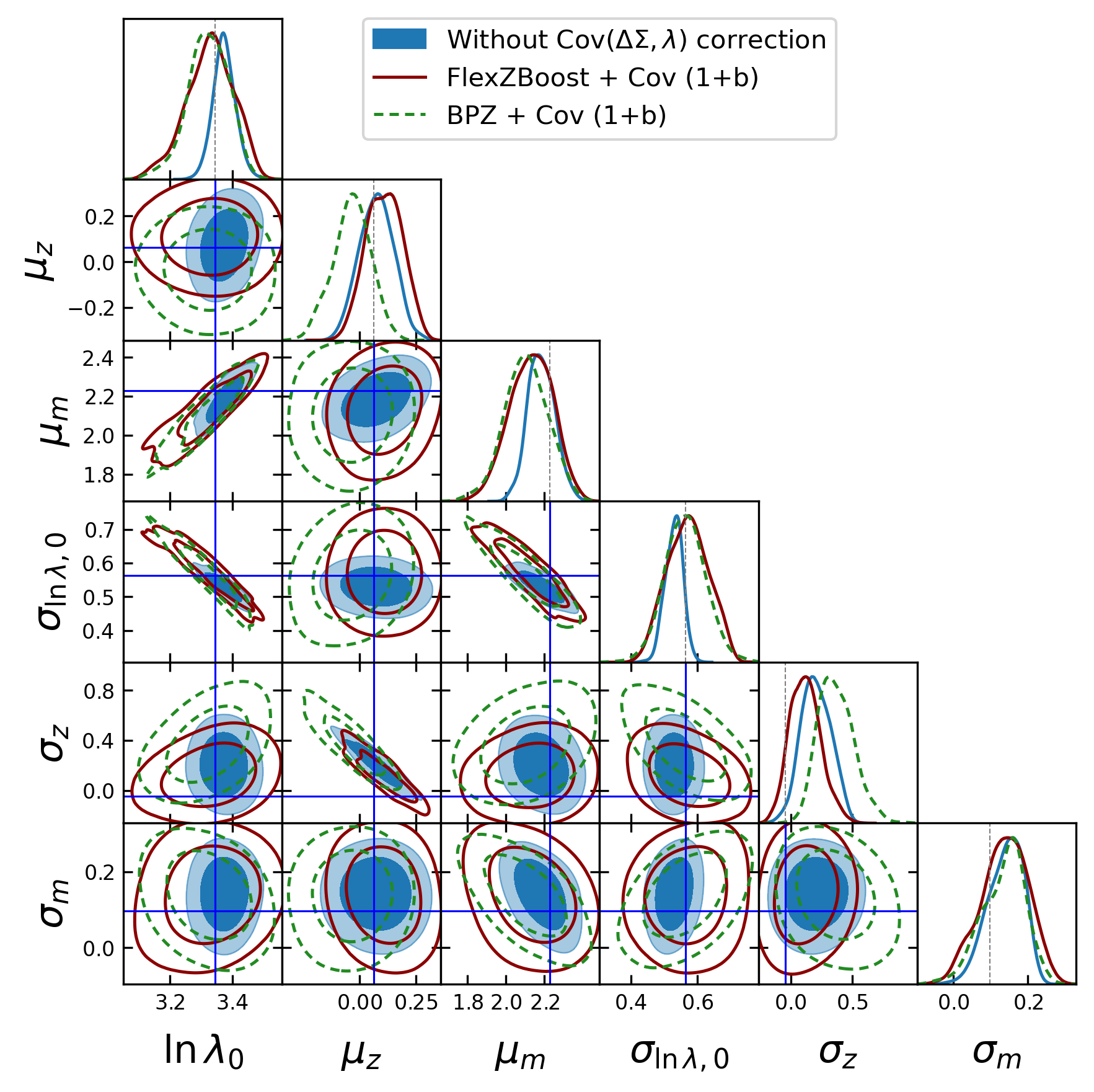}
    \caption{Impact of shear-richness covariance on scaling relation parameter estimation. Left: Posterior distribution of the scaling relation parameters with (dashed blue) and without (filled blue) the shear-richness covariance correction in \eqreff{eq:1st_order_correction}, considering true source galaxy redshifts. The red (resp. black) unfilled contours are obtained considering the shear-richness covariance and using FlexZBoost (resp. BPZ) photometric redshifts for source galaxies. Right: the blue filled contours are the same as in the left panel. The purple (resp. dashed green) unfilled contours are obtained considering the shear-richness covariance and using FlexZBoost (resp. BPZ) photometric redshifts, and letting the bias $(1+b)$ as a free parameter.}
\label{fig:impact_shear_richness_cov_posterior_scaling_relation}
\end{figure*}

To constrain the mass-richness relation while accounting for this effect, we model the stacked excess surface density profile using \eqreff{eq:1st_order_correction}. We propagate the error of the shear-richness covariance (represented by the error bars in \figreff{fig:covariance_redshift_binned}) by adding the associated variance to the diagonal element of the total excess surface density profile covariance matrices, scaled by $[\log(10)\beta_1/\mu_m]^2$. We use the stacked profiles estimated with true background source redshifts and employ the inference setup defined in our baseline analysis, which involves a joint likelihood between counts and the stacked lensing profile.
In \figreff{fig:impact_shear_richness_cov_posterior_scaling_relation} (left panel), the filled (resp. dashed) contours show the posterior of the scaling relation parameters when accounting for (resp. not accounting for) the shear-richness covariance correction as calculated in \eqreff{eq:1st_order_correction}. We observe that the "corrected" posterior is shifted by approximately $0.5\sigma$ compared to the baseline analysis.
From this test, we find that the impact of property covariance significantly influences the scaling parameter inference. The size of this bias is larger than the one obtained using only FlexZBoost redshifts (see \secreff{sec:impact_photoz}), but smaller than that from using only BPZ redshifts. Posterior shifts of roughly the same amplitude as the FleXZBoost-only (resp. BPZ-only) are obtained when using FlexZBoost (resp. BPZ) photometric source redshifts in addition to shear-richness covariance, as shown in the left panel of \figreff{fig:impact_shear_richness_cov_posterior_scaling_relation}. Finally, we show in the right panel of \figreff{fig:impact_shear_richness_cov_posterior_scaling_relation} all sources of observational systematic effects that we tested in this work, namely the combination of (i) source photometric redshifts (either FlexZBoost or BPZ) (ii) the inclusion of the shear-richness covariance, and (iii) the addition of a free correcting photometric factor $(1+b)$ in the fit, to be compared with the baseline case (blue filled contours: true source redshift, and no shear-richness covariance).

As mentioned above, the amplitude of the shear–richness covariance term is primarily influenced by the galaxy formation history within clusters and, secondarily, by the method used to populate halos in the simulation. As mentioned in \cite{Korytov2019cosmoDC2}, after defining the number of galaxies per halo, galaxy positions are drawn randomly in the halo environment (according to a generic NFW profile), thus not following the halo complex shape. As a result, the simulated shear–richness correlation is likely suppressed relative to what would be expected in real data. SkySim5000 \citep{Abolfathi2021DC2}, the extended version of cosmoDC2 covering 5,000 deg$^2$, features improved ray-tracing resolution and a more realistic model for galaxy population in halos. In particular, galaxies are distributed based on the triaxial configuration of halo dark matter particles, better capturing the underlying mass distribution. Leveraging SkySim5000 in future work will allow for a more realistic characterization of shear–richness covariance, especially at smaller scales and over a larger cluster sample.

Second, the shear–richness covariance also depends on the performance of the cluster finder. In our analysis, we use the redMaPPer cluster catalog derived from mock datasets, where the detection process is rather idealized compared to what would be expected from real observational data.

These two effects explain somewhat the low amplitude (about $1$ to $10\%$) of the shear-richness covariance\footnote{We also note that the $\ln(10) \beta_1/\mu_m$ factor in \eqreff{eq:1st_order_correction} is the order of unity; it does not change our conclusions.}.

\section{Summary and conclusions}
\label{sec:conclusion}
We have conducted several analyses based on the lensing profiles and abundance of redMaPPer-detected clusters in the DC2 simulation to constrain their mass-richness relation. We have defined a baseline analysis, with state-of-the-art modeling choices and estimators. From that, we have tested the impact of other modeling choices and observational systematics on the inferred mass-richness relation, also comparing it to a "fiducial" scaling relation. To do that, we used either stacked cluster lensing profiles or mean cluster lensing masses in combination with cluster count to explore several analysis setups (fixing or varying the cluster concentration, changing the dark matter density profile, adding shear-richness covariance). 

\subsection{Summary}
\label{sec:summary}

For the various study cases explored in this paper, the best-fit values of the six-parameter redMaPPer mass–richness relation are summarized in \tabreff{tab:params_WLN_mass_richness}, with the corresponding study cases and their associated paper sections listed in \tabreff{tab:summary_table}.

\begin{figure*}
    \centering
\includegraphics[width=1\textwidth]{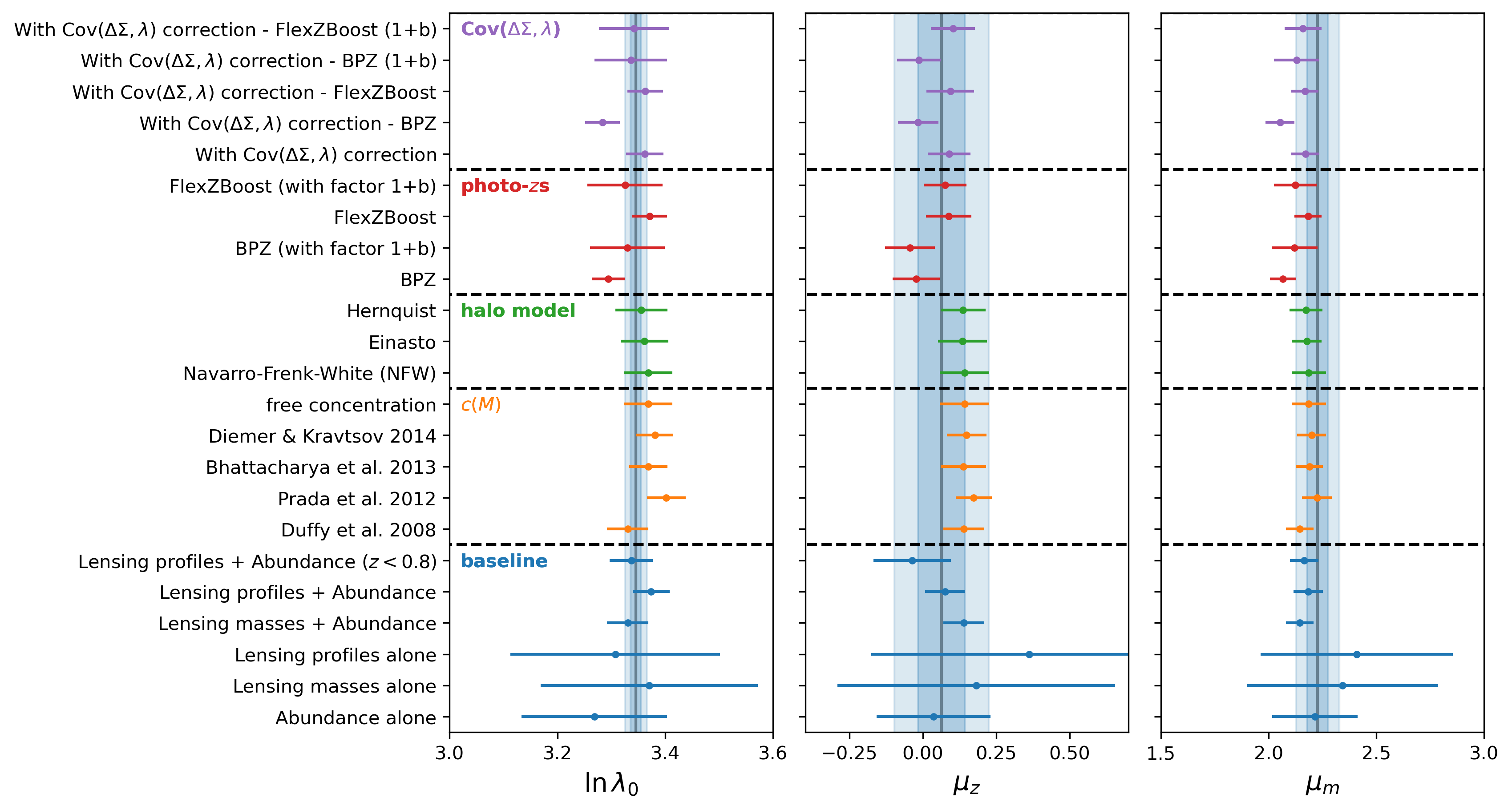}
    \caption{Summary of the constraints on the scaling relation parameters (only $\ln \lambda_0$, $\mu_z$ and $\mu_m$) obtained in \secreff{sec:results}. The vertical shaded region in each subplot represents the fiducial constraints presented in \secreff{sec:fiducial_relation}. For clarity, each color used for the plot corresponds to a subsection in \secreff{sec:results}.}
    \label{fig:summary}
\end{figure*}

In \figreff{fig:summary}, we illustrate the constraints on the three key parameters of the mean mass–richness relation defined in \eqreff{eq:richness_mass}: $\ln \lambda_0$, $\mu_z$, and $\mu_m$. The vertical bands correspond to the fiducial constraints reported in \tabreff{tab:best_fits_scaling_relation_matching}, indicating the 1$\sigma$ and 2$\sigma$ confidence intervals. We first performed the analysis under ideal conditions, using the true shapes of background galaxies (i.e., their intrinsic ellipticities sheared by the local shear field) and their true redshifts. While the limited ray-tracing resolution of the DC2 simulation affects the innermost cluster regions ($R < 1$ Mpc), we found that combining cluster counts and lensing observables—whether via mean mass measurements or stacked profiles—yields tighter constraints on the scaling relation than using each probe independently. This improvement arises from the complementary degeneracy directions of the two observables. Using a Navarro–Frenk–White (NFW, \citealt{Navarro1997nfw}) profile with the concentration–mass relation of \citet{Duffy2008cM}, and restricting the radial fit to the one-halo regime ($1 < R < 3.5$ Mpc), we obtain constraints that are consistent at the $1\sigma$ level with the fiducial values derived from the match between redMaPPer clusters and dark matter halos. This consistency validates our joint count-and-lensing approach as a reliable method for constraining the mass–richness relation. We further tested the robustness of our results by varying the modeling assumptions of the stacked weak lensing profiles. In particular, we showed that the inferred posteriors are stable when adopting different concentration-mass relations, and remain consistent within $1\sigma$ with the results of the free-concentration case. Similarly, using alternative dark matter density profiles instead of NFW—such as Einasto or Hernquist—also results in robust and consistent constraints on the scaling relation parameters.

On the data side, we tested the impact of photometric redshift uncertainties on the inference of the mass–richness scaling relation by using outputs from two photometric redshift algorithms applied to cosmoDC2 galaxy magnitudes: BPZ \citep{Benitez2011BPZ} and FlexZBoost \citep{IzBicki2017Flexzboost}. To accommodate the performance of each algorithm, we adapted the source galaxy selection accordingly. We found that FlexZBoost yields results that are fully consistent with the true-redshift case, indicating no significant bias. In contrast, the use of BPZ introduces a bias greater than $1\sigma$ in the normalization of the scaling relation. However, this bias can be mitigated by introducing a multiplicative correction factor, fitted jointly with the standard scaling relation parameters. In addition, we investigated the effect of shear–richness covariance, which arises from selection biases in the cluster-finding process. We found that, within the radial range considered for fitting the cluster lensing profiles, the impact of covariance alone can shift $\mu_m$ by 0.5$\sigma$. This is comparable to the shift from the different $c(M)$ models or using BPZ redshifts.

We complement the summary plot by quantifying the level of tension between the results of various analyses and the fiducial constraints, as well as with respect to the baseline analysis. To this end, we employ the Gaussian tension metric \citep{Raveri2021tension,Leizerovich2024tension}, which expresses the tension between a Gaussian posterior $\mathcal{P}$ and a reference posterior $\mathcal{P}_{\rm fid}$ for a parameter $\theta$ in units of standard deviations. The tension is defined as
\begin{equation}
    n_\sigma(\theta) = \frac{\langle \theta \rangle_{\mathcal{P}} - \langle \theta \rangle_{\mathcal{P}_{\rm fid}}}{\sqrt{\sigma^2(\theta)_{\mathcal{P}} + \sigma^2(\theta)_{\mathcal{P}_{\rm fid}}}},
\end{equation}
where $\langle \theta \rangle_{\mathcal{P}}$ (resp. $\langle \theta \rangle_{\mathcal{P}_{\rm fid}}$) is the mean of the parameter posterior $\mathcal{P}(\theta)$ (resp. $\mathcal{P}_{\rm fid}(\theta)$), and $\sigma^2(\theta)_{\mathcal{P}}$ (resp. $\sigma^2(\theta)_{\mathcal{P}_{\rm fid}}$) is the associated variance. For the different study cases, the level of tension with respect to the fiducial constraints (resp. the baseline case: weak lensing profiles combined with counts) is shown as circular dots (resp. star symbols) in \figreff{fig:summary_tension}. This approach allows us to quantitatively assess the Bayesian distance between the outcomes of various analysis setups and the fiducial model.

We find that combining probes (counts and lensing) does not significantly increase the level of tension, while it does enhance the statistical precision of the constraints. This demonstrates that our counts+lensing pipeline yields a consistent and reliable description of the observables. Furthermore, across all tested analysis configurations, the tension remains below the $2\sigma$ threshold. This highlights the robustness of our baseline analysis: variations in modeling assumptions or data choices result in posterior shifts that consistently stay within $2\sigma$, underscoring the stability of the inferred scaling relation parameters.

\subsection{Discussions and conclusions}

To fully harness the cosmological constraining power of galaxy clusters, achieving high precision on the cluster scaling relation is essential. This can be accomplished by jointly analyzing cluster counts and cluster weak gravitational lensing, enabling the inference of the mass-richness relation and its intrinsic scatter. For such analyses to be reliable, constraints from both lensing and counts must be robust against systematic effects. In this paper, we have focused specifically on systematics related to cluster weak lensing.

In light of the first LSST data, we investigate the mass-richness relation of redMaPPer-detected galaxy clusters within the redshift range $0.2 < z < 1$ and richness range $20 < \lambda < 200$, using the DESC Data Challenge 2 simulated dataset. We infer a six-parameter log-normal cluster scaling relation—accounting for mass and redshift dependence in both the mean and scatter—through a joint analysis of stacked redMaPPer weak lensing profiles and cluster counts, enhancing the overall constraining power. A series of robustness tests were performed, exploring the impact of cluster lensing profile modeling choices, photometric redshift uncertainties, and shear-richness covariance effects. We find that our constraints remain stable when lensing profiles are fitted in the one-halo regime ($1 < R < 3.5$ Mpc), assuming either true or photometric redshifts, and perfectly known source shapes. Furthermore, our results yield unbiased estimates of the fiducial redMaPPer mass-richness relation. While the constraining power in this analysis is limited compared to LSST expectations—to be $\sqrt{\Omega_{\rm LSST}/\Omega_{\rm DC2}} \approx 6.4$ times larger—this work provides a solid foundation for scaling relation inference with future LSST cluster samples.

Moreover, the code work allows us to demonstrate the use of some of the software tools developed within DESC in concrete cluster analysis, representing a significant step in the calibration, validation, and methodological development. This paper thoroughly addresses all major systematics, except for shape measurement, which is discussed in detail hereafter. The framework employed in this analysis is well integrated into the CLMM and CCL pipelines, paving the way for the development of cluster-based analyses in other official DESC pipeline codes such as \texttt{TJPCov}\footnote{\url{https://github.com/LSSTDESC/tjpcov}} (official DESC covariance calculator interface), \texttt{Firecrown}\footnote{\url{https://github.com/LSSTDESC/firecrown}} (providing the DESC framework for implementing likelihoods and sampling parameters), and \texttt{TXPipe}\footnote{\url{https://github.com/LSSTDESC/txpipe}} \citep[DESC library for summary statistics measurement, see][]{Prat2023txpipe}.

Due to the lack of ray-tracing resolution that motivated our conservative cut, we excluded the innermost regions, where the cluster lensing signal is the strongest. These scales are, however, crucial to enhancing the precision of weak lensing cluster masses. Analyzing such radial ranges requires particular attention to modeling the dilution of the cluster lensing signal by member galaxies \citep[contamination, see e.g.][]{Varga2019contamination} and mis-centering due to uncertainties in cluster finding methods \citep{Zhang2019redmappermiscentering,Kohlinger2015lensingbias}, which can be significant when referring to stacks of galaxy clusters. 

Our analysis could benefit from further improvement to infer the redMaPPer cluster mass-richness relation. First, we have considered the fiducial halo mass function of \cite{Despali2015hmf}, but propagating calibration uncertainties of this relation in our pipeline could have a non-negligible impact \citep{Artis2021HMF,Kugel2024systematiccosmomodel}. Due to the noise level, we have neglected the off-diagonal terms in the covariances of the stacked lensing profiles. However, rather close radial bins are known to be somewhat correlated due to the intrinsic variation of the halo density profile and large-scale structures \citep{Wu2019covarianceDeltaSigma}, which could degrade the precision of our results, even though they are much smaller than the galaxy shot/shape noise contribution.

Moreover, for the mean mass (two-step) approach, the link between the "true" mean mass of a stack and the inferred mean weak lensing mass may not be trivial \citep{Becker2011modeling}, contrary to what we used in \eqreff{eq:M_lensing}, i.e. a standard average mass over the redshift-richness cluster distribution. In reality, the excess surface density profile does not simply scale with the underlying cluster mass. For instance, \citet{Melchior2017logslope} has proposed a simple model where the excess surface density follows $\Delta\Sigma(R|M) \propto M^{\Gamma(R|M)}$ where $\Gamma$ is the logarithmic slope of the excess surface density\footnote{ \citet{Melchior2017logslope} found a typical value of $\Gamma = 0.74$ for an NFW profile.}, such that the mean weak lensing mass $\langle \widehat{M}_{\rm WL}^{\rm stack}\rangle$ has to be modeled by $\sim \langle m^\Gamma\rangle^{1/ \Gamma}$. Such non-linear dependence can be accounted for in the mean mass modeling in \eqreff{eq:M_lensing}. Beyond this rather simple correction to link the mean lensing mass to the true average mass of the stack, \citet{McClintock2019masscalibration} have used simulations to reconstruct the stack lensing signal around clusters with masses drawn from the known $P(M|\lambda)$ relation (as well as for concentrations and miscentering offsets drawn from specific distributions). They found that the bias between the true average mass and mean weak lensing mass may reach $10\%$ (see their Figure 9) and has a strong richness and redshift dependence, denoting a strong impact of intrinsic scatter (in richness, concentration, mis-centering, etc.).

Even if not sensitive to mis-centering in our analysis (due to our restrictive low radial cut), we have considered a fixed concentration-mass relation. However, it is known that cluster concentrations are scattered around a mean $c(M)$ relation \citep[see e.g.][]{Bullock2001cMr,Darragh2023cMr}. This effect is particularly important in stacked analysis, especially at small radii. Accounting for this effect should degrade the precision of our obtained constraints by lowering the sensitivity of our model.

Finally, noise in galaxy shapes is one of the current main sources of scatter in weak lensing cluster mass estimation. First, as already mentioned, it originates from the intrinsic scatter of galaxy shapes, which is more and more mitigated by the increasing statistics of ongoing and future optical surveys. Second, it originates from the galaxy shape measurement algorithms, returning dispersed and biased estimations of intrinsic galaxy ellipticities. Shape reconstruction methods generally need to be calibrated from simulation or internally. Dispersed measured shapes as well as calibration uncertainties in these algorithms inevitably propagate to the cluster mass estimates, downgrading their precision. For that, the DC2 object catalog \citep{Abolfathi2021DC2} was built using the Rubin LSST science pipeline by identifying galaxies on cosmoDC2-based realistic images of the sky, including several realistic observational effects, such as atmospheric turbulence, telescope optics, and some detector effects. Several codes are used to measure shapes of detected galaxies, such as HSM \citep{HIRATA2003HSM,Mandelbaum2005HSM} used on the HSC data \citep{Mandelbaum2018HSMHSC}, and {\sc Metacalibration} \citep{Sheldon2017metacalibration}, used in the DES-Y3 cosmic shear analysis in \citet{Abbott2022COSMICSHEARdesY3}. Ongoing efforts in DESC aim at studying the impact of galaxy detection and shape measurement in cluster fields, on the accuracy and the precision of cluster mass estimates (Ramel et al. in prep), which are essential to handle the full error budget on the mass-richness relation, in preparation for LSST cluster cosmology.

\begin{acknowledgements}
The authors thank the anonymous reviewer for their insightful comments and suggestions. This paper has undergone internal review by
the LSST Dark Energy Science Collaboration. The
authors thank the internal reviewers Tomomi Sunayama and Shenming Fu for their valuable
comments. The authors thank Fabien Lacasa for useful discussions and for their help in using the PySSC code. The authors thank Dominique Boutigny for his help in using Qserv to access the cosmoDC2 simulated datasets at CC-IN2P3.  

CP conceptualized the project, developed the code, led the analysis, contributed to the paper's structure and text, and oversaw the overall writing process. ZZ contributed to the shear-richness covariance analysis, the early development of the analysis framework and code, and the paper's structure and text. MA developed the DESC matching library ClEvaR and contributed to the paper's structure and text. CC assisted with project conceptualization and advising, participated in the early development and review of the analysis framework and code, and contributed to the paper's structure and text. TG provided the parametrization of the redMaPPer selection function. MR prepared the matched cosmoDC2-redMaPPer catalog, reviewed various aspects of the analysis, and contributed to the paper's structure and text. The remaining authors generated the data used in this work, contributed to the project's infrastructure, and participated in discussions interpreting the results.

The DESC acknowledges ongoing support from the Institut National de 
Physique Nucl\'eaire et de Physique des Particules in France; the 
Science \& Technology Facilities Council in the United Kingdom; and the 
Department of Energy and the LSST Discovery Alliance
in the United States.  DESC uses the resources of the IN2P3 
Computing Center (CC-IN2P3--Lyon/Villeurbanne - France) funded by the 
Centre National de la Recherche Scientifique; the National Energy 
Research Scientific Computing Center, a DOE Office of Science User 
Facility supported by the Office of Science of the U.S.\ Department of
Energy under Contract No.\ DE-AC02-05CH11231; STFC DiRAC HPC Facilities, 
funded by UK BEIS National E-infrastructure capital grants; and the UK 
particle physics grid, supported by the GridPP Collaboration.  This 
work was performed in part under DOE Contract DE-AC02-76SF00515. MA is supported by the PRIN 2022 project EMC2 - Euclid Mission Cluster Cosmology: unlock the full cosmological utility of the Euclid photometric cluster catalog (code no. J53D23001620006). CA acknowledges support from DOE grant DE‐SC0019193 and the Leinweber Center for Theoretical Physics. AF acknowledges support from the National Science Foundation under Cooperative Agreement 2421782 and the Simons Foundation award MPS-AI-00010515.

We thank the developers and maintainers of the following software tools used in this work: \texttt{NumPy} \citep{vanderWaltnumpy}, \texttt{SciPy} \citep{jonesscipy}, \texttt{Matplotlib} \citep{Hunter2007matplotlib}, \texttt{GetDist} \citep{Lewis2019getdist}, \texttt{emcee} \citep{ForemanMackey2013emcee}, \texttt{Astropy} \citep{Astropy20213}, \texttt{Jupyter} \citep{jupyter}.
\end{acknowledgements}
\bibliographystyle{aa}
\bibliography{aa54107-25.bib} 

\begin{appendix}
\section{Stability of the fiducial cluster scaling relation}
\label{app:fiducial}
\begin{figure}
    \centering
    \includegraphics[width=0.48\textwidth]{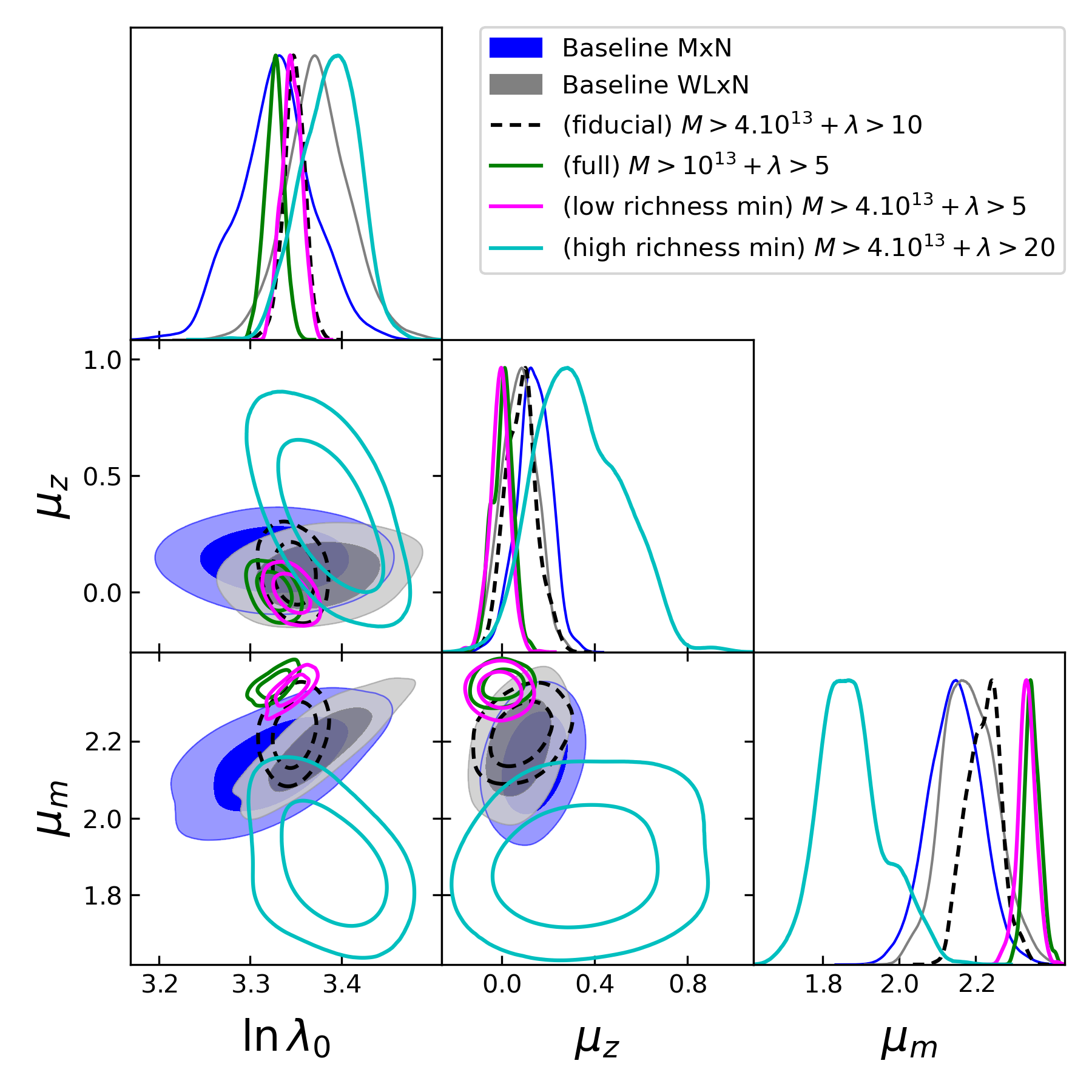}
    \caption{Left: Posterior distribution of the mean scaling relation parameters from our count+lensing baselines (filled contours) and the matched redMaPPer-cosmoDC2 cluster-halo match catalog, for different richness and mass cuts.}
    \label{fig:app_fid}
\end{figure}

The fiducial scaling relation used throughout this paper was derived from the matched catalog between redMaPPer clusters and cosmoDC2 halos, restricted to $\lambda > 5$ and $M_{\rm 200c} > 10^{13} M_{\odot}$. In this appendix, we detail the various selection cuts applied to the matched catalog before inferring the fiducial relation using the likelihood defined in \eqreff{eq:fiducial_likelihood}. For illustration, the posterior distributions of the mean scaling relation parameters, obtained using the full matched catalog (i.e., before applying any cuts), are shown as green unfilled contours in \figreff{fig:app_fid}.

Initially, we assume that the matched catalog is unaffected by the purity of the redMaPPer cluster finder, as false detections have been excluded after matching. However, the completeness of redMaPPer still impacts the catalog, particularly at the low-mass end where some halos are not detected. Ideally, the likelihood in \eqreff{eq:fiducial_likelihood} should account for such missing halos. To partially mitigate this effect, we apply a lower mass cut of $M_{\rm 200c} > 4 \times 10^{13} M_{\odot}$, a regime where redMaPPer’s completeness exceeds $50\%$. The resulting posterior, reflecting this correction, is shown as magenta unfilled contours in \figreff{fig:app_fid}.

Additionally, our cluster scaling relation model in \eqreff{eq:p_lambda_m} does not account for Poisson noise in redMaPPer richness determination. This noise can be incorporated by revising \eqreff{eq:p_lambda_m}:
\begin{equation}
    P(\lambda | m,z) = \int\limits_0^{+\infty}\mathcal{P}(\lambda|\lambda_{\rm int})P(\lambda_{\rm int}|m,z)d\lambda_{\rm int},
\end{equation}
where $\mathcal{P}(\lambda|\lambda_{\rm int})$ is the Poisson distribution, and $P(\lambda_{\rm int}|m,z)$ represents the probability density function for the intrinsic richness, which can be expressed with the same equation as in \eqreff{eq:p_lambda_m}, but with intrinsic variance only. This updated $P(\lambda | m,z)$ can be approximated by using the same $P(\lambda_{\rm int}|m,z)$ while updating the variance by
\begin{equation}
    \sigma_{\ln \lambda|m,z}^2 \approx [\sigma_{\ln \lambda|m,z}]_{\rm int}^2 + \frac{\exp(\langle \ln\lambda|m,z\rangle)-1}{\exp(2\langle \ln\lambda|m,z\rangle)},
\end{equation}
where $[\sigma_{\ln \lambda|m,z}]_{\rm int}^2$ is provided in \eqreff{eq:sigma_forward_modelling}, now only describing the intrinsic variance of richness. The impact of the Poisson noise, represented by the second term \citep[see e.g.][]{Zhang2023triax}, diminishes for large richness. Consequently, we apply a low richness cut $\lambda > 10$ to align with the Poisson-free distribution in \eqreff{eq:p_lambda_m}.

Combining this with the mass cut $M_{\rm 200c} > 4\times10^{13} M_{\odot}$, the constraints are represented as black dashed unfilled contours in \figreff{fig:app_fid} and are identified as our fiducial choice. We observe that the three fiducial constraints agree within $< 2\sigma$ of the baseline count+weak lensing posteriors. We also explore a stricter cut, $\lambda > 20$, combined with $M_{\rm 200c} > 4\times10^{13} M_{\odot}$. This $\lambda > 20$ scenario results in larger error bars (still compatible with count+weak lensing constraints) but biases the mass slope $\mu_m$ lower than other choices. In fact, from the left panel in  \figreff{fig:fiducial_relation}, the $\lambda > 20$ cut over-weights high-richness clusters in the fiducial fit, even after correcting the distribution $P(\lambda|m,z)$ using a truncated Gaussian to account for the minimum richness. While we adopt the black dashed contours as our fiducial constraints, further investigations are warranted.

\section{Impact of the radial fitting range on the scaling relation parameters}
\label{app:impact_rmax}
In this appendix, we test the impact of the fitting radial range $[1, R_{\rm max}]$ of the stacked excess surface density profiles on the scaling relation parameters. Our baseline adopts a conservative choice of $R_{\rm max} = 3.5$ Mpc for fitting the stacked excess surface density profiles, ensuring sensitivity to the one-halo regime only. At larger radii, the influence of the two-halo term is expected to become more significant, and modeling the stacked profiles using only the one-halo regime may be insufficient. From \citet{Oguri2011lensing}, the two-halo excess surface density is given by
\begin{equation}
    \Delta\Sigma_{\rm 2h}(R) = \frac{\rho_m(z)b_h(M_{\rm 200c},z)}{(1 + z)^3D_A(z)^2} \int\frac{ldl}{(2\pi)} P_{m}(k_l, z)J_2\left(l\theta\right),
    \label{eq:two_halo_term_ds}
\end{equation}
where $k_l = l/[D_A(z)(1+z)]$, $J_2$ is the second order Bessel function of the first kind, $\theta=R/D_A(z)$ is the separation angle, $\rho_m(z)$ is the matter density at redshift $z$, $b_h(M,z)$ is the halo bias at mass $M$ and redshift $z$ and $P_{m}(k, z)$ is the matter power spectrum at the halo redshift $z$. Finally, the total excess surface density profile is given by 
\begin{equation}
    \Delta\Sigma(R) = \Delta\Sigma_{\rm 1h}(R)+\Delta\Sigma_{\rm 2h}(R).
    \label{eq:full_ds_1h+2h}
\end{equation}

\begin{figure}
    \centering
    \includegraphics[width=0.48\textwidth]{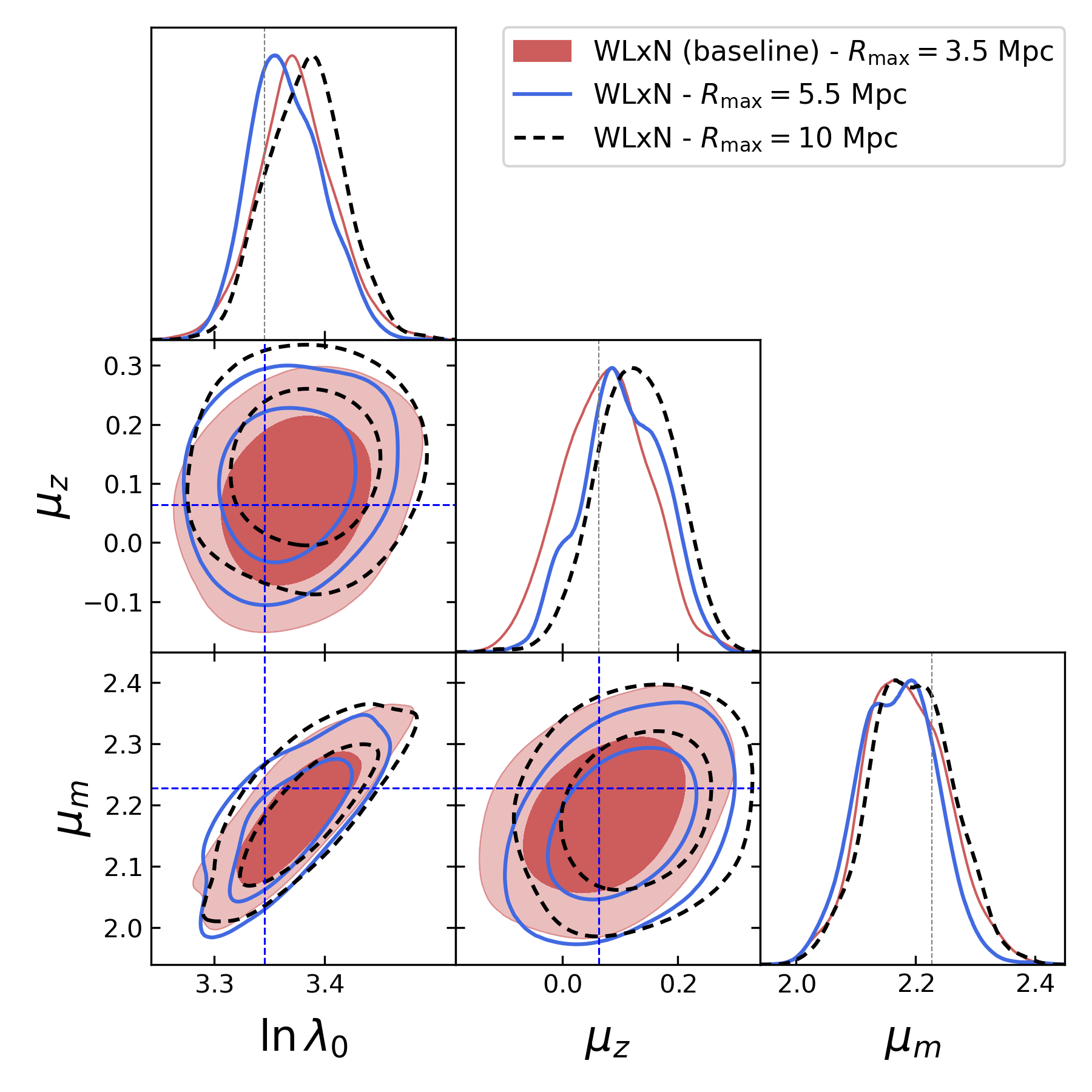}
    \caption{Posterior distribution of the mean scaling relation parameters from the combination of count and lensing profiles, for three different values of $R_{\rm max}$, used in the stacked lensing profile fitting range $[1, R_{\rm max}]$.}
    \label{fig:app_rmax}
\end{figure}

We repeat the scaling relation parameter fit using $R_{\rm max} = 5.5$ and $R_{\rm max} = 10$ Mpc, respectively. The mean scaling relation parameters for these different fitting radial ranges are shown in \figreff{fig:app_rmax}. Both new cuts yield conclusions consistent with the baseline, with shifts of less than $< 1\sigma$ compared to the baseline constraints.

Interestingly, the results for $R_{\rm max} = 10$ Mpc are close to the baseline constraints, despite the absence of modeling for the two-halo term in the stacked lensing profile. We attribute this to the lower signal-to-noise ratio of measurements at large scales, which reduces the impact of the two-halo term. However, with larger datasets, the two-halo term must be accurately modeled beyond $R = 4$ Mpc \citep{Murata2019HSCrichnessmassrelation, McClintock2019masscalibration, Melchior2017logslope}.

\section{Some other observational systematics}
\subsection{Miscentering}
\label{app:miscentering}
When the inferred cluster center differs from the one of its underlying halo's gravitational potential -- which happens when deriving the cluster center from the intra-cluster galaxy distribution -- the lensing profile modeling must account for a mis-centering term. For a single mis-centered cluster with offset radius $R_{\rm mis}$, its radially averaged projected mass density is given by
\begin{equation}
\Sigma_{\rm mis}(R, R_{\rm mis}) = \frac{1}{2\pi}\int\limits_0^{2\pi} d\theta\  \Sigma \left(\sqrt{R^2+R_{\rm mis}^2 + 2RR_{\rm mis}\cos\theta}\right). 
\end{equation}
For a stack of clusters, the above equation can be averaged over the distribution of miscentering $P(R_{\rm mis})$, such as
\begin{equation}
    \Sigma_{\rm mis}^{\rm stack}(R) = \int\limits_0^{+\infty} dR_{\rm mis} P(R_{\rm mis})\Sigma_{\rm mis}(R, R_{\rm mis}).
\end{equation}
From this equation, we can obtain $\Delta\Sigma_{\rm mis}$ by injecting it into \eqreff{eq:dsigma_diff}. The total one-halo term depends on the fraction $f_{\rm mis}$ of miscentered clusters, as shown in, e.g., \citet{Giocoli2021stackedlensingAMICO}, and is given by
\begin{equation}
    \Delta\Sigma_{\rm 1h}(R) = f_{\rm mis}\Delta\Sigma_{\rm 1h, mis}(R) + (1 - f_{\rm mis}) \Delta\Sigma_{\rm 1h, cen}(R).
    \label{eq:Delta_Sigma_full}
\end{equation}
where $\Delta\Sigma_{\rm 1h, cen}(R)$ denotes the lensing profile of a perfectly centered stack of clusters. To test the impact of miscentering of redMaPPer clusters, we use the matched catalog between redMaPPer clusters and cosmoDC2 dark matter halos obtained in \secreff{sec:matched_catalog}. In each cluster-halo pair, we consider the halo-member galaxy flagged as \texttt{is}$\_$\texttt{central==True} in the cosmoDC2 extra-galactic catalog to trace the center of the halo's gravitational potential, from which we compute the corresponding projected offset radius $R_{\rm mis}$ of the matched redMaPPer cluster. We found that the fraction of perfectly centered clusters (for which $R_{\rm mis} = 0$) is 85$\%$ so $f_{\rm mis} = 0.15$. For pairs with $R_{\rm mis} \neq 0$, the distribution of offsets between cosmoDC2 halo and redMaPPer cluster centers is well described by a Gamma distribution \citep{McClintock2019masscalibration} given by
\begin{equation}
    P(R_{\rm mis}) = \frac{R_{\rm mis}}{\sigma_{\rm mis}^2}\exp\{-\frac{R_{\rm mis}}{\sigma_{\rm mis}}\},
\end{equation} 
with $\sigma_{\rm mis} = 0.12$ Mpc. We use the \texttt{cluster$\_$toolkit} package\footnote{\url{https://github.com/tmcclintock/cluster_toolkit}} to compute the miscentered profile evoked in \secreff{sec:formalism} over a wide range of mass and redshift, and we found that the bias compared to a perfectly centered profile is at most 1$\%$ at $R = 1$ Mpc, and 0.25$\%$ at $R = 2$ Mpc. For simplicity, since the computation of the miscentering term is computationally demanding, we will neglect this contribution.

\subsection{Contamination}
\label{app:contamination}
The background source selection may cause contamination of the source sample by cluster member galaxies \citep{Varga2019contamination}, which dilutes the signal in the innermost region. The observed contaminated lensing profile accounting for this dilution is given by
\begin{equation}
    \Delta\Sigma_{ij}^{\rm cont}(R) = [1-f_{\rm cl}(R)]\Delta\Sigma_{ij}(R),
\end{equation}
where $\Delta\Sigma_{ij}(R)$ is given in \eqreff{eq:DS_stack_th}, $f_{\rm cl}(R)$ is the fraction of 'source-selected' member galaxies at a distance $R$ from the cluster center, and is a decreasing function of $R$. At fixed $R$, the fraction $f_{\rm cl}$ increases with mass, since massive clusters have more member galaxies, and thus are more subject to this effect \citep{Varga2019contamination}. 

For each redMaPPer cluster, we consider the corresponding matched dark matter halo in the matched catalog presented in \secreff{sec:DC2dataset}. We applied the photo-$z$ source selection and then identified the remaining halo 'member galaxies', as labeled in the simulation \citep[which are randomly drawn at the HOD level in the DC2 workflow to populate halos, see][]{Korytov2019cosmoDC2}. This procedure enables us to recover $f_{\rm cl}(R)$ for a variety of masses and redshifts, and we found that the correcting factor $[1 - f_{\rm cl}(R = 1 \text{ Mpc})] \sim 1$ after combining the two cuts, and for both BPZ and FlexZBoost algorithms. We found that our source selection enables us to discriminate efficiently between true sources and member galaxies (the effect is less than $0.1\%$).
\section{Derivation of the selection bias}
\label{app:selection_bias}
In this appendix, we use the notation:
\begin{equation}
    N(m,z) =\frac{dn(m,z)}{dm}\frac{d^2V(z)}{dzd\Omega}.
\end{equation}
We can consider that the measured shear and richness are correlated, such that they can be modeled as correlated random variables via $P(\Delta\Sigma, \lambda | m)$. Then, the average excess surface density profile within the $i$-th redshift bin and the $j$-th richness bin is given by
\begin{align}
    \langle \Delta\Sigma \rangle 
    \propto&  \int\limits_{z_i}^{z_{i+1}} dz \int\limits_{m_{\rm lim}}^{m_{\rm max}} dm N(m,z)\int\limits_{\lambda_{j}}^{\lambda_{j+1}} d\lambda P(\lambda|m,z)\langle \Delta\Sigma |m, \lambda, z\rangle.
\end{align}
From \citet{Wu2022selectionbiases}, we consider $P(\Delta\Sigma, \lambda | m)$ as a multivariate Gaussian distribution, with means $\langle \Delta\Sigma | m, z \rangle$ and $\langle \ln \lambda | m, z \rangle$, and with covariance matrix
\begin{equation}
\mathrm{C}= \begin{pmatrix}
\sigma_{\ln \lambda}^2 & r\sigma_{\ln \lambda}\sigma_{\Delta\Sigma} \\
r\sigma_{\ln \lambda}\sigma_{\Delta\Sigma}  & \sigma_{\Delta\Sigma}^2\\
\end{pmatrix}.
\end{equation}
From \citet{Wu2019covarianceDeltaSigma} (see their Eq. 8), the conditional mean of $\Delta\Sigma$ is given by
\begin{equation}
    \langle \Delta\Sigma |m, \lambda, z \rangle = \langle \Delta\Sigma |m, z \rangle + r\frac{\sigma_{\Delta\Sigma}}{\sigma_{\ln \lambda}}[\ln \lambda - \langle \ln \lambda|m, z\rangle].
    \label{eq:ds_selection_bias_m_z}
\end{equation}
From \citet{Farahi2022kllr}, we can approximate the halo mass function to the form
\begin{equation}
N(m,z) \approx A(z) \exp\left[- \beta_1 \ln m  \right].
\end{equation} 
By using the simplified mean mass-richness relation:
\begin{equation}
    \langle \ln \lambda|m, z\rangle = \pi(z) + \bar{\mu}_m \ln m,
\end{equation}
where $\bar{\mu}_m$ is linked to the $\mu_m$ in \eqreff{eq:richness_mass} by $\bar{\mu}_m = \mu_m/\ln(10)$, and the results in \cite{Evrard2014massobservable} (see their Eq. 3 and 4), we have that
\begin{equation}
    \ln \lambda - \langle \ln \lambda|m, z\rangle = \frac{\sigma_{\ln\lambda}^2}{\bar{\mu}_m}\left(\frac{\langle\ln M| \lambda,z\rangle}{\sigma_{\ln M}^2} + \beta_1\right) - \bar{\mu}_m \ln M.
\end{equation}
Using that $\sigma_{\ln \lambda} = \bar{\mu}_m \sigma_{\ln M}$ \citep{Evrard2014massobservable}, we have
\begin{equation}
    \ln \lambda - \langle \ln \lambda|m, z\rangle =  \frac{\sigma_{\ln\lambda}^2}{\bar{\mu}_m}\beta_1 + \bar{\mu}_m\langle\ln M| \lambda,z\rangle- \bar{\mu}_m \ln M.
    \label{eq:dlnlambda}
\end{equation}
So, by combining \eqreff{eq:ds_selection_bias_m_z} and \eqreff{eq:dlnlambda}, we get
\begin{align}
    \langle \Delta\Sigma |m, \lambda, z \rangle &= \langle \Delta\Sigma |m, z \rangle+ \frac{\beta_1}{\bar{\mu}_m} \mathrm{Cov}(\Delta\Sigma,\ln\lambda|m,z) \\+& \frac{\bar{\mu}_m}{\sigma_{\ln\lambda}^2}\mathrm{Cov}(\Delta\Sigma,\ln\lambda|m,z)\left(\langle\ln M| \lambda,z\rangle- \ln M\right).
\end{align}
From above, we have that the stacked shear decomposes as $\langle \Delta\Sigma \rangle = \langle \Delta\Sigma \rangle_1 + \langle \Delta\Sigma \rangle_2 + \langle \Delta\Sigma \rangle_3$. The first term is given by
\begin{equation}
    \langle \Delta\Sigma \rangle_1 = \int\limits_{z_i}^{z_{i+1}} dz \int\limits_{m_{\rm lim}}^{m_{\rm max}} dm N(m, z) \int\limits_{\lambda_{j}}^{\lambda_{j+1}} d\lambda P(\lambda|m,z)\langle \Delta\Sigma |m, z \rangle.
    \label{eq:ds_1_naive}
\end{equation}
The second term is given by
\begin{align}
    \langle \Delta\Sigma \rangle_2 &= \frac{\beta_1}{\bar{\mu}_m}\int\limits_{z_i}^{z_{i+1}} dz\int\limits_{m_{\rm lim}}^{m_{\rm max}} dm N(m,z) \int\limits_{\lambda_{j}}^{\lambda_{j+1}} d\lambda   P(\lambda|m,z)\mathrm{Cov}(\Delta\Sigma,\ln\lambda|m,z)\\
    &\approx \frac{\beta_1}{\bar{\mu}_m}\int\limits_{z_i}^{z_{i+1}} dz \int\limits_{\lambda_{j}}^{\lambda_{j+1}} d\lambda P(\lambda, z) \mathrm{Cov}(\Delta\Sigma,\ln\lambda|\langle m|\lambda,z\rangle,z)\\
    &\approx \frac{\beta_1}{\bar{\mu}_m}\langle\mathrm{Cov}(\Delta\Sigma, \ln\lambda|m,z)\rangle_{ij}
\end{align}
where the second equation is obtained by considering that $\mathrm{Cov}(\Delta\Sigma, \ln\lambda | m, z)$ evolves linearly with mass $m$. The term $\langle \Delta\Sigma \rangle_3$ is negligible and not considered in this analysis \citep[see e.g.][]{Wu2022selectionbiases, Zhang2023cov}. Then, from the above equation, we see that a possible shear-richness correlation (e.g., $r \neq 0$, or equivalently $\mathrm{Cov}(\Delta\Sigma, \ln\lambda | m, z) \neq 0$), arising from selection bias in cluster finder algorithms, may lead to a misinterpretation of the stacked cluster profile as a naive average as given in \eqreff{eq:ds_1_naive}.
\onecolumn
\section{Constraints of the scaling relation: summary table}
\label{app:table_scale_rel}
\begin{figure*}[h!]
    \centering
    \includegraphics[width=0.85\textwidth]{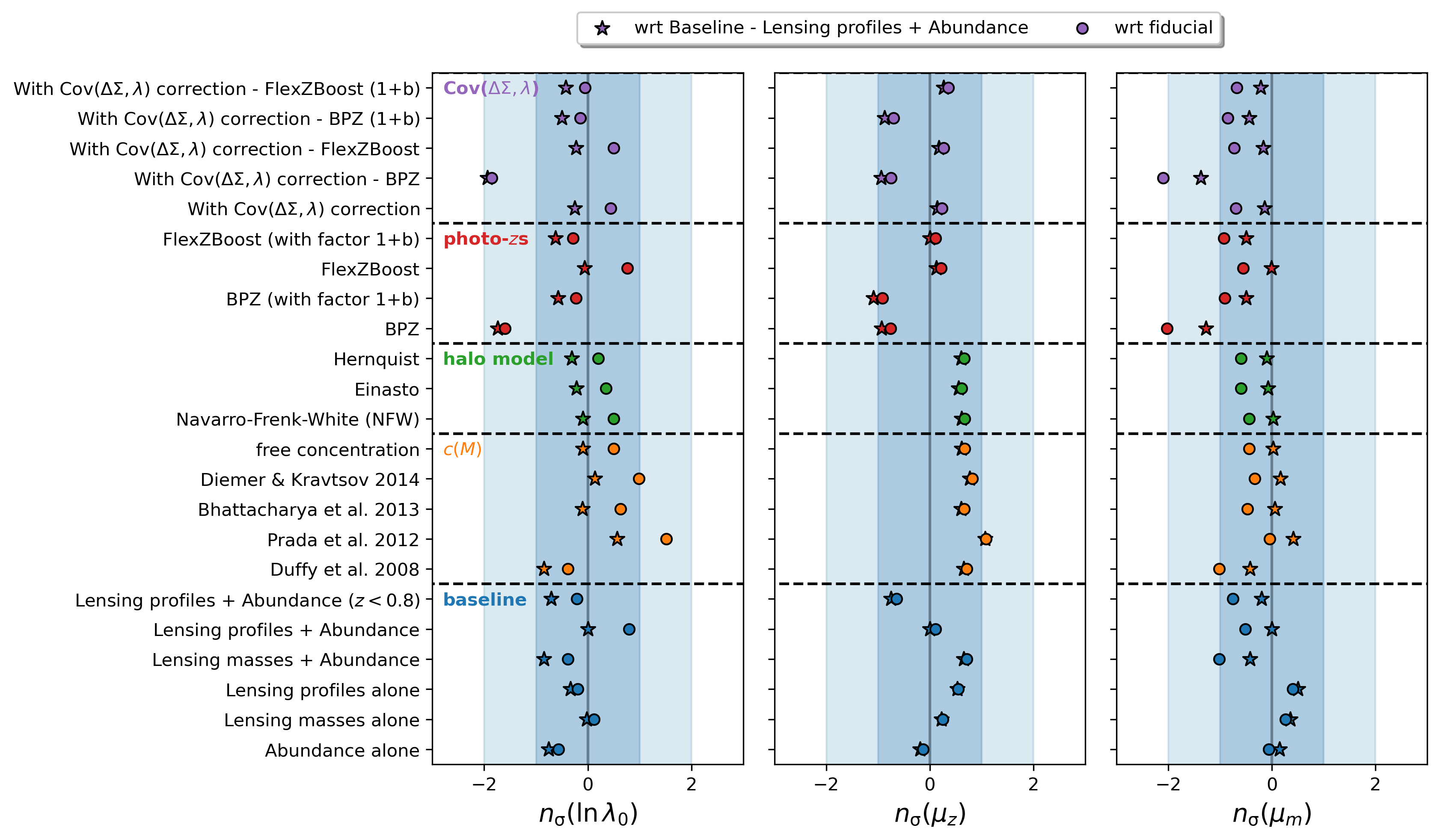}
    \caption{Tension in numbers of $\sigma$ between the different weak lensing/count analyses and the fiducial constraints (circle dots) for the parameters listed in \figreff{fig:summary}, and between the different analyses and the baseline weak lensing/count analysis (star dots).}
    \label{fig:summary_tension}
\end{figure*}

We list in \tabreff{tab:params_WLN_mass_richness} the values of the scaling relation parameters obtained from cluster count and/or cluster lensing in \secreff{sec:baseline_analysis} (baseline analysis), \secreff{sec:impact_cM_relation} (impact of $c(M)$ relation), \secreff{sec:impact_halo_model} (impact of density profile), \secreff{sec:impact_photoz} (impact of source galaxy photometric redshifts), and \secreff{sec:impact_shear_richness} (impact of shear-richness covariance). The \figreff{fig:summary_tension} displays the tension metric for the different study cases with respect to (i) the fiducial constraints and (ii) the count+weak lensing profiles analysis, labeled as 'baseline' in \secreff{sec:baseline_analysis}.
\begin{table*}[h!]
\begin{center}
\caption{Best fit parameters of the cluster scaling relation in \eqreff{eq:p_lambda_m}.}
\label{tab:params_WLN_mass_richness}
\resizebox{0.85\textwidth}{!}{%
\begin{tabular}{ ccccccc } 
  Parameters & $\ln\lambda_0$ & $\mu_z$ & $\mu_m$ & $\sigma_{\ln\lambda_0}$ & $\sigma_z$ & $\sigma_m$\\
  \hline
  \multicolumn{7}{c}{\textbf{Baseline analysis, \secreff{sec:baseline_analysis}}} \\
  \hline
Abundance alone & $3.27 \pm 0.14$  & $0.04 \pm 0.19$  & $2.22 \pm 0.20$  & $0.53 \pm 0.11$  & $0.06 \pm 0.25$  & $-0.11 \pm 0.14$ \\
Lensing masses alone & $3.37 \pm 0.20$  & $0.18 \pm 0.47$  & $2.34 \pm 0.44$  & $0.52 \pm 0.10$  & $-0.03 \pm 0.32$  & $-0.00 \pm 0.15$ \\
Lensing profiles alone & $3.31 \pm 0.19$  & $0.36 \pm 0.54$  & $2.41 \pm 0.45$  & $0.54 \pm 0.08$  & $-0.05 \pm 0.33$  & $-0.03 \pm 0.15$ \\
Lensing masses + Abundance & $3.33 \pm 0.04$  & $0.14 \pm 0.07$  & $2.14 \pm 0.06$  & $0.58 \pm 0.02$  & $0.06 \pm 0.11$  & $0.13 \pm 0.07$ \\
Lensing profiles + Abundance & $3.37 \pm 0.03$  & $0.08 \pm 0.07$  & $2.18 \pm 0.07$  & $0.53 \pm 0.03$  & $0.20 \pm 0.11$  & $0.14 \pm 0.05$ \\
Lensing profiles + Abundance ($z < 0.8$) & $3.34 \pm 0.04$  & $-0.04 \pm 0.13$  & $2.16 \pm 0.07$  & $0.54 \pm 0.02$  & $0.17 \pm 0.15$  & $0.08 \pm 0.08$ \\
  \hline
  \multicolumn{7}{c}{\textbf{Impact of the concentration-mass relation, \secreff{sec:impact_cM_relation}}} \\
  \hline
  Duffy et al. 2008 & $3.33 \pm 0.04$  & $0.14 \pm 0.07$  & $2.14 \pm 0.06$  & $0.58 \pm 0.02$  & $0.06 \pm 0.11$  & $0.13 \pm 0.07$ \\
Prada et al. 2012 & $3.40 \pm 0.04$  & $0.17 \pm 0.06$  & $2.22 \pm 0.07$  & $0.52 \pm 0.03$  & $0.02 \pm 0.10$  & $0.16 \pm 0.05$ \\
Bhattacharya et al. 2013 & $3.37 \pm 0.04$  & $0.14 \pm 0.08$  & $2.19 \pm 0.06$  & $0.55 \pm 0.03$  & $0.07 \pm 0.12$  & $0.14 \pm 0.05$ \\
Diemer Kravtsov 2014 & $3.38 \pm 0.03$  & $0.15 \pm 0.07$  & $2.20 \pm 0.07$  & $0.54 \pm 0.03$  & $0.06 \pm 0.12$  & $0.15 \pm 0.05$ \\
free concentration & $3.37 \pm 0.04$  & $0.14 \pm 0.08$  & $2.19 \pm 0.08$  & $0.54 \pm 0.03$  & $0.08 \pm 0.14$  & $0.15 \pm 0.06$ \\
  \hline
  \multicolumn{7}{c}{\textbf{Impact of the dark matter density profile, \secreff{sec:impact_halo_model}}} \\
  \hline
Navarro-Frenk-White (NFW) & $3.37 \pm 0.04$  & $0.14 \pm 0.08$  & $2.19 \pm 0.08$  & $0.54 \pm 0.03$  & $0.08 \pm 0.14$  & $0.15 \pm 0.06$ \\
Einasto & $3.36 \pm 0.04$  & $0.13 \pm 0.08$  & $2.18 \pm 0.07$  & $0.55 \pm 0.03$  & $0.08 \pm 0.14$  & $0.14 \pm 0.06$ \\
Hernquist & $3.36 \pm 0.05$  & $0.14 \pm 0.08$  & $2.17 \pm 0.08$  & $0.55 \pm 0.03$  & $0.08 \pm 0.13$  & $0.13 \pm 0.06$ \\
  \hline
  \multicolumn{7}{c}{\textbf{Impact of source photometric redshifts, \secreff{sec:impact_photoz}}} \\
  \hline
true source redshifts & $3.37 \pm 0.03$  & $0.08 \pm 0.07$  & $2.18 \pm 0.07$  & $0.53 \pm 0.03$  & $0.20 \pm 0.11$  & $0.14 \pm 0.05$ \\
BPZ & $3.29 \pm 0.03$  & $-0.02 \pm 0.08$  & $2.07 \pm 0.06$  & $0.59 \pm 0.02$  & $0.32 \pm 0.11$  & $0.14 \pm 0.06$ \\
BPZ (with factor 1+b) & $3.33 \pm 0.07$  & $-0.04 \pm 0.09$  & $2.12 \pm 0.11$  & $0.55 \pm 0.06$  & $0.40 \pm 0.15$  & $0.13 \pm 0.06$ \\
FlexZBoost & $3.37 \pm 0.03$  & $0.09 \pm 0.08$  & $2.18 \pm 0.06$  & $0.54 \pm 0.02$  & $0.17 \pm 0.12$  & $0.14 \pm 0.05$ \\
FlexZBoost (with factor 1+b) & $3.33 \pm 0.07$  & $0.08 \pm 0.07$  & $2.12 \pm 0.10$  & $0.57 \pm 0.06$  & $0.17 \pm 0.13$  & $0.14 \pm 0.06$ \\
  \hline
  \multicolumn{7}{c}{\textbf{Impact of source photometric redshifts, \secreff{sec:impact_shear_richness}}} \\
  \hline
Without Cov($\Delta\Sigma,\lambda$) correction & $3.37 \pm 0.03$  & $0.08 \pm 0.07$  & $2.18 \pm 0.07$  & $0.53 \pm 0.03$  & $0.20 \pm 0.12$  & $0.14 \pm 0.04$ \\
With Cov($\Delta\Sigma,\lambda$) correction & $3.36 \pm 0.03$  & $0.09 \pm 0.07$  & $2.17 \pm 0.06$  & $0.54 \pm 0.02$  & $0.17 \pm 0.12$  & $0.14 \pm 0.05$ \\
With Cov($\Delta\Sigma,\lambda$) correction - BPZ & $3.28 \pm 0.03$  & $-0.02 \pm 0.07$  & $2.05 \pm 0.07$  & $0.61 \pm 0.02$  & $0.30 \pm 0.10$  & $0.14 \pm 0.06$ \\
With Cov($\Delta\Sigma,\lambda$) correction - FlexZBoost & $3.36 \pm 0.03$  & $0.09 \pm 0.08$  & $2.17 \pm 0.06$  & $0.55 \pm 0.02$  & $0.16 \pm 0.13$  & $0.15 \pm 0.05$ \\
With Cov($\Delta\Sigma,\lambda$) correction - BPZ (1+b) & $3.34 \pm 0.07$  & $-0.01 \pm 0.08$  & $2.13 \pm 0.10$  & $0.55 \pm 0.06$  & $0.36 \pm 0.14$  & $0.14 \pm 0.06$ \\
With Cov($\Delta\Sigma,\lambda$) correction - FlexZBoost (1+b) & $3.34 \pm 0.07$  & $0.10 \pm 0.08$  & $2.16 \pm 0.09$  & $0.56 \pm 0.06$  & $0.13 \pm 0.14$  & $0.13 \pm 0.06$ 
\end{tabular}}
\end{center}
\end{table*}
\end{appendix}
\end{document}